\documentclass[review,3p,12pt]{elsarticle}

\usepackage{lineno}

\usepackage{enumitem}
\usepackage{lipsum}
\usepackage[utf8]{inputenc}
\usepackage[T1]{fontenc}
\usepackage{amsmath}
\usepackage{amsfonts}
\usepackage{amssymb}
\usepackage{graphicx} %pour afficher des images
\usepackage{float}	%pour forcer le placement des images
\usepackage{geometry} %pour la modification des marges
\usepackage{fancyhdr} %pour modification des pieds de page
\usepackage{longtable}
\usepackage{color} %pour les textes en gris
\usepackage{listings}
\usepackage{subfigure}
\usepackage[pdfborder={0 0 0}]{hyperref} %pour que les références soient des liens hypertextes
\usepackage{url}
\usepackage{eurosym}
\usepackage{icomma} %pour enlever l'espace après une virgule dans les nombres décimaux
\usepackage{multirow,array}
\usepackage{color}
\usepackage{textcomp}
\usepackage{gensymb}
\usepackage{lscape}
\usepackage{bm}
\usepackage{cases}

\let\oldequation\equation
\let\oldendequation\endequation

\renewenvironment{equation}
  {\linenomathNonumbers\oldequation}
  {\oldendequation\endlinenomath}

\journal{J. Mech. Phys. Sol.}

\begin{document}

\begin{frontmatter}
 
\title{Propagation of a plane-strain hydraulic fracture accounting for a rough cohesive zone}

\author[a]{Dong Liu}
\author[a]{Brice Lecampion\corref{brice.lecampion@epfl.ch}}

\address[a]{Geo-Energy Laboratory-Gaznat Chair on Geo-Energy, Ecole Polytechique Fédérale de Lausanne, EPFL-ENAC-IIC-GEL, Lausanne CH-1015, Switzerland}

\begin{abstract}
The quasi-brittle nature of rocks challenges the basic assumptions of linear hydraulic fracture mechanics (LHFM): namely, linear elastic fracture mechanics and smooth parallel plates lubrication fluid flow inside the propagating fracture. We relax these hypotheses and investigate in details the growth of a plane-strain hydraulic fracture in an impermeable medium accounting for a rough cohesive zone and a fluid lag. In addition to a dimensionless toughness  and  the time-scale $t_{om}$ of coalescence of the fluid and fracture fronts governing the fracture evolution in the LHFM case, the solution now also depends on the ratio between the in-situ and material peak cohesive stress $\sigma_o/\sigma_c$ and the intensity of the flow deviation induced by aperture roughness (captured by a dimensionless power exponent). We show that the solution is appropriately described by a nucleation time-scale $t_{cm}=t_{om}\times (\sigma_o/\sigma_c)^3$, which delineates the fracture growth into three distinct stages: a nucleation phase ($t\ll t_{cm}$), an intermediate stage ($t \sim t_{cm}$) and late time ($t \gg t_{cm}$) stage where convergence toward LHFM predictions finally occurs. A highly non-linear hydro-mechanical coupling  takes place as the fluid front enters the rough cohesive zone which itself evolves during the nucleation and intermediate stages of growth. This coupling leads to significant additional viscous flow dissipation. As a result, the fracture evolution deviates from LHFM predictions with shorter fracture lengths, larger widths and  net pressures. These deviations from LHFM ultimately decrease at late times ($t \gg t_{cm}$) as the ratios of the lag and cohesive zone sizes with the fracture length both become smaller. The deviations increase with larger dimensionless toughness and larger $\sigma_o/\sigma_c$  ratio, as both have the effect of further localizing viscous dissipation near the fluid front located in the small rough cohesive zone. The convergence toward LHFM can occur at very late time  compared to the nucleation time-scale $t_{cm}$ (by a factor of hundred to thousand times) for realistic values of $\sigma_o/\sigma_c$ encountered at depth. The impact of a rough cohesive zone appears to be prominent for laboratory experiments and short in-situ injections  in quasi-brittle rocks with ultimately a larger energy demand compared to LHFM predictions. 
\end{abstract}

\end{frontmatter}

\section{Introduction}

\begin{figure}
\centering
\begin{tabular}{cc}
     \includegraphics[width=0.45 \linewidth]{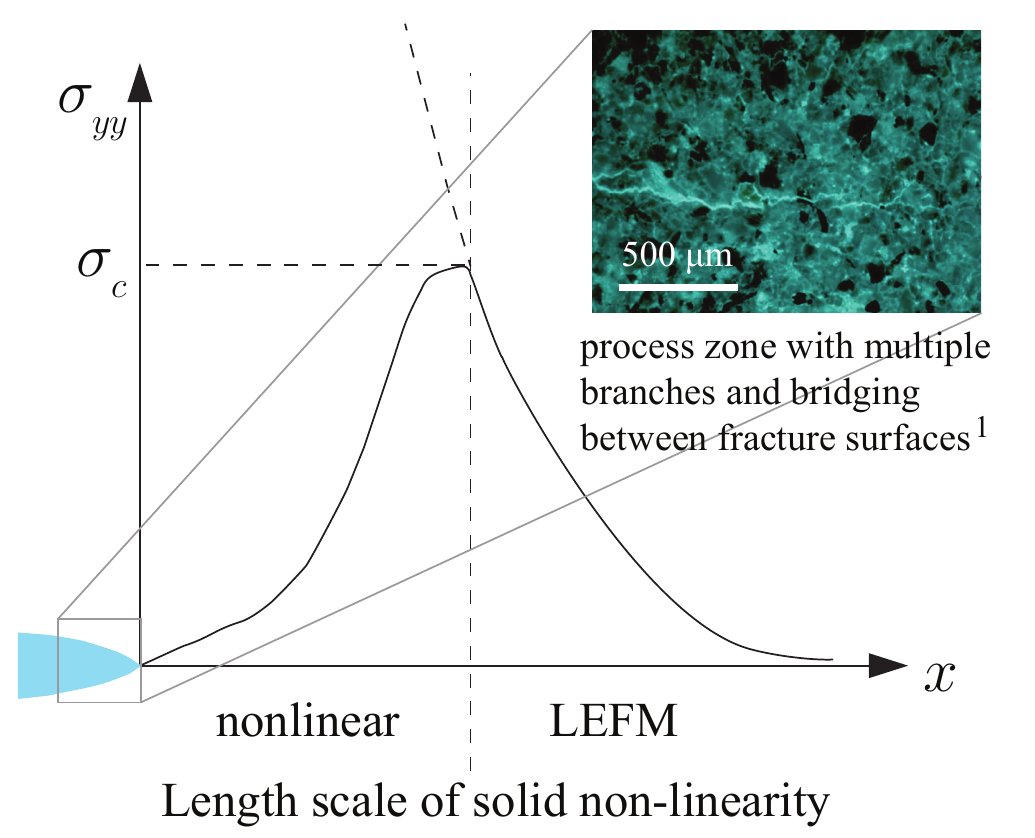} &  
    \includegraphics[width=0.45 \linewidth]{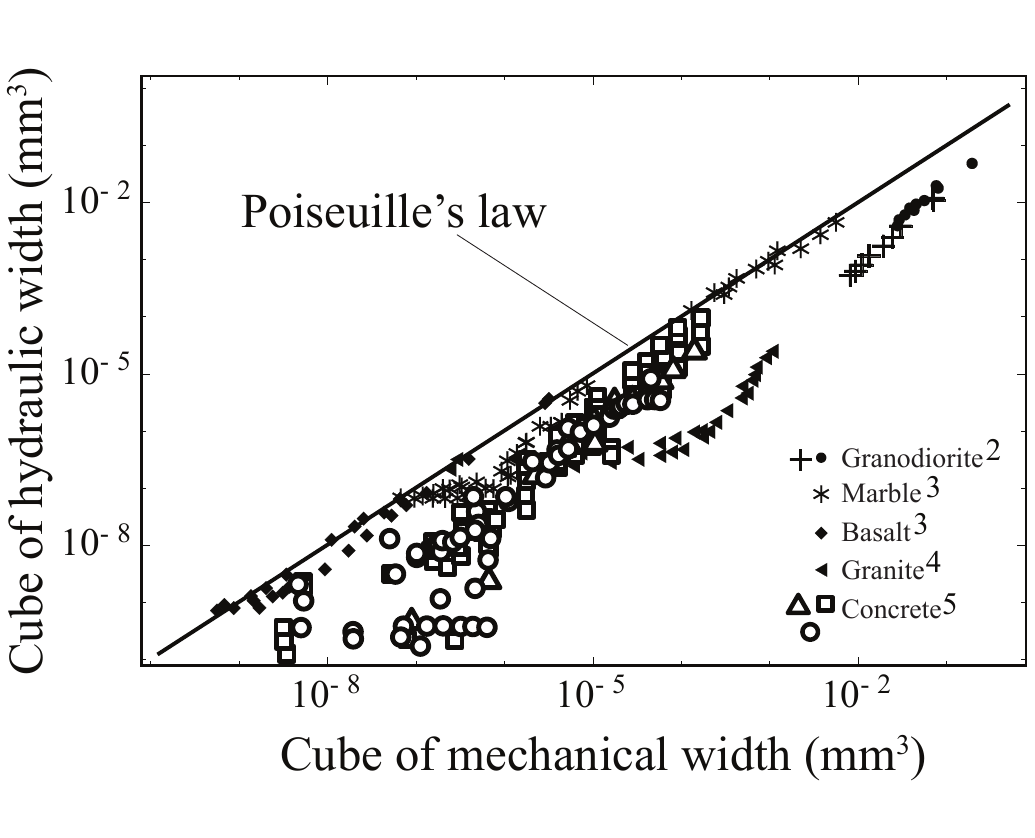} \\
     a)&b) 
\end{tabular}
\caption{Illustration of a) the length scale of solid non-linearity and b) deviated fluid flow from Poiseuille's law (cubic law). Figure b) is adapted from figure 5 in \cite{Rens1995} with additional data. The superscript indicates the source of the image and data: 1. \cite{Lhom2005}; 2. \cite{ScEv1986}; 3. \cite{WiWa1980}; 4. \cite{RaGa1985}; 5. \cite{Gara2015,BrGe1997b}.}
\label{fig:lengthscalerelation}
\end{figure}

The growth of a hydraulic fracture (HF) in an impermeable linear elastic solid is now relatively well understood, 
in particular the competition between the energy dissipated in the creation of new fracture surfaces and the one dissipated in viscous fluid flow. Such a competition leads to distinct propagation regimes depending on the main dissipative mechanism \citep{Deto2016}. Linear elastic fracture mechanics (LEFM) combined with lubrication theory (linear hydraulic fracture mechanics - LHFM for short) have successfully predicted experimental observations for the growth of a simple planar fracture in model materials such as PMMA and glass \citep{BuDe2008,LeDe2017,XiYo2017}. However, some observations on rocks at the laboratory \citep{ThHo1993,VaPa99} and field scales  \citep{Shly1985,ShWo1988} are not consistent with some indicating that linear hydraulic fracture mechanics (LHFM)  underestimates the observed fluid pressure and overestimates the fracture length. 
These observations hint toward a possibly larger energy demand compared to LHFM predictions and challenge two of its basic assumptions: i) fracture process governed by LEFM and ii) lubrication  flow between two smooth parallel surfaces resulting in Poiseuille's law. 
A non-linear process zone always exists in the vicinity of the fracture tip (Fig.~\ref{fig:lengthscalerelation}). This is especially true for quasi-brittle materials like rocks. 
The stresses are capped by a finite peak strength in the fracture process zone while the aperture roughness is no longer negligible and decreases the fracture permeability. How such non-linearities affect the solid-fluid coupling inside the fracture and as a result its growth is the main goal of this paper. We focus on the propagation of a plane-strain hydraulic fracture from nucleation to the late stages of growth where the process zone is  inherently much smaller than the fracture length.

A number of previous investigations have dealt with the relaxation of the LEFM assumption on HF growth: either using theories accounting for bulk plastic dissipation around the tip \citep{Papa1997,Papa1999,PaAt2006,SaPa2013}, or with an increasing apparent fracture toughness with length embedding different toughening mechanisms \citep{LiLG2019}, or/and adopting cohesive zone models (CZM) as a propagation criterion (see \cite{LeBu18} for review). Among these approaches, cohesive zone models are the most widely used due to their simplicity: the fracture growth is simply tracked via a cohesive traction-separation law. Studies of hydraulically driven fracture using  CZM  \citep{ChBu2009,Chen2012,Leca2012,YaLi2015} all show that the numerical solutions can be well approximated with LEFM/LHFM solutions. 
However these conclusions just follow from the fact that these simulations fall in the small-scale-yielding limit where the cohesive zone only takes up a small fraction of the whole fracture. In addition, in all these contributions, the existence of a fluid lag is  neglected as well as the effect of roughness on flow. The assumption of a negligible fluid lag is often claimed to be valid for sufficiently deep fractures (where the confining stress is large) on the basis of the LHFM results.

However, the existence of  a fluid lag is to lubrication flow what the process zone is to fracture mechanics. It removes the negative fluid pressure singularities at the fracture tip associated with suction resulting from the elasto-hydrodynamics coupling \citep{DeDe1994,GaDe2000}. In fact, the presence of a fluid lag is necessary if accounting for the presence of a cohesive zone in order to ensure that the stresses remain finite.
\cite{Rubi1993} has pioneered studies accounting for a cohesive zone and  a fluid lag by investigating the stress field around a plane-strain HF. 
The obtained results are, however, restricted to the particular case where the fluid lag is always larger than the cohesive zone.  \cite{Rubi1993} argues that the fluid lag increases with the fracture length and thus possibly influences the off-plane inelastic deformation. Recently, \cite{Gara2019} has derived the complete solution of a steadily moving semi-infinite smooth cohesive fracture with a fluid lag. The results demonstrate the strong influence of the ratio between the minimum in-situ compressive stress and the material peak cohesive stress $\sigma_o/\sigma_c$ on the near tip asymptotes. Such a semi-infinite fracture solution is obviously valid only when the process zone has fully nucleated and is smaller than the fracture length. 
These investigations assume smooth fracture surfaces in the cohesive zone (and thus Poiseuille's law). The effect of roughness on the interplay between the fluid front and cohesive zone growth still calls for further investigation.

In this paper, we investigate the growth of a finite plane-strain hydraulic fracture from nucleation to the late stage of growth
accounting for the presence of both a cohesive zone and a fluid lag. We also investigate  the impact of a  decreased hydraulic conductivity in the rough cohesive zone using existing phenomenological approximations. 
After a description of the model, we discuss the overall structure of the solution thanks to a scaling analysis. We then explore the coupled effect of the fluid lag, cohesive zone and roughness numerically using a specifically developed numerical scheme. We then discuss  implications for the HF growth both at the laboratory and field scales.  

%%%%%%%%%%%%%%%%%%%%%%%%%%%%%%%%%%%%%%%%%%
\section{Problem Formulation}
%%%%%%%%%%%%%%%%%%%%%%%%%%%%%%%%%%%%%%%%%%

\begin{figure}
\centering
\begin{tabular}{cc}
\multicolumn{2}{c}{
\includegraphics[width=0.9 \linewidth]{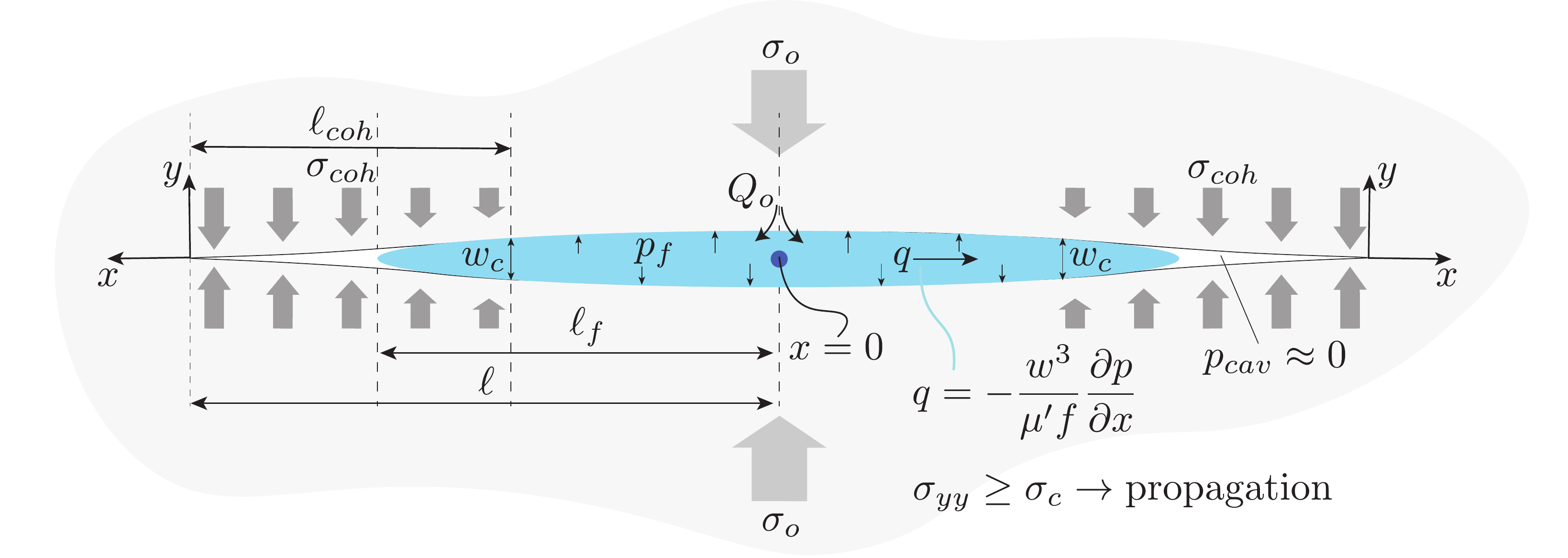} 
}\\
\multicolumn{2}{c}{a)} \\
\includegraphics[width=0.45 \linewidth]{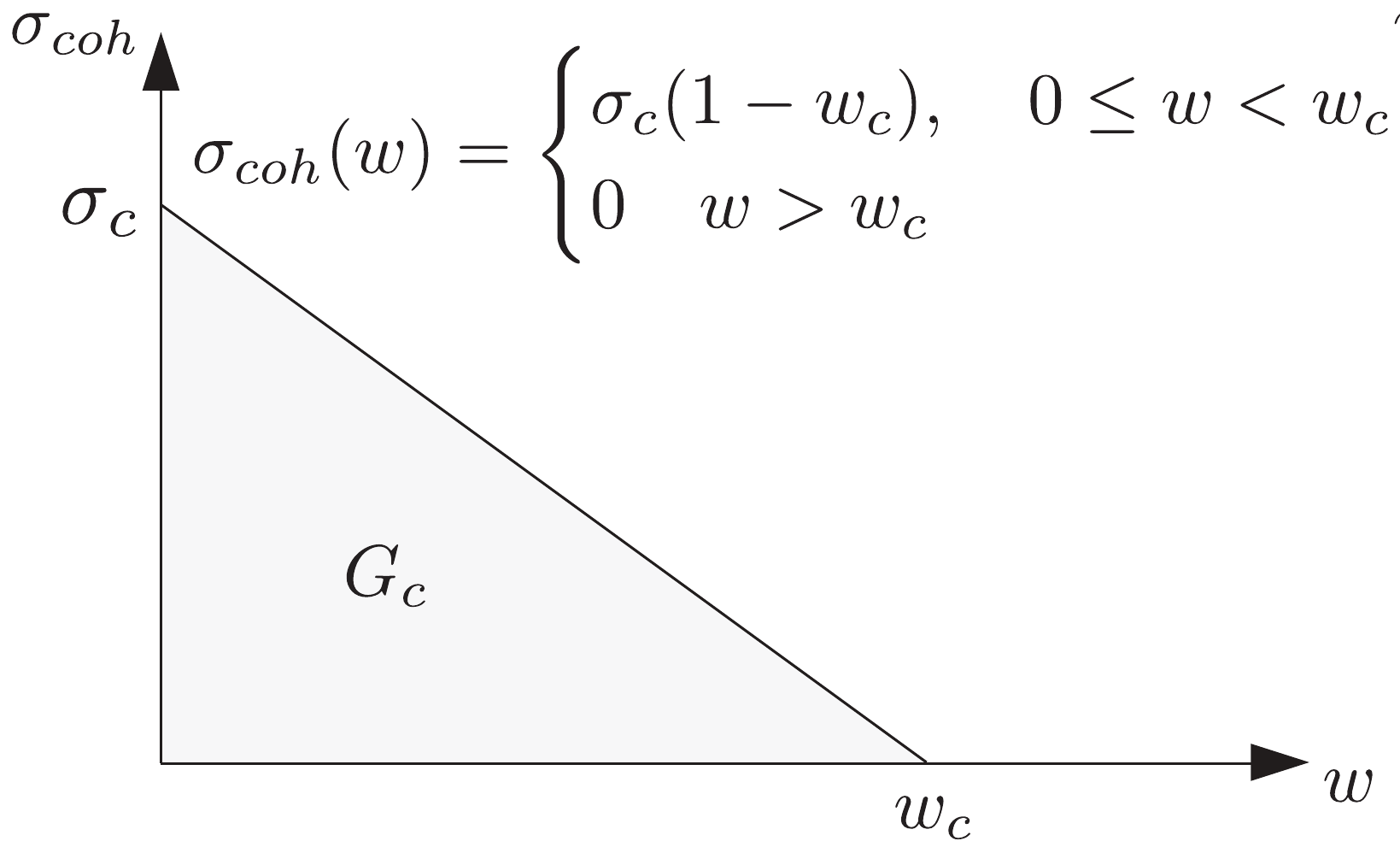}&
\includegraphics[width=0.45 \linewidth]{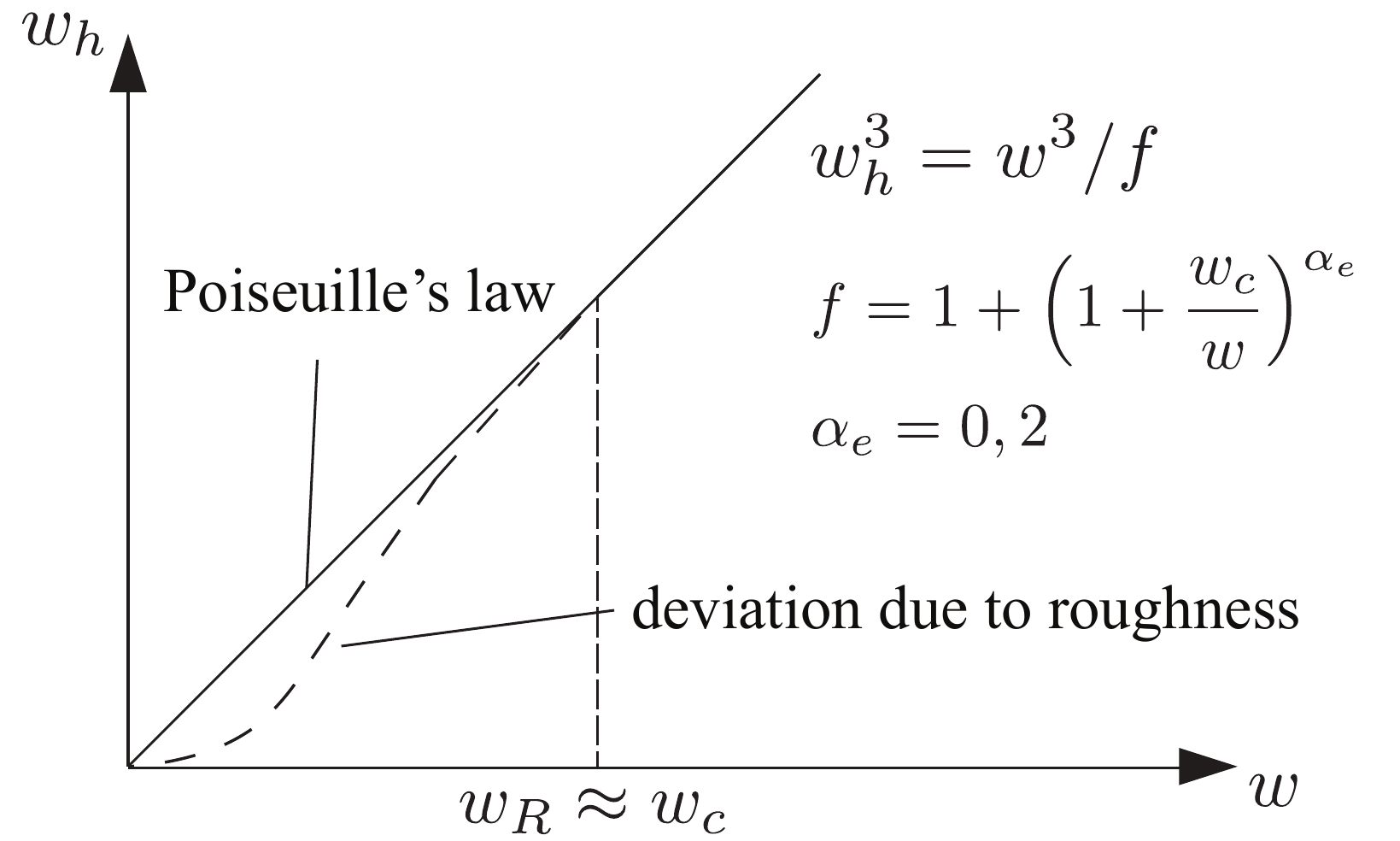}\\
b) & c)\\ 
\end{tabular}
\caption{Illustration of a) a propagating plane-strain hydraulic fracture accounting for a cohesive zone and a fluid lag, b) a linear softening cohesive zone model, and c) roughness-induced deviation of the fluid flow inside the cohesive zone.}
\label{fig:problemstatement}
\end{figure}

We consider a plane-strain hydraulic fracture of half-length $\ell$ propagating in an infinite homogeneous impermeable quasi-brittle isotropic medium (Fig.~\ref{fig:problemstatement}). We denote $\sigma_o$ as the minimum in-situ compressive stress acting normal to the fracture plane. 
The fracture growth occurs in pure tensile mode and is driven by the injection of an incompressible Newtonian fluid at a constant rate $Q_o$ in the fracture center. 
We account for both the existence of a cohesive zone (of length $\ell_{coh}$) and a fluid-less cavity (of length $\ell-\ell_f$) near the tips of the propagating fracture as described in Fig.~\ref{fig:problemstatement}.

\subsection{Solid mechanics}
    
\subsubsection{Cohesive zone model}
We adopt for simplicity a linear-softening cohesive zone model to simulate the fracture process zone, where cohesive traction decreases linearly at the tip from the peak cohesive stress $\sigma_c$ to zero at a critical aperture $w_c$, as illustrated in Fig.~\ref{fig:problemstatement}. Such a traction separation law can be simply written as:
\begin{equation}
\begin{aligned}
\sigma_{coh}(w)= 
\begin{cases}
\sigma_c(1-w_c) \quad 0 \leq w<w_c  \\
0 \quad w>w_c
\end{cases}
\end{aligned}
\label{eq:CZMlaw}
\end{equation}
where $\sigma_c$ is the material peak strength (the maximum cohesive traction). The length of the cohesive zone $\ell_{coh}$ is given by the distance from the fracture tip where the critical opening is reached: $w(\ell_{coh})=w_c$. For such a linear weakening model, the fracture energy is given by: 
\begin{equation}
G_c=\frac{1}{2}\sigma_c w_c %=\frac{K_{Ic}^2}{E'}
\label{eq:fracEnergy}
\end{equation}
Note that in linear elastic fracture mechanics in pure mode I, the fracture energy is related to the material fracture toughness  $K_{Ic}$ by Irwin's relation for co-planar growth $G_c^{LEFM}=K_{Ic}^2/E^\prime$, where $E^\prime$ is the plane-strain elastic modulus. Equalizing the quasi-brittle fracture energy with the LEFM expression allows to define an equivalent fracture toughness $K_{Ic}=\sqrt{G_c E^\prime}$ thus allowing comparison with known results for HF growth under the LEFM assumption.

\subsubsection{Elastic deformation}
For a purely tensile plane-strain fracture, in an infinite elastic medium, the quasi-static balance of momentum reduces to the following boundary integral equation (see for example \cite{HiKe1996}): 
\begin{equation}
\frac{E^\prime}{4\pi}\int_{-\ell}^{\ell}  \frac{\partial w (x^\prime, t)}{\partial x^\prime}  \frac{\text{d} x^\prime}{x-x^\prime}  =p_f(x, t)-\sigma_o-\sigma_{coh}(w(x, t)), \quad x, x^\prime  \in[-\ell, \ell]
\end{equation}
where $E^\prime=E/(1-\nu^2)$ is the plane-strain modulus, $\nu$ the Poisson's ratio of the material. In view of the problem symmetry, the previous integral equation can be conveniently written for on one-wing of the fracture:
\begin{equation}
 \frac{E^\prime}{4\pi}\int_0^{\ell}  \left(\frac{1}{x-x^\prime} -\frac{1}{x^\prime+x}\right)\frac{\partial w (x^\prime, t)}{\partial x^\prime} \text{d}x^\prime=p_f (x, t)-\sigma_o-\sigma_{coh}(w (x, t)), \quad x, x^\prime  \in[0, \ell]
 \label{eq:Elasticity}
\end{equation}
Due to the presence of cohesive forces and the traction separation law, this boundary integral equation is non-linear. 

Using a cohesive zone model, the fracture advance $\ell(t)$ is based on the stress component  $\sigma_{yy}$ perpendicular to the fracture plane  ahead of the current fracture tip. In other words, the fracture  propagates when 
\begin{equation}
\sigma_{yy}(x=\ell)=\sigma_c
\end{equation}
It is worth pointing out that at any given time, the cohesive forces cancel the stress singularity at the fracture tip that would be otherwise present. The stress intensity factor $K_I$ must thus be zero at all times. For a pure mode I crack, the stress intensity factor is obtained via the weight function approach directly from the profile of the net loading \citep{Buec1970,Rice1972}:
\begin{equation}
K_I=\frac{2 \sqrt{\ell}}{\sqrt{\pi}} \int_{0}^{\ell} \frac{p_f(x, t)-\sigma_o-\sigma_{coh}(w (x, t))}{(\ell^2-x^2)^{1/2}}\text{ d}x =0
\label{eq:zerosingularity}
\end{equation}   
The requirement $K_I=0$ can be altenatively used as a propagation condition, or checked a posteriori as an error estimate.

\subsection{Lubrication flow in a rough tensile fracture}

Under the lubrication approximation, for an incompressible fluid and an impermeable medium (negligible leak-off), 
the fluid mass conservation in the deformable fracture reduces to
\begin{equation}
\frac{\partial w}{\partial t}+\frac{\partial q}{\partial x}=0 \text{ in the fluid filled part }  x \in [0,\ell_f]
\label{eq:continuity} % should be continuity
\end{equation}
where $q(x, t)$ is the local fluid flux inside the fracture and $\ell_f(t)$ denotes the current fluid front position.

As the aperture is small near the tip and especially in the cohesive zone, it can not be considered as much larger than its small scale spatial variation - i.e. its roughness. The rough surfaces in possibly partial contact in the cohesive zone results in a decrease of the hydraulic transmissivity of the fracture compared to the cubic law. This has been observed in a large number of flow experiments in joints under different normal stress 
(Fig.~\ref{fig:lengthscalerelation}). 
A number of empirical approximations have been put forward in literature to describe such a deviation from the cubic Poiseuille's law. A typical approach consists in introducing a friction/correction factor $f$ in Poiseuille's law relating fluid flux to the pressure gradient:
\begin{equation}
q= -\frac{w^3}{\mu^\prime f}\frac{\partial p_f}{\partial x }, \qquad 0<x<\ell_f, \quad f=1+\alpha_c\times\left(\frac{w_R}{w}\right)^{\alpha_e}
\label{eq:frictionfactor}
\end{equation}
where $\mu^\prime=12\mu$ is the effective fluid viscosity. $w_R$ is a critical opening below which the fluid flow deviates from the cubic law. $\alpha_c$ and $\alpha_e$ are two material-dependent coefficients.
Table~\ref{tab:frictionfactor} lists experimentally derived values of $\alpha_c$ and $\alpha_e$ for fractures in different materials. They are closely related to the fractal properties of the self-affine rough fracture surfaces \citep{TaAu2010,JiDo2017}.
Interestingly, these roughness properties are also related to the size of the process zone, above and below which the off-plane height variation may present different roughness exponents \citep{MoMo2005,BoPo2006,PoAu2007,MoBo2008}. Moreover, 
a process zone length scale can be extracted from the spatial correlations of the slopes of a rough fracture surface \citep{VePo2015}. 
Fracture roughness therefore appears to correlate with both the process zone $w_c$ and the fluid flow deviation $w_R$ width scales. 
On the account that $w_R, w_c \sim 1-100 \mu$m in most rocks \citep{Rens1995,Gara2015,Gara2019}, we assume $w_R \approx w_c$ in what follows. %From Eq.~(\ref{eq:frictionfactor}), the misfit of $w_R$ due to the this approximation $w_R\approx w_c$ would result in a change in the choice of $\alpha_c$. 
Interested in the general effect of the interplay between the cohesive zone and the roughness induced flow on the fracture growth, we further simplify the friction factor $f$ by assuming that $\alpha_c=1$. The friction factor therefore reduces to
\begin{equation}
f=1+\left(\frac{w_c}{w}\right)^{\alpha_e} %\alpha_c=1, w_R \approx w_c \rightarrow 
\label{eq:cohdev}
\end{equation}
where $\alpha_e=0$ in the smooth fracture limit and $\alpha_e=2$ for a rough fracture. The resulting deviation between mechanical and hydraulic aperture for such a simplified fluid flow deviation model is illustrated in Fig.~\ref{fig:problemstatement}.

\begin{table}
\centering
\begin{tabular}{c|c|c|c}
\hline
Reference & $\alpha_e$ & $\alpha_c$ & Materials\\
\hline
\cite{Lomi1951} &  1.5 & 6.0 & Sand-coated glass\\
\cite{Loui1969} & 1.5 & 8.8 & Concrete \\
\cite{Rens1995} & 2 & 1.5 & Basalt, granite, granodiorite, marble, quartzite\\
\cite{ZiBo1996} & 2 & 1.5 & Granite \\%and expoxied replicas from fractures in granite\\
%Qudros & 1.5 & 20.5 & \\
\cite{XiLi2011} & 1 & 1& Plaster surfaces copied from granite and sandstone\\
\cite{Gara2015} & 1 & 1 & Concrete\\
%RaHo, 2011 \cite{RaHo2011} & 1 & 2.25 & Numerical generation\\
%AmIl, 1994 \cite{AmIl1994} & 1.2 & 0.6 & Numerical generation\\
\cite{XiGa2015} & 2 & 5.65 & Sandstone\\
\cite{ZhNe2015} & 1.12 & $10^{-3}$ & Granite, limestone\\
\hline
\end{tabular}
\caption{Different empirical models suggested in literature for the friction factor (Eq.~(\ref{eq:frictionfactor})).}
\label{tab:frictionfactor}
\end{table}

 \subsection{Boundary and initial conditions}
The fluid is injected  at the fracture center at a constant injection rate $Q_o$ (in $m^2/s$ under plane-strain conditions), such that the flow entering one-wing of the fracture is:
\begin{equation}
q(x=0^+, t) = Q_o/2
\end{equation}
which can be alternatively be accounted by the global fluid volume balance, integrating the continuity equation (\ref{eq:continuity}) for the fluid:
\begin{equation}
2 \int_0^{\ell_f(t)} w(x, t) \text{d}x =Q_o t 
\end{equation}
In the fluid lag near the fracture tip, the fluid is vaporized and its pressure is equal to the cavitation pressure $p_{cav}$, which is typically much smaller than the liquid pressure $p_f$ in the fluid filled part and the in-situ confining stress $\sigma_o$. We thus have the following pressure boundary condition in the lag:
\begin{equation}
    p_f(x, t)=p_{cav}\approx 0, \quad x \in [\ell_f(t), \ell(t)]
\end{equation}
The fluid front velocity $\dot{\ell}_f$ is equal to the mean fluid velocity $q/w$ at that fluid front location $x=\ell_f$ (Stefan condition):
\begin{equation}
    \dot{\ell}_f = -\left.\frac{w^2}{\mu^\prime f(w)} \frac{\partial p_f}{\partial x}\right|_{x=\ell_f} 
\end{equation}
The fracture opening is zero at the fracture tip taken as the beginning of the cohesive zone: 
\begin{equation}
  w(x=\ell, t)=0
  \label{eq:BCwidthtip}
\end{equation} 

\paragraph{Initial conditions}

We model the nucleation process, and the coupled developments of the cohesive zone and the fluid lag. We start from a negligibly small fracture in which cohesive forces have not completely vanishes: the fracture length equals the cohesive zone length initially. Upon the start of injection, this initially static flaw is fully filled with fluid at a pressure slightly larger than the in-situ stress $\sigma_o$.

\subsection{Energy balance}
The energy balance for a propagating hydraulic fracture can be constructed by combining two separate energy balance equations, one for the viscous fluid flow and the other one for the quasi-brittle medium deformation of an advancing crack \citep{LeDe2007}. The external power $P_e=Q_o p_{f0}$ (where $p_{f0}=p_f(x=0, t)$ is the fluid pressure at the inlet)  provided by the injecting fluid is balanced by five terms: 
\begin{itemize}
    \item the rate of work done to overcome the in-situ confining stress: $\dot{W}_o=Q_o \sigma_o$
    \item the rate of change of the elastic energy stored in the solid $\dot{W}_e$:
    \begin{equation}
        %D_e
        \dot{W}_e=\int_0^{\ell_f} p \frac{\partial w}{\partial t} \text{d}x+\int_0^{\ell_f}  w \frac{\partial p}{\partial t} \text{d}x- \sigma_o \int_{\ell_f}^{\ell}  \frac{\partial w}{\partial t}\text{d}x
    \end{equation}
  \item   a power associated with the rate of the change of the fluid lag cavity volume times the in-situ far-field stress  $\dot{W}_l$:
    \begin{equation}
        %D_c
        \dot{W}_l=2\sigma_o \frac{\text{d}}{\text{d}t} \int_{\ell_f}^{\ell} w \text{d}x
    \end{equation}
    \item the viscous dissipation rate in the fluid filled region of the fracture $D_v$:
    \begin{equation}
        D_v=-2 \int_0^{\ell_f}  q \frac{\partial p}{\partial x} \text{d}x
    \end{equation}
    \item  the energy rate  associated with the debonding of cohesive forces and the creation of new fracture surfaces $D_k$:
    \begin{equation}
        D_k=-\int_0^{\ell}  w \frac{\partial \sigma_{coh}}{\partial t}  \text{d}x+\int_0^{\ell} \sigma_{coh} \frac{\partial w}{\partial t}  \text{d}x.
    \end{equation}
\end{itemize}
 Accounting for the symmetry of the problem, we can define an apparent fracture energy
\begin{equation}
G_{c,app}=\frac{D_k}{2\dot{\ell}}
\label{eq:AppEnergy}
\end{equation}
In the coordinate system of the moving tip, we can rewrite Eq.~(\ref{eq:AppEnergy}) for the linear weakening cohesive zone model as follows (see more details in \ref{sec:energy}):
\begin{equation}
    G_{c,app}=\frac{1}{2}\sigma_c  w(\hat{x}=\ell_{coh})+\frac{1}{2\dot{\ell}}\sigma_c \int_0^{\ell_{coh}}\left. \frac{\partial w}{\partial t}\right|_{\hat{x}} \text{d}\hat{x}, \quad \hat{x}=\ell-x
\label{eq:GappRelation}
\end{equation}
When the fracture has already nucleated and the cohesive zone size is negligible compared to the fracture length ($\ell \gg \ell_{coh}$), the first term in Eq.~(\ref{eq:GappRelation}) equals the real fracture energy $G_c$ with $w(\hat{x}=\ell_{coh})=w_c$.
For a large fracture, where the cohesive zone is nearly constant, the second term tends to zero as the material time derivative of width is 
negligible for fracture with slow variation of velocity: more precisely, in the tip reference frame the convective derivative $\dot{\ell}\frac{\partial \cdot }{\partial \hat{x}}$ (which leads when integrated to the first term) dominates over the material time derivative $\frac{\partial \cdot }{\partial t}|_{\hat{x}}$.
 As a result, the apparent energy tends to equal to the real fracture energy $G_{c,app} \approx G_c$ at large time. However, it does not necessarily imply that the fracture width asymptote in the near tip region follows the LEFM limit. It only results from the fact that the convective derivative dominates - and as such the travelling semi-infinite fracture solution of \cite{Gara2019} applies (where different tip asymptotes emerge as function of the ratio $\sigma_o/\sigma_c$).
 However the equivalence $G_{c,app}=G_c$ does not hold when the fracture length is comparable to the cohesive zone $\ell \geq \ell_{coh}$.
 The first term increases with time until $w(\hat{x}=\ell_{coh})$ reaches the critical opening $w_c$ at nucleation while the second term results from the competition between the fracture velocity and the material rate change of the volume embedded inside the cohesive zone. As a result of this second term, the evolution of the apparent fracture energy may not be necessarily monotonic in an intermediate phase as we shall see later from our numerical simulations.

%%%%%%%%%%%%%%%%%%%%%%%%%%%%%%%%%%%%%%%%%%%%%%%%%%%%%%%%%%%
\section{Structure of the solution\label{sec:structure}}
%%%%%%%%%%%%%%%%%%%%%%%%%%%%%%%%%%%%%%%%%%%%%%%%%%%%%%%%%%%

Before investigating the problem numerically, we discuss the evolution of such a quasi-brittle HF at the light of dimensional analysis. We notably highlight the difference brought upon the existence of a process zone compared to the linear elastic fracture mechanics case \citep{GaDe2005,Gara2006,LeDe2007}.
Following previous work on hydraulic fracturing \citep{Gara2000,Deto2004}, we scale the fracture width, fluid pressure, flux, fracture length, extent of the liquid filled part of the fracture and the extent of the cohesive zone introducing corresponding  width $W$, pressure $P$ and length $L$ characteristic scales:
\begin{eqnarray}
&w(x,t)=W(t) \Omega (\xi, \mathcal{P}),\quad p_{f}(x,t)-\sigma_o =P(t) \Pi (\xi,\mathcal{P}), \quad q(x,t)=Q_o \Psi(\xi,\mathcal{P})\\
&\quad\ell(t)=L(t)\gamma (\mathcal{P}), \quad \ell_f(t)=L_f(t)\gamma_f (\mathcal{P})
%,\quad \ell_{coh}(t)=L_{coh}(t)\gamma_c (\mathcal{P})
\label{eq:scaledef}
\end{eqnarray}
 where $\xi=x/\ell$ is a dimensionless coordinate. The dimensionless variables also depend on one or more dimensionless number $\mathcal{P}$ and time. 
 Introducing such a scaling in the governing equations of the problem allows to isolate different dimensionless groups associated with the different physical mechanisms at play (elasticity, injected volume, viscosity, fracture energy) and define relevant scalings. 
    
Before going further, we briefly list the dimensionless form of the governing equations and the expression of the different dimensionless groups appearing in the governing equations (\ref{eq:CZMlaw})-(\ref{eq:BCwidthtip}). 
    \begin{itemize}
        \item The elasticity equation can be re-written as: 
\begin{equation}
    \Pi-\Sigma_{coh}(\Omega(\xi))=\mathcal{G}_e \frac{1}{4\pi}\frac{1}{\gamma}\int_0^{1} \frac{\partial \Omega}{\partial \xi} \left(\frac{1}{{\xi}-{\xi}^\prime}-\frac{1}{{\xi}+{\xi}^\prime}\right)\text{d}{\xi}^\prime, \quad 0<{\xi}, {\xi}^\prime<1
\end{equation} with $\mathcal{G}_e=\dfrac{W E^\prime}{P L}$ and the dimensionless traction-separation law as 
\begin{equation}
\Sigma_{coh}=\mathcal{G}_t\times \left(1-\frac{\Omega}{\mathcal{G}_{w}}\right), \quad \Omega<\mathcal{G}_{w}
\end{equation} with $\mathcal{G}_t=\sigma_c/P$ and $\mathcal{G}_{w}=w_c/W$.

\item The dimensionless fluid continuity and roughness corrected Poiseuille's law are better expressed by
scaling the spatial coordinate with the fluid front position - thus introducing the ratio of scales $\mathcal{G}_l=L_f/L$ -
 $\hat{\xi}=x/\ell_f = \xi \times (\gamma/\gamma_f) / \mathcal{G}_l$:
\begin{eqnarray}
&t\dfrac{\partial \Omega}{\partial t}+t \dfrac{\dot{W}}{W}\Omega + \mathcal{G}_v \dfrac{1}{\gamma_f} \dfrac{\partial \Psi}{\partial \hat{\xi}} =0  \\
&\Psi = -
\dfrac{1}{\mathcal{G}_m} \dfrac{\Omega^3}{f\times \gamma_f} \dfrac{\partial \Pi }{\partial \hat{\xi}} 
\end{eqnarray}
with $\mathcal{G}_v=\dfrac{Q_o t}{W L_f} $ related to the fracture volume, and $\mathcal{G}_m=\dfrac{\mu^\prime Q_o L_f}{P W^3}$ related to fluid viscosity, while the friction roughness correction $f$ 
can be simply re-written as 
$f=1  +\left(\mathcal{G}_{w}/\Omega\right)^{\alpha_e} $.

\item the entering flux boundary conditions becomes 
\begin{equation}
    \Psi(\xi=0^+,t) = 1/2
\end{equation}
while the dimensionless net pressure $\Pi$ in the fluid lag is 
\begin{equation}
    \Pi(\xi\le\xi_f=\ell_f/\ell) = - \mathcal{G}_o=- \frac{\sigma_o}{P}
\end{equation} 
    \end{itemize}
    
It is worth noting that for the linear weakening law the dimensionless fracture energy is simply $\mathcal{G}_c = (w_c \sigma_c) / ( 2 P W) = \dfrac{1}{2} \mathcal{G}_w \mathcal{G}_t$. In addition, in order to make the link with the LEFM scalings that use a reduced fracture toughness defined as $K^\prime = \sqrt{32/\pi} K_{Ic}$, we use the equivalent dimensionless fracture toughness: $\mathcal{G}_k= \sqrt{32/\pi}\sqrt{\mathcal{G}_e \mathcal{G}_c}=\sqrt{16/\pi}\sqrt{\mathcal{G}_e \mathcal{G}_w \mathcal{G}_t}$ in the following.
     
 The well-known scalings under the LHFM assumptions for the case of negligible lag ($\mathcal{G}_l=L_f/L=1$) are obtained by recognizing that elasticity is always important ($\mathcal{G}_e=1$), and the fact that without fluid leak-off the fracture volume equals the injected volume at all time ($\mathcal{G}_v=1$). The viscosity and  toughness scalings are then obtained by either setting $\mathcal{G}_m$ (M/viscous scaling) or  $\mathcal{G}_k$ (K/toughness scaling) to unity. Alternatively, the fluid lag dominated scaling (O-vertex) is obtained by recognizing that viscous effects are necessary for cavitation to occur ($\mathcal{G}_m=1$) and  the lag covers a significant part of the fracture such that the pressure scale is given by the in-situ stress  ($\mathcal{G}_o=1$). Similarly elasticity ($\mathcal{G}_e=1$) and fluid volume  ($\mathcal{G}_v=1$) are driving mechanisms. These well-known scalings for the different limiting propagation regimes are recalled in Table~\ref{tab:scalings}.
     
Under the assumption of linear elastic fracture mechanics, as discussed in \cite{Gara2006,LeDe2007}, a plane-strain HF evolves from an early-time solution where the fluid lag is maximum to a late solution where the fluid and fracture front coalesces (zero lag case) over a time-scale 
\begin{equation}
         t_{om} =\frac{E^{\prime 2} \mu^\prime }{\sigma_o^3}
         \label{eq:tom}
\end{equation}
This time-scale directly emerges as the time it takes for the dimensionless in-situ stress $\mathcal{G}_o$ to reach unity in the zero lag scalings.  
In addition, the solution also depends on a dimensionless toughness $\mathcal{K}_m$ (or alternatively dimensionless viscosity) independent of time. The fluid lag is the largest for small dimensionless toughness and is negligible at all time for large dimensionless toughness. The propagation can thus be illustrated via a triangular phase diagram, whose three vertices (O-M-K) corresponds to three limiting regimes. The O-vertex corresponds to the limiting case of a large lag / negligible toughness, the M-vertex corresponds to viscosity dominated propagation with a negligible fluid lag while the K-vertex corresponds to a toughness dominated propagation where viscous effects are always negligible and as a result no fluid lag exists. 
     
The introduction of a cohesive zone modifies partly this propagation diagram. One can define a cohesive zone scaling (which will be coined with the letter C) by  setting the pressure scales $P$ to the peak cohesive stress $\sigma_c$  ($\mathcal{G}_t=1$), the opening scale $W$ to the critical opening $w_c$ ($\mathcal{G}_w=1$). We then readily obtain from elasticity ($\mathcal{G}_e=1$) that the fracture characteristic length  $L$ equals the classical cohesive characteristic length scale \citep{Rice1968,HiMo1976}:
\begin{equation}
    L_{coh} = \frac{E^\prime w_c}{\sigma_c}
    \label{eq:Lcoh}
\end{equation}
Such a scaling is relevant at early time when the cohesive zone scale is of the order of the fracture length. We know from the LHFM limit that the fluid lag is also important at early time. From lubrication flow, combining fluid continuity and Poiseuille's law to obtain the Reynolds equation enables to define the corresponding fluid front scale $L_f$ as 
\begin{equation}
 L_{f,c}=w_c  \frac{\sqrt{\sigma_c t}}{\sqrt{\mu^\prime}} 
\end{equation}
(by setting the resulting dimensionless group $\mathcal{G}_v/\mathcal{G}_m$ in the Reynolds equation to one).
Another time-scale $t_{cm}$ thus emerges as the characteristic time for which the fluid front in that cohesive scaling is of the same order of magnitude than the characteristic fracture / cohesive length:
\begin{equation}
    t_{cm}=\frac{E^{\prime 2} \mu^\prime}{\sigma_c^3} = t_{om} \times \left(\frac{\sigma_o}{\sigma_c} \right)^3.
     \label{eq:tcm}
\end{equation} 
This time-scale quantifies the time required for the cohesive zone to develop
in relation to the penetration of the fluid. It is worth noting that the ratio of time-scales $t_{cm}/t_{om} $ related to the fluid lag in the cohesive (C) and LHFM (O) scalings is directly related to the ratio between the in-situ confining stress and the peak cohesive stress.

Three stages of growth can thus be delineated as function of the evolution of the cohesive zone.
\begin{itemize}
\item Stage I for early time ($t\ll t_{cm}$): the whole fracture length is embedded inside the cohesive zone.  The latter develops yet is not fully nucleated. We will refer to this stage as the nucleation stage in the following.
\item Stage II for intermediate times (of the order of $t_{cm}$): the cohesive zone has now fully nucleated and part of the fracture surfaces are completely separated without cohesion ($w>w_c$ in the central part of the fracture). The cohesive zone remains important compared to the whole fracture length and may be not yet stabilized.  We will refer to this stage as the intermediate propagation stage.
\item Stage III ($t\gg t_{cm}$): the cohesive zone now only takes up a very small fraction of the whole fracture such that the small-scale-yielding assumption becomes valid. We will refer to this stage as the late time propagation stage.
\end{itemize}

From the different scalings in Table \ref{tab:scalings}, we see that using $t_{cm}$ as a characteristic time-scale, the evolution of a HF in a quasi-brittle material depends only on i) a dimensionless toughness $\mathcal{K}_m$, ii) the ratio between the confining stress and material strength $\sigma_o/\sigma_c$ and iii) the dimensionless roughness exponent $\alpha_e$.  

\begin{figure}
\centering
\includegraphics[width=0.8\linewidth]{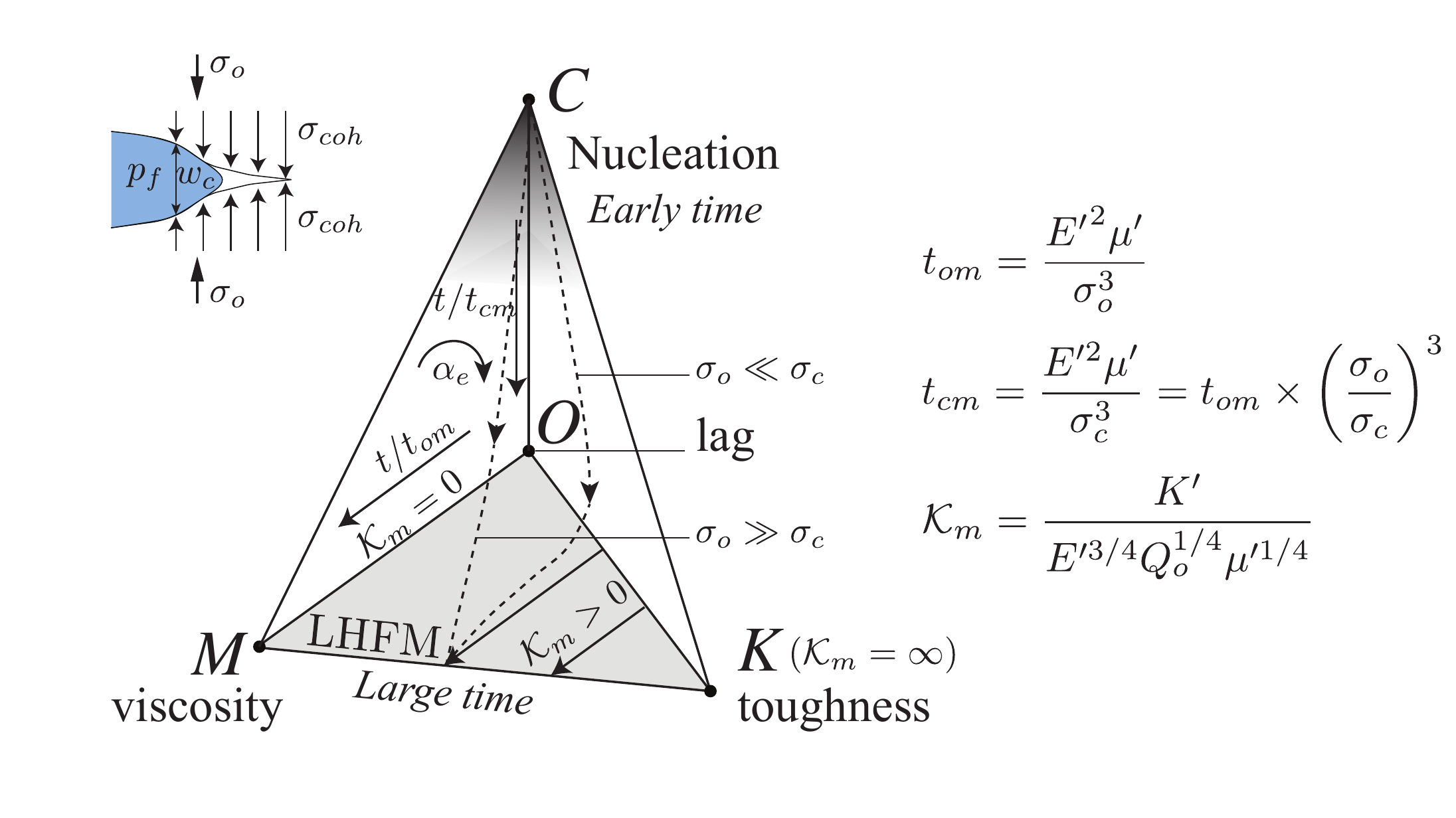}
\caption{Propagation diagram of a plane-strain hydraulic fracture with a rough cohesive fracture tip. The bottom $O-M-K$ triangle corresponds to the LHFM limit.}
\label{fig:ScalingKGD}
\end{figure}

For a quasi-brittle impermeable material, the propagation can be schematically grasped by the propagation diagram depicted on Fig.~\ref{fig:ScalingKGD}. The propagation starts in a cohesive / nucleation regime (vertex C) and ultimately ends up at large time on the M-K edge (LHFM / small-scale-yielding limit) at a point depending on the (time-independent) dimensionless toughness $\mathcal{K}_m$. How the fracture evolves from the nucleation (vertex C) stages to the large time LHFM limits is function of the ratio $\sigma_o/\sigma_c$ as well as the roughness exponent. 
When $\sigma_o\ll\sigma_c$ ($t_{cm}\ll t_{om}$), the cohesive zone develops faster than the time required for the fluid front to coalesce with the fracture front. In that case, the small-scale-yielding assumption may become valid early in conjunction with the presence of a fluid lag (O-K edge) - the fluid front will lie outside of the cohesive zone for some time.
On the other limit, for  $\sigma_o > \sigma_c$, the fluid front tends to remain inside the cohesive zone which develops slower than the fluid front progress.

How exactly, the growth of the HF is influenced by the interplay between the cohesive zone and lag evolution 
for different values of $\sigma_o/\sigma_c$, dimensionless toughness and fracture roughness intensity
will  be now investigated numerically.

\begin{table}
    \centering
    % here the idea is to express the characteristic scales dependent only on problem parameters / so that readers can see on what they depend on
    % for dimneionless numbers, we use relations K_m, \sigma_o/\sigma_c  and t/t_om t/t_cm
    \begin{tabular}{c|c|c|c|c}
    \hline
             & C & O & M & K  \\ \hline 
         $L$ & $L_{coh}=\dfrac{E^\prime w_c}{\sigma_c}$  & 
         $\dfrac{E^{\prime} Q_o^{1/2} \mu^{\prime 1/4} t^{1/4}}{\sigma_o^{5/4}}$ &
         $\dfrac{E^{\prime 1/6} Q_{o}^{1/2} t^{2/3}}{\mu^{\prime 1/6}}$  &
            $\dfrac{E^{\prime 2/3}Q_o^{2/3}t^{2/3}}{K^{\prime 2/3} }$    \\
%         $\mathcal{K}_{m}^{-2/3} L_{m}$ \\
         $P$ & $\sigma_c$ & $\sigma_o$ & $\dfrac{E^{\prime 2/3} \mu^{\prime 1/3}}{t^{1/3}}$  
        & $\dfrac{K^{\prime 4/3}}{E^{\prime 1/3}Q_o^{1/3}t^{1/3}} $
        % &  $\mathcal{K}_{m}^{4/3} P_{m}$
         \\ 
         $W$  & $w_c$  & $\dfrac{Q_o^{1/2}\mu^{\prime 1/4} t^{1/4}}{\sigma_o^{1/4}} $ &
         $\dfrac{Q_o^{1/2}\mu^{\prime 1/6} t^{1/3}}{E^{\prime 1/6}} $ 
         & $\dfrac{K^{\prime 2/3} Q_o^{1/3} t^{1/3}}{E^{\prime 2/3}} $
         %&  $\mathcal{K}_{m}^{2/3} W_{m}$ 
         \\ 
         $L_f/L$ & $ (t/t_{cm})^{1/2}$  &  $(t/t_{om})^{1/2} $ & 1 &  1 \\ \hline 
   $\mathcal{G}_m$ & $\left({16}/{\pi}\right)^2 \mathcal{K}_m^{-4} \left({t}/{t_{cm}}\right)^{1/2} $ 
   & 1 & 1 & $\mathcal{K}_m^{-4}$  \\ 
    $\mathcal{G}_k$ & $(16/\pi)^{1/2}$ & $\mathcal{K}_m=\dfrac{K^\prime}{E^{\prime 3/4}Q_o^{1/4} \mu^{\prime 1/4}}$ &$\mathcal{K}_m$ & 1 \\
    $\mathcal{G}_o$ & $\sigma_o/\sigma_c$ & 1  & $(t/t_{om})^{1/3}$ & $ (t/t_{om})^{1/3} \mathcal{K}_m^{-4/3}$  \\
        $\mathcal{G}_w$ & 1 & $ \left(\dfrac{t_{cm}}{t} \dfrac{\sigma_o}{\sigma_c}\right)^{1/4}\dfrac{\pi}{16} \mathcal{K}_m^2 $ &
    $  \left(\dfrac{t_{cm}}{t}\right)^{1/3} \dfrac{\pi}{16} \mathcal{K}_m^{2}$ & $ \left(\dfrac{t_{cm}}{t}\right)^{1/3} \dfrac{\pi}{16} \mathcal{K}_m^{4/3}  $ \\
    \hline
    \end{tabular} 
\caption{Characteristic scales and dimensionless numbers governing the evolution of a plane-strain quasi-brittle HF in the different limiting regimes: C - lag/cohesive/nucleation, O - lag/viscous/ LHFM, M - fully filled/viscous/LHFM, K - fully filled/toughness/LHFM. The evolution of the HF is also function of the dimensionless roughness exponent $\alpha_e$.
    The time-scales $t_{om}$ and $t_{cm}$ defined in Eqs~(\ref{eq:tom}), (\ref{eq:tcm}) are related as  $t_{cm}/t_{om}=(\sigma_o/\sigma_c)^3$.
    The $16/\pi$ factors appearing in the dimensionless numbers are due to the use of $K^\prime=\sqrt{32/\pi} K_{Ic}$ in the LHFM based scalings \citep{Deto2004,Gara2006} and the fact that for the linear weakening cohesive law $G_c=w_c \sigma_c /2$.}
    \label{tab:scalings}
\end{table}

% Here   it is better to display G_w (instead of G_t) which decays with time - highlight the shrinking of the cohesive zone length.

%%%%%%%%%%%%%%%%%%%%%%%%%%%%%%%%%%%%%%%%%%%%%%%%%%%%%%%%%%%%%%
\section{Numerical scheme \label{sec:numericaldifficulty}} %%%
%%%%%%%%%%%%%%%%%%%%%%%%%%%%%%%%%%%%%%%%%%%%%%%%%%%%%%%%%%%%%%

In order to decipher the interplay between the fluid front and cohesive zone, it is necessary to account for the nucleation of both the cohesive zone and the fluid lag. Previous numerical investigation using LHFM either tracks explicitly the fluid front in addition to the fracture front
 \citep{LeDe2007,ZhJe2005,GoDe2011} or  uses a cavitation algorithm introducing a fluid state variable $\theta\in [0,1]$ (1 for the liquid phase, 0 for the vapour phase)  \citep{Shen2014,MoSh2018} in a similar way than thin-film lubrication cavitation models (see for example \cite{Szer2010}).

The cavitation approach enables the spontaneous nucleation of the fluid lag but adds another variables and additional inequalities conditions ($p_f \geq 0,\, 0\leq \theta \leq 1,\, p_f (1-\theta)=0$) in each element.
The computational cost of such cavitation schemes increases significantly as quadratic programing problem needs to be solved at each time-step. 
We therefore propose here an algorithm taking advantage of both the cavitation scheme at early time (when the fluid lag nucleates from an initially fully liquid filled flaw) and a fluid-front-tracking scheme at later times. 

Our algorithm consists of the use of two successive schemes, both based on a fixed regular grid with constant mesh size. At the beginning of the simulation, we adopt an Elrod-Adams type scheme similar to the one described in \cite{MoSh2018}. This scheme automatically captures the appearance of the fluid lag in the most accurate manner \citep{LiLe2019}. In a second stage of the simulation, we use the results of the previous algorithm to initialize a scheme similar to  \cite{GoDe2011} where the fluid front position is tracked explicitly via the introduction of a filling fraction variable in the partly filled element at the lag boundary. We discretize respectively the elasticity and fluid mass conservation using a displacement discontinuity method with piece-wise constant elements and finite difference. We use an implicit time-integration scheme to solve iteratively for the fluid pressure and the associated opening. An additional outer loop solves for the time-step increment corresponding to a fixed increment of fracture length. More details are given in \ref{sec:algorithm}.

\begin{figure}
\centering
\includegraphics[width=0.5\linewidth]{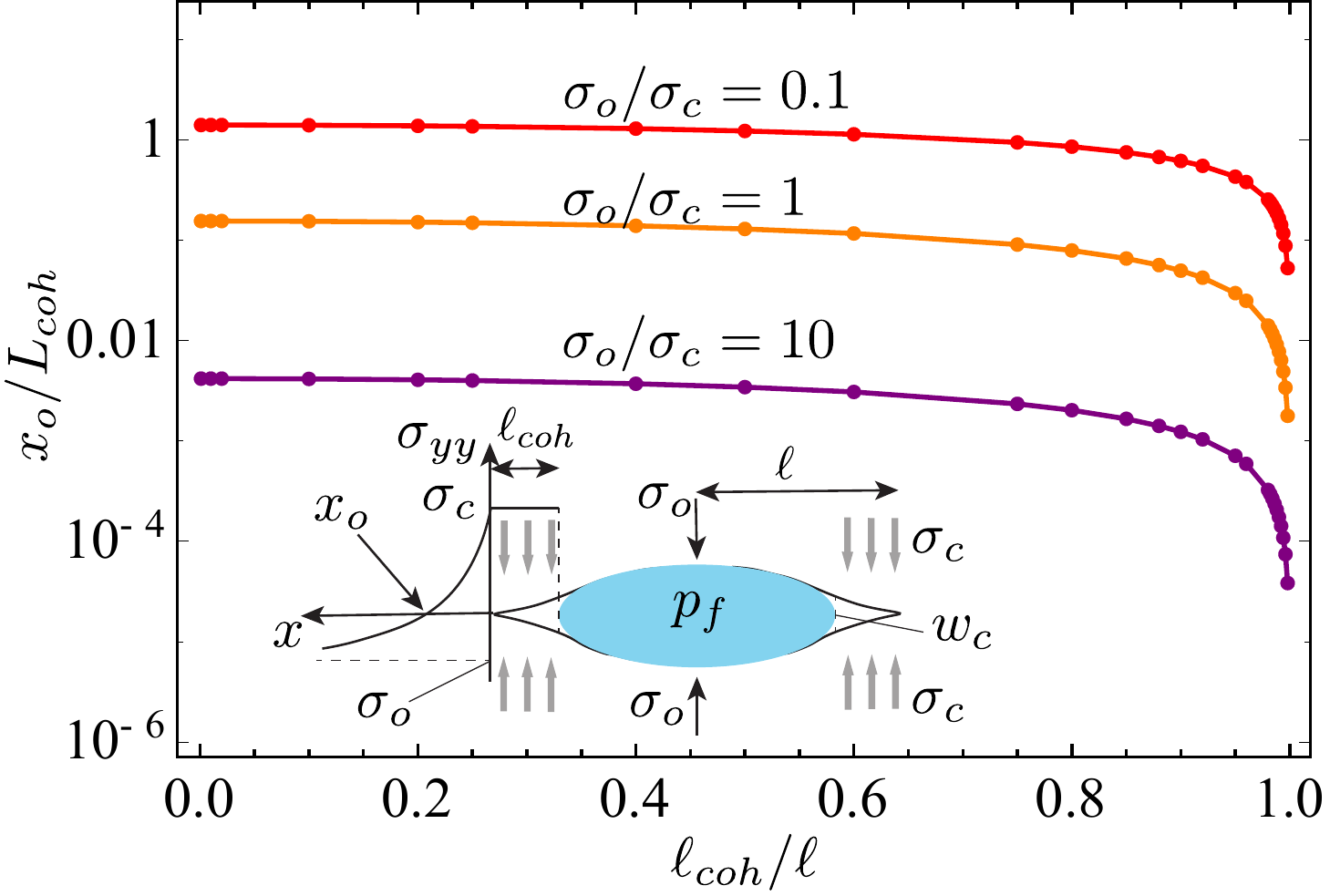}
\caption{Evolution of the size of the tensile zone ahead of the fracture tip with the cohesive fraction for different confining to peak cohesive stress ratios. The pressure is uniform everywhere inside the fracture but no fluid is allowed to enter the cohesive zone (Dugdale cohesive zone model).}
\label{fig:stressAccuracy}
\end{figure}

\paragraph{Mesh requirements}
A sufficient number of cohesive elements is necessary to ensure the stress accuracy ahead of the fracture tip and the resolution of the fracture propagation condition. A minimum of three elements are suggested to mesh the cohesive zone to ensure sufficient accuracy of the near tip stress field \citep{FaNe2001,MoBe2002,TuDa2007}. 
In dry fracture mechanics, the technique of artificially enlarging the cohesive zone length while keeping the fracture energy constant is often used (increasing $w_c$ and decreasing  $\sigma_c$ accordingly) \citep{BaPl1997,TuDa2007} thus allowing the use of coarser meshes.
Unfortunately, such a technique is not adequate for cohesive hydraulic fractures. It is only valid when the confining $\sigma_o$ is adjusted together with $\sigma_c$ in order to keep the ratio of time scales $t_{cm}/t_{om}$   unchanged (see Eq.~(\ref{eq:tcm})). If not, this will change the physics of the fluid front-cohesive zone coupling. 
Another important difference with dry fracture mechanics is the fact that the fracture propagates in a medium under initially compressive state of stresses, as such the tensile region ahead of the fracture is inherently smaller as the confinement increases. 
Assuming a fluid lag the same size as the cohesive zone, Fig.~\ref{fig:stressAccuracy} displays the evolution of the tensile zone ahead of the fracture tip as the uniformly pressurized HF grows under different confinements. The tensile zone significantly shrinks as the confining stress increases, and therefore requires for a finer mesh. Such a confinement-related mesh requirement has been seldomly discussed in previous studies \citep{ChBu2009,Chen2012,SaPa2012,CaGr2012,SaKh2015,Wang2015,LiDe2017} where the fluid front-cohesive zone coupling is often neglected (zero fluid lag, small cohesive zone) and the simulation performed under zero confinement. 
In this paper, we release the confinement-related requirement by adapting the time-step for a given fixed fracture increment to fulfill the propagation stress propagation condition. We also check a posteriori that the stress intensity factor is indeed null using Eq.~(\ref{eq:zerosingularity}).
We obtain an absolute error on Eq.~(\ref{eq:zerosingularity}) of about 5\% (in a range between 0.1 and 8\%) for all the reported simulations. % with at least tens of cohesive elements during the fracture propagation.

Apart from the tensile zone ahead of the fracture tip, one also needs to resolve the fluid lag which shrinks tremendously as the fracture grows but still influences the solution \citep{Gara2019}. At least one partially-filled (lag) element is necessary to account for the influence of the fluid cavity on the tip stress field. At large time, the fluid lag becomes negligible compared to the cohesive zone. This is ultimately  the bottle-neck governing the computational burden due to the  mesh requirement of at minima one element in the fluid lag.
For all the results presented in the following, 
we actually stop the simulations when the fluid fraction $\xi_f=\ell_f/\ell$ reached $0.99$ or when the fracture length was already within five percent of the LHFM solutions.

\section{Results}

We now  numerically explore the propagation diagram described in Fig.~\ref{fig:ScalingKGD}.
We perform a series of simulations covering dimensionless toughness from 1 to 4  and different level of confining to peak cohesive stress ratio $\sigma_o/\sigma_c$ from 0.1 to 10 for either a smooth ($\alpha_e=0$) or rough ($\alpha_e=2$) fracture. These conditions span the transition from viscosity to toughness dominated growth regimes, as well as laboratory ($\sigma_o/\sigma_c=0.1-1$) and field conditions ($\sigma_o/\sigma_c=10$).

%-------------------
\subsection{A smooth cohesive fracture ($\alpha_e=0$)}
%-------------------

\begin{figure}
\centering
\begin{tabular}{cc}
     \includegraphics[width=0.46\linewidth]{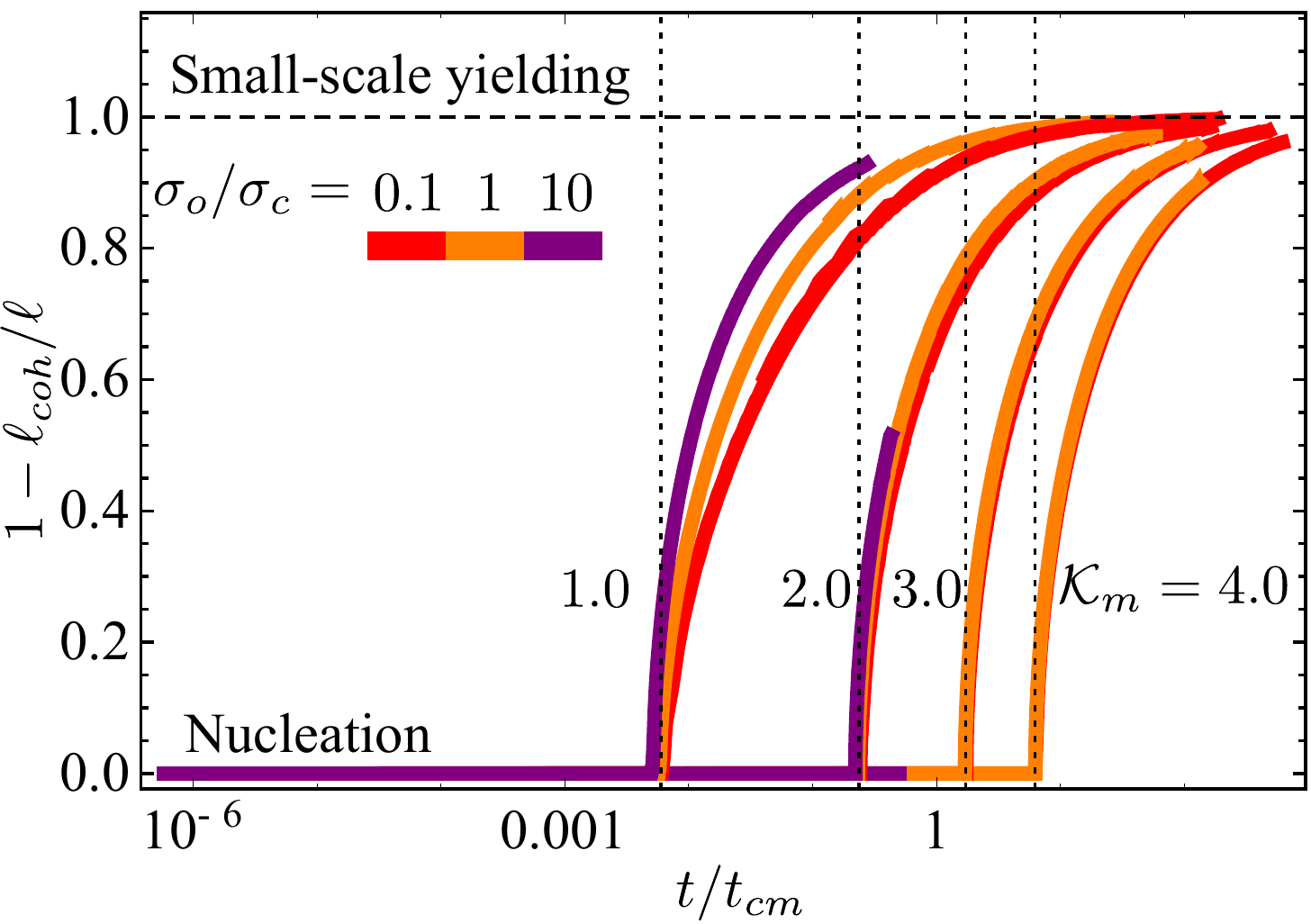}&  
     \includegraphics[width=0.46\linewidth]{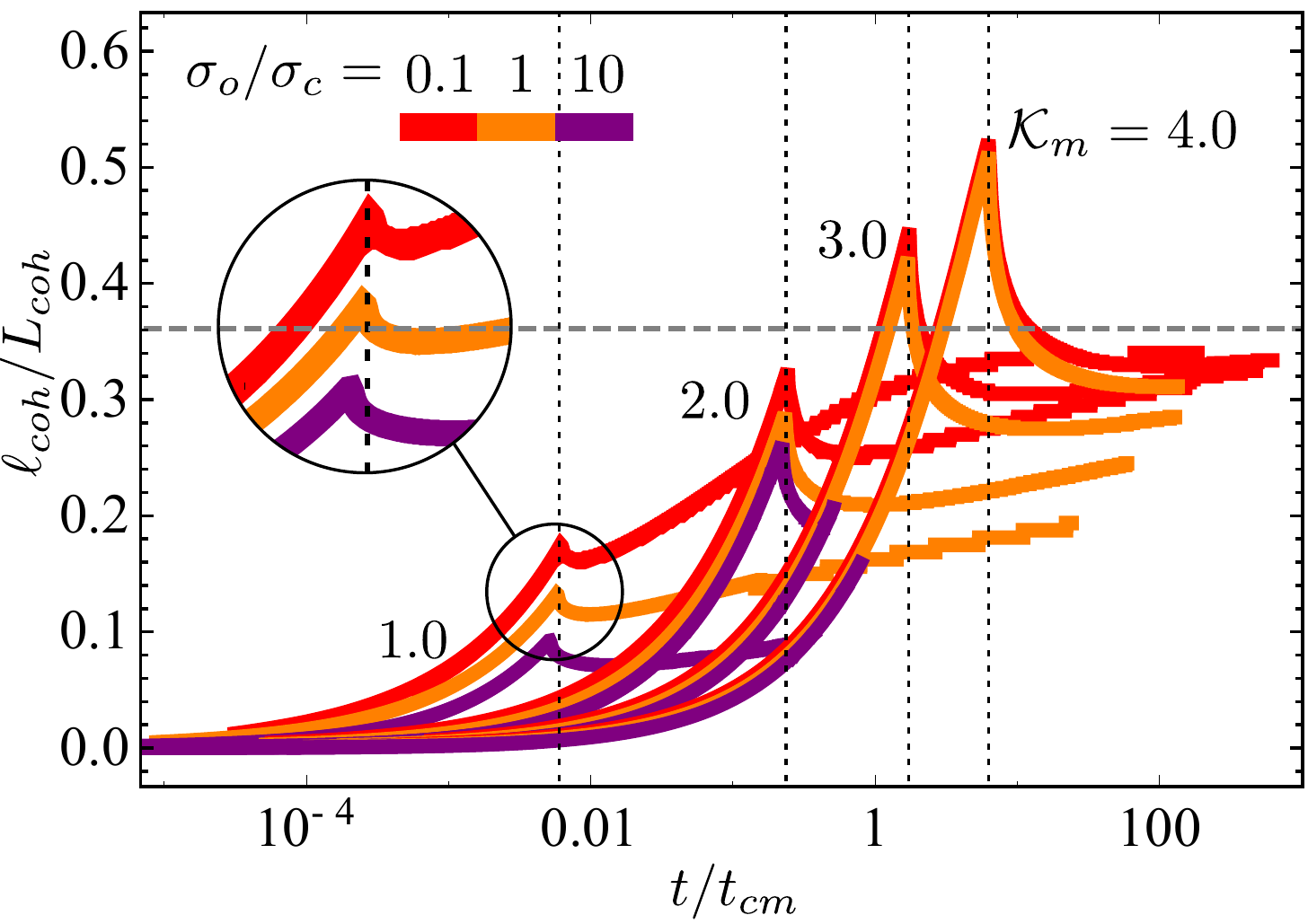}\\
     a)& b) 
\end{tabular}
\caption{Evolution of a) the non-cohesive fraction $1-\ell_{coh}/\ell$ and b) dimensionless cohesive length $\ell_{coh}/L_{coh}$ with $t/t_{cm}$ for $\mathcal{K}_m = 1-4$. The red, orange, and purple curves correspond to $\sigma_o/\sigma_c=0.1, 1.0, 10$ respectively. The dotted vertical lines indicate the cohesive zone nucleation period for $\sigma_o/\sigma_c=0.1$, $\mathcal{K}_m=1-4$. The dashed horizontal line represents the small-scale yielding asymptote ($\approx 0.115 \pi$) of the cohesive zone length for the linear-softening cohesive model \citep{DeTa2010}.}
% an isolated cohesive zone in tension by load or displacement control (fixed grip), they all tend to the same asymptote.
\label{fig:cohlengthcms}
\end{figure}

\begin{figure}
\centering
\includegraphics[width=0.5\linewidth]{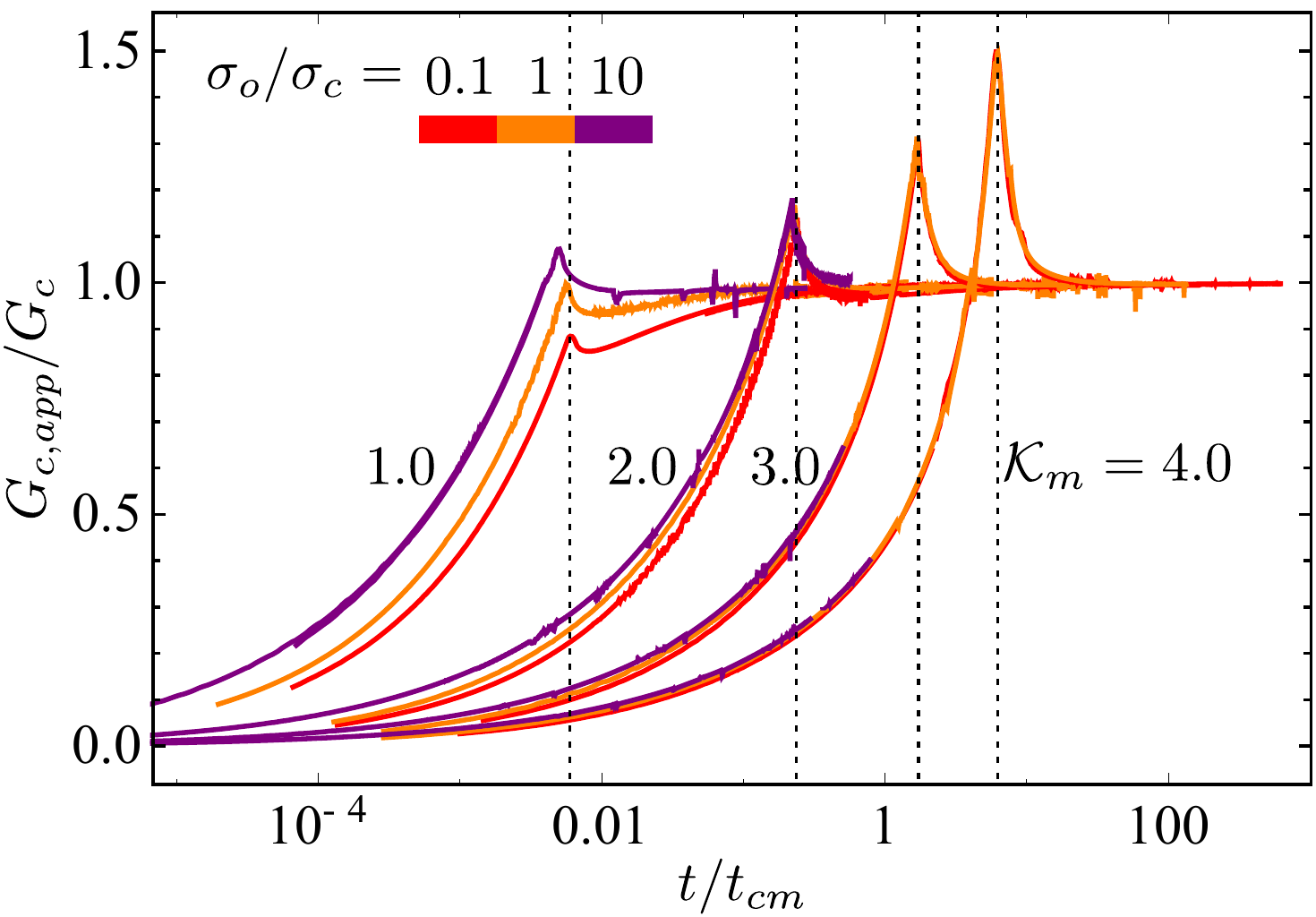} 
\caption{Smooth cohesive fracture tip: evolution of the apparent fracture energy $G_{c,app}/G_c$ with $t/t_{cm}$ for $\mathcal{K}_m = 1-4$. The red, orange, and purple curves correspond to $\sigma_o/\sigma_c=0.1, 1.0, 10$ respectively. The dotted vertical lines indicate the cohesive zone nucleation period for $\sigma_o/\sigma_c=0.1$, $\mathcal{K}_m=1-4$. }
\label{fig:apparentdissipation}
\end{figure}

The three stages related to nucleation, intermediate and late time propagation are well visible on the time evolution of the dimensionless cohesive length (Fig.~\ref{fig:cohlengthcms}), apparent fracture energy (Fig.~\ref{fig:apparentdissipation}), fracture length (Fig.~\ref{fig:lengthcm}), as well as inlet width (Fig.~\ref{fig:widthcm}) and net-pressure (Fig.~\ref{fig:pressurecm}).

\paragraph{Cohesive zone growth}
The scaled cohesive length $\ell_{coh}/L_{coh}$ evolves non-monotonically with time (Fig.~\ref{fig:cohlengthcms}). This evolution is dependent on both the dimensionless toughness $\mathcal{K}_m $ and $\sigma_o/\sigma_c$.
At early time during the nucleation phase, when the fracture length is completely embedded inside the cohesive zone ($1-\ell_{coh}/\ell=0$), the cohesive zone increases monotonically (Fig.~\ref{fig:cohlengthcms}).
We define the time $t_c$ as the end of the nucleation phase, when here after $1-\ell_{coh}/\ell>0$.
From our simulations, we found that $t_c$ follows approximately an exponential relation $t_c/t_{cm} \sim \mathcal{K}_m^{5.17}$ for $\mathcal{K}_m \in [1-4]$. This exponent is consistent with the range of exponents in the viscosity ($t_c/t_{cm} \sim \mathcal{K}_m^6$) and toughness ($t_c/t_{cm} \sim \mathcal{K}_m^4$) dominated regimes which can be obtained by setting $\mathcal{G}_w=W/w_c=1$ in respectively the M- and K-scaling in Table~\ref{tab:scalings}. In addition, $t_c$ also slightly depends on the dimensionless confinement $\sigma_o/\sigma_c$, see the inset on Fig.~\ref{fig:cohlengthcms}. Larger confinement slightly reduces this nucleation period for a given $\mathcal{K}_m$. 
The cohesive zone length at nucleation are larger for larger dimensionless toughness and then decreases with time after nucleation. 

At large time, we observe that - at least for low dimensionless confinement - the cohesive zone length tends to a similar value for all dimensionless toughness. 
Unfortunately, this is less observable for larger dimensionless confinement which leads to prohibitive computational cost such that the simulations were stopped prior to stabilization of the cohesive zone length. However, the trend for $\sigma_o/\sigma_c=1$ hints that a similar behavior holds for larger confinement albeit possibly much later in time.

\paragraph{Associated energy dissipation}
The energy spent in debonding cohesive forces (apparent fracture energy) increases similarly to the growth of the cohesive zone length (Fig.~\ref{fig:apparentdissipation}). This is due to the fact that $\dot{\ell}\approx \dot{\ell}_{coh}$ during the nucleation stage. Interestingly, the apparent fracture energy may even go above the fracture energy $G_c$ at nucleation for large dimensionless toughness / large dimensionless confinement as illustrated in Fig.~\ref{fig:apparentdissipation}. At large time, the apparent fracture energy converges to the fracture energy $G_c$, confirming the fact that the material derivative of width (in the moving tip frame) becomes negligible in Eq.~(\ref{eq:GappRelation}). 
This confirms that at large time (when $1-\ell_{coh}/\ell \sim 1$) one can use the solution of a steadily moving semi-infinite hydraulic fracture solution accounting for cohesive forces \citep{Gara2019}.  
However, care must be taken to use such a semi-infinite fracture solution when the cohesive zone length is of the same order than the overall fracture length. For example, the results obtained in \cite{Gara2019} based on the use of an equation of motion and the semi-infinite cohesive HF solution lead to cohesive zone length larger than the finite fracture length under the premises of the constant apparent fracture energy. This ultimately leads to an over-estimation of fracturing energy dissipation and larger deviation from LEFM solutions as it neglects the evolution of the apparent fracture energy associated with the nucleation phase.

\paragraph{Comparisons with linear hydraulic fracture mechanics (LHFM)}

The time evolution of dimensionless fracture length (scaled by the viscosity dominated LHFM growth length scale $L_m(t)$ - see Table.~\ref{tab:scalings}) is displayed as dashed curves on  Fig.~\ref{fig:lengthcm}. The corresponding inlet net-pressure and width evolution for the smooth cohesive zone are displayed as dashed curves on Fig.~\ref{fig:widthcm} and \ref{fig:pressurecm} respectively. 
Our results indicate that the CZM solutions converge toward the LHFM ones (for the corresponding dimensionless toughness) at large times $t\gg t_{cm}$. The exact dimensionless time for such a convergence toward the LHFM solution is larger for larger dimensionless toughness, and smaller for larger $\sigma_o/\sigma_c$.

\begin{figure}
\centering
\includegraphics[width=0.46\linewidth]{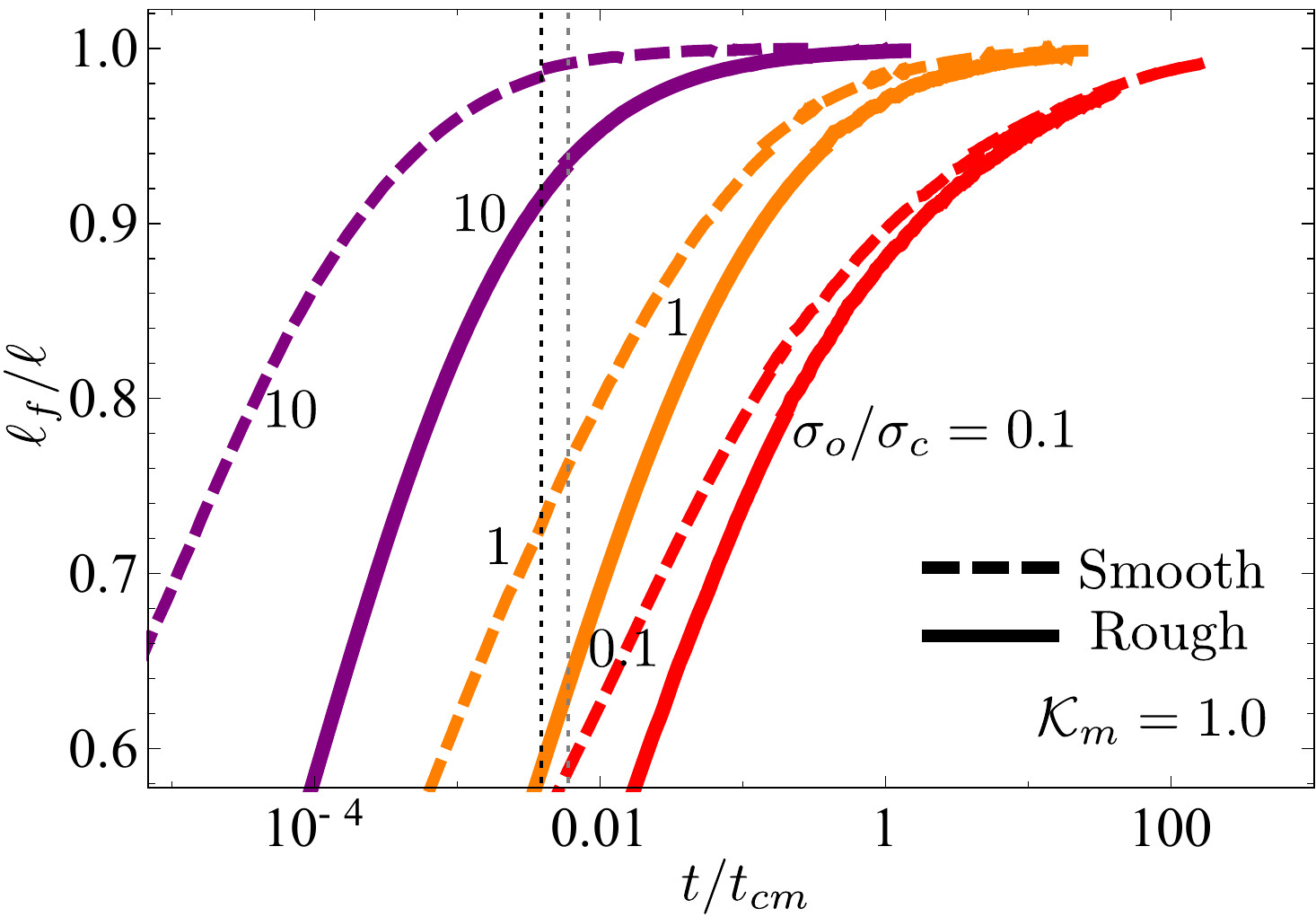}
\includegraphics[width=0.46\linewidth]{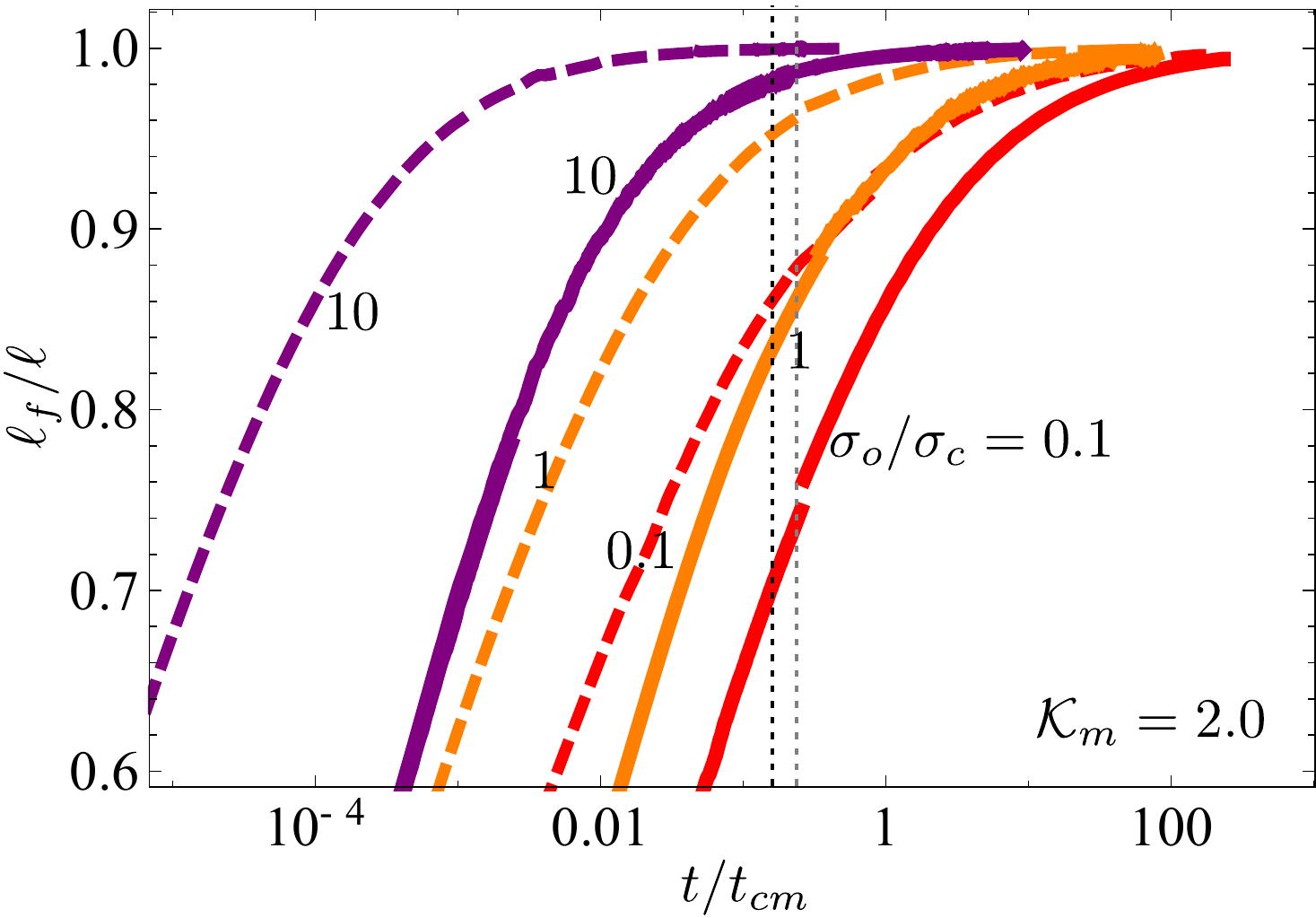}\\
\includegraphics[width=0.46\linewidth]{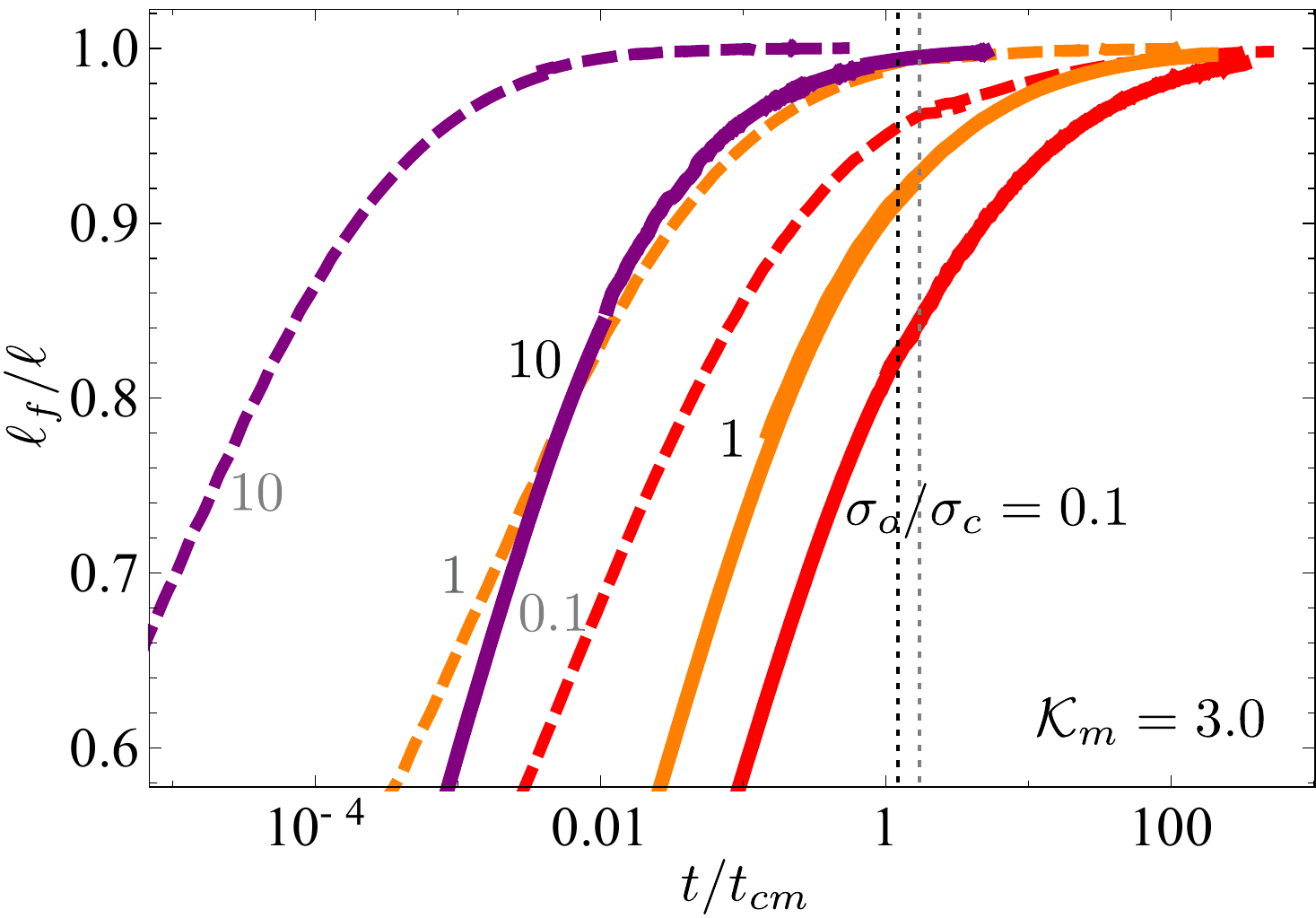}
\includegraphics[width=0.46\linewidth]{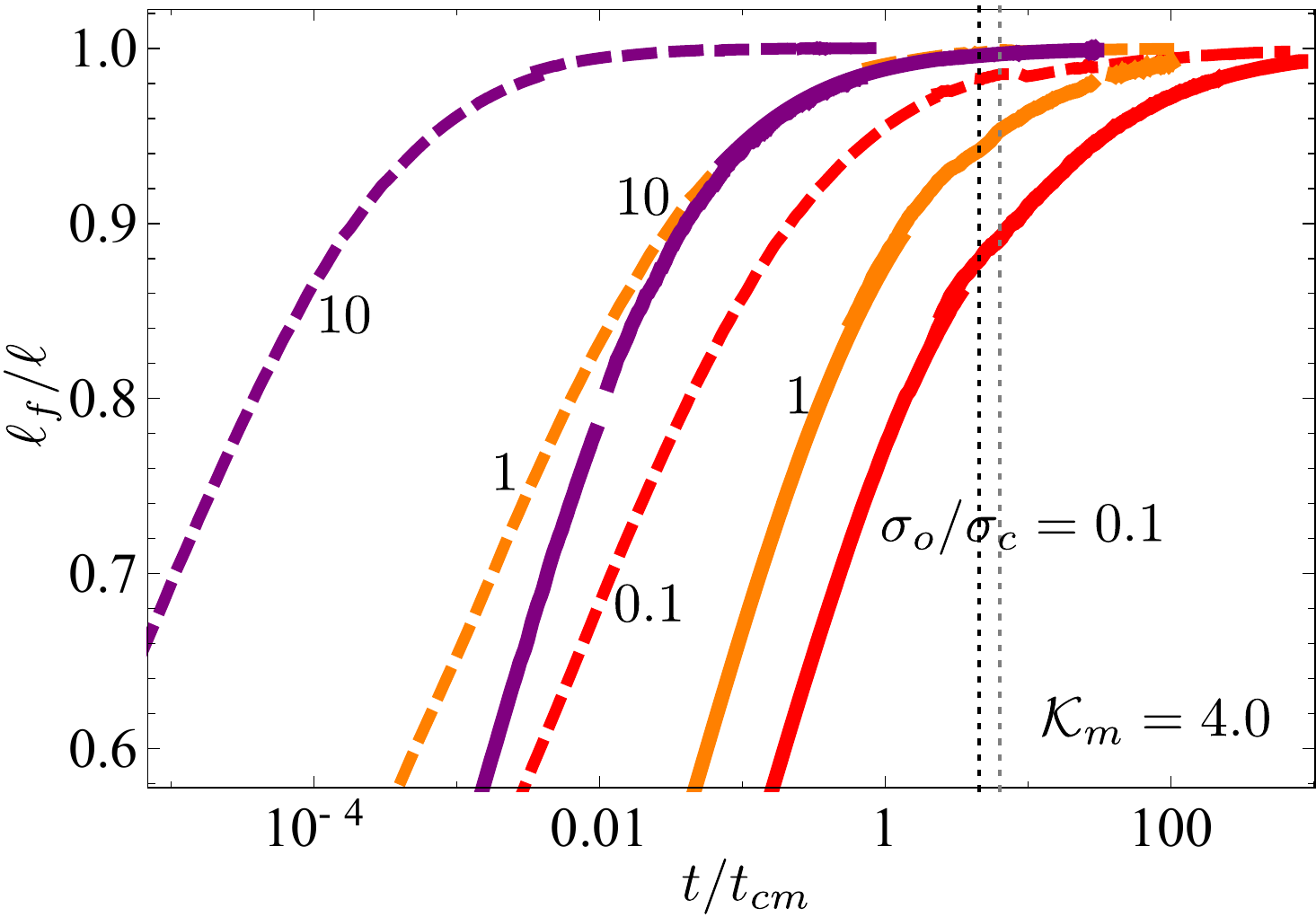}
\caption{Evolution of the fluid fraction $\xi_f=\ell_f/\ell$ with $t/t_{cm}$ for $\mathcal{K}_m = 1-4$. The red, orange, purple curves correspond to $\sigma_o/\sigma_c=0.1, 1.0, 10$ and the solid and dashed curves correspond respectively to a rough ($\alpha_e=2$) and smooth fracture ($\alpha_e=0$). The dotted vertical lines indicate the cohesive zone nucleation period of $\sigma_o/\sigma_c=0.1$ for a smooth (gray) and a rough (black) fracture.}
\label{fig:xifcm}
\end{figure}

Interestingly, the fracture length is larger at the early stage of growth compared to the LHFM estimate while the inlet opening and pressure are smaller. 
These differences directly result from the fact that the cohesive forces 
greatly increases the fluid lag size and impacts its evolution during the nucleation and intermediate stages of growth. 
 Indeed, in the LHFM case, the fluid lag is negligible at all times for dimensionless toughness $\mathcal{K}_m$ larger than $\sim 1.5$  as reported in \cite{Gara2006,LeDe2007}. For $\mathcal{K}_m=1$, the fluid fraction in the LHFM case is already small at early time: it evolves  from 0.9 (when $t\ll t_{om}$ ) to 1 (for $t\approx t_{om}$ (see Fig.~\ref{fig:GammaXif} in appendix). For the same dimensionless toughness, the fluid fraction is lower than 0.6 at early time when accounting for the cohesive zone (see Fig.~\ref{fig:xifcm}). The large extent of the fluid lag is similarly found for larger dimensionless toughness - a striking difference with the LHFM case for which no fluid lag is observed for $\mathcal{K}_m>1.5$.
The cohesive forces significantly enhance the suction effect and thus lag size during nucleation. For the same value of $\mathcal{K}_m$, a larger  confinement compared to peak strength (larger $\sigma_o/\sigma_c$) decreases the lag size. 
Larger $\sigma_o/\sigma_c$ results in steeper fluid pressure gradient and accelerates the penetration of the fluid front into the cohesive zone during the nucleation and intermediate phase (see Fig.~\ref{fig:penetrationcm}) . 

As the dimensionless toughness increases, the effect of $\sigma_o/\sigma_c$ becomes limited to the nucleation phase (see the length, inlet pressure, width evolution on Figs.~\ref{fig:lengthcm},~\ref{fig:pressurecm},~\ref{fig:widthcm}). After nucleation, the solutions appear independent of $\sigma_o/\sigma_c$ for $t>t_{cm}$ for the $\mathcal{K}_m=3$ and $4$ cases. 
The fact that $\sigma_o/\sigma_c$ does not influence the growth after nucleation for large toughness  can be traced back to the fact that the fluid lag cavity is very small in comparison to the cohesive zone length as can be seen on Fig.~\ref{fig:penetrationcm}. 

\begin{figure}
    \centering
    \begin{tabular}{cc}
         \includegraphics[width=0.46\linewidth]{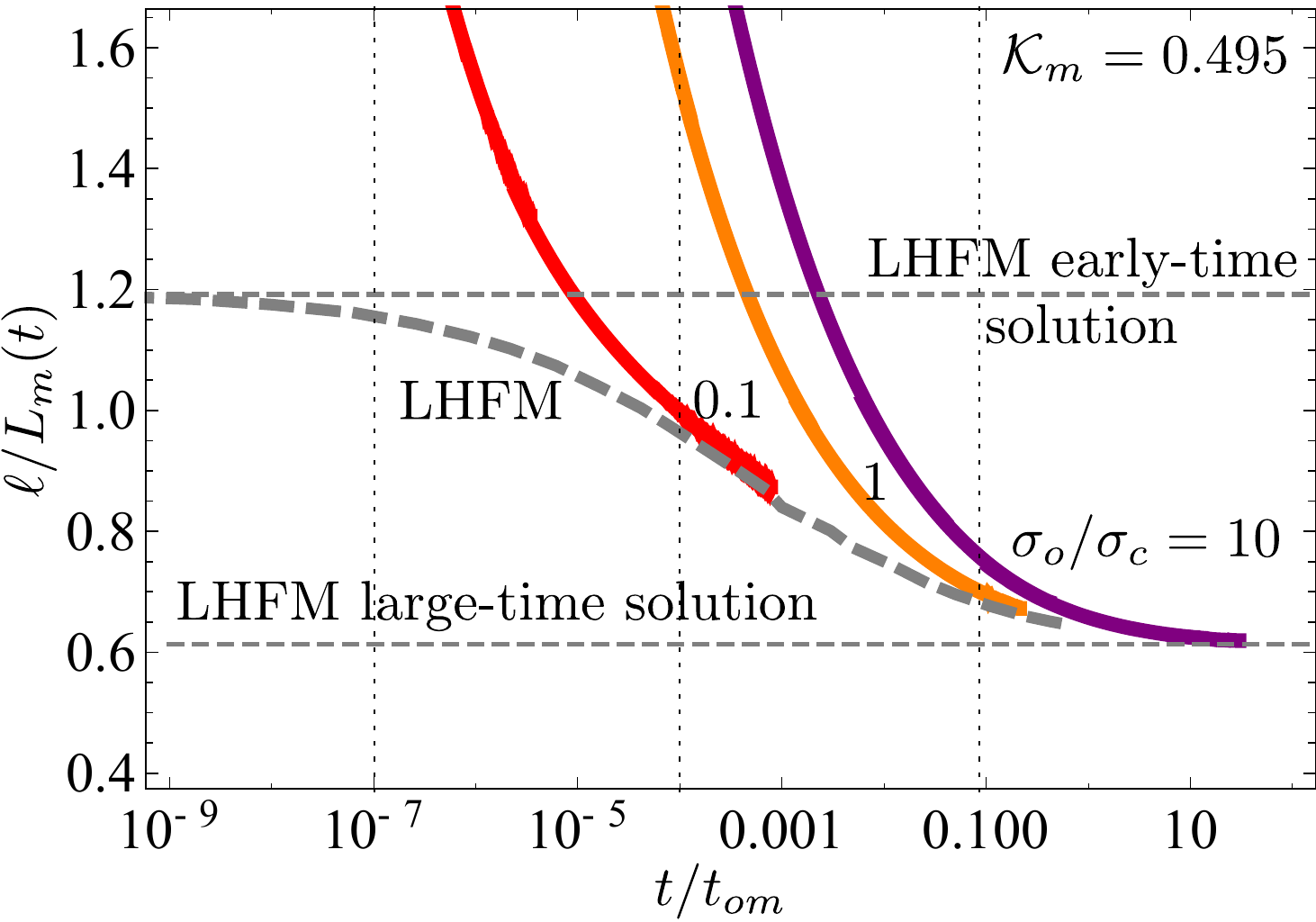}&
         \includegraphics[width=0.46\linewidth]{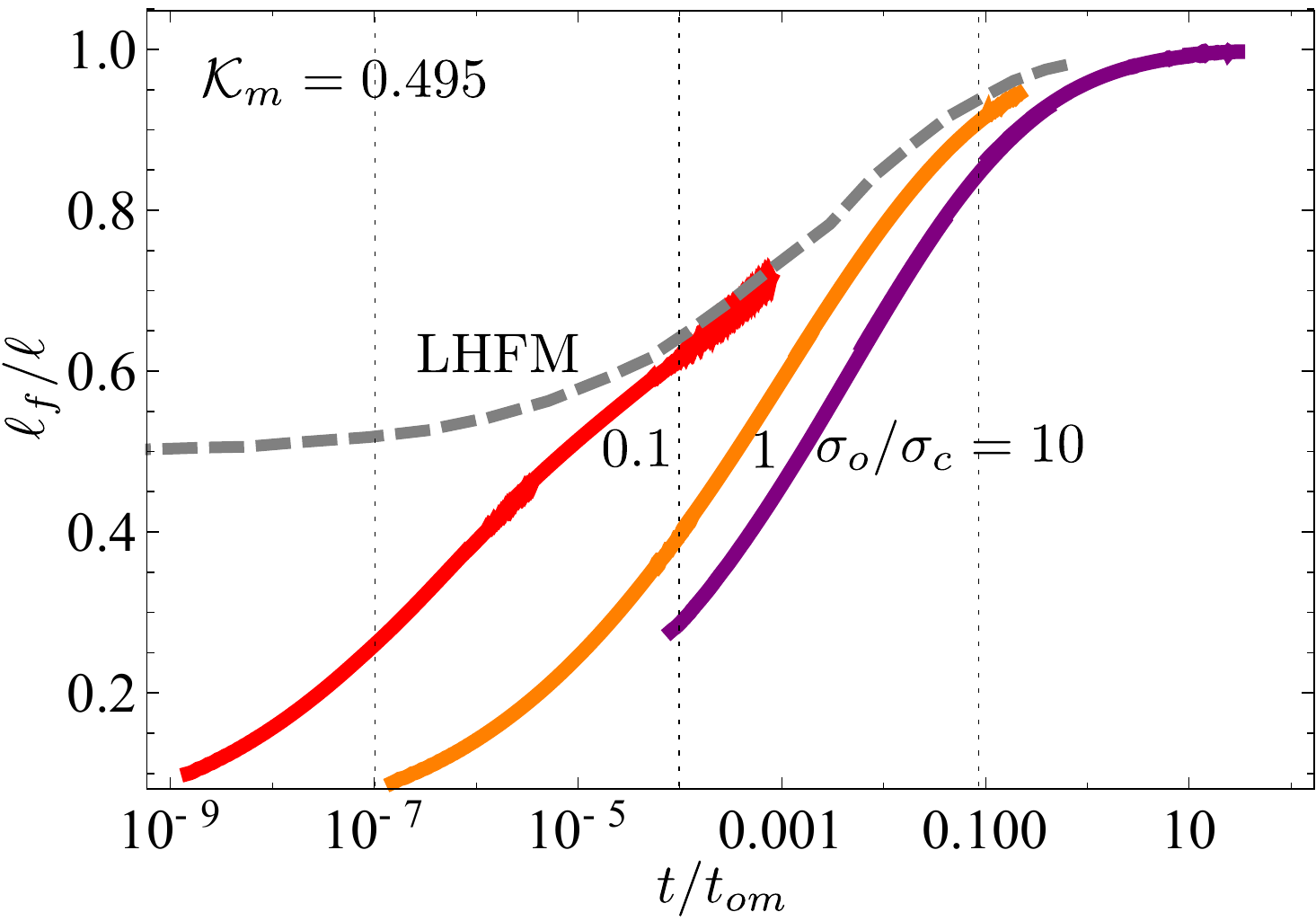}\\
         a) & b)\\
         \includegraphics[width=0.46\linewidth]{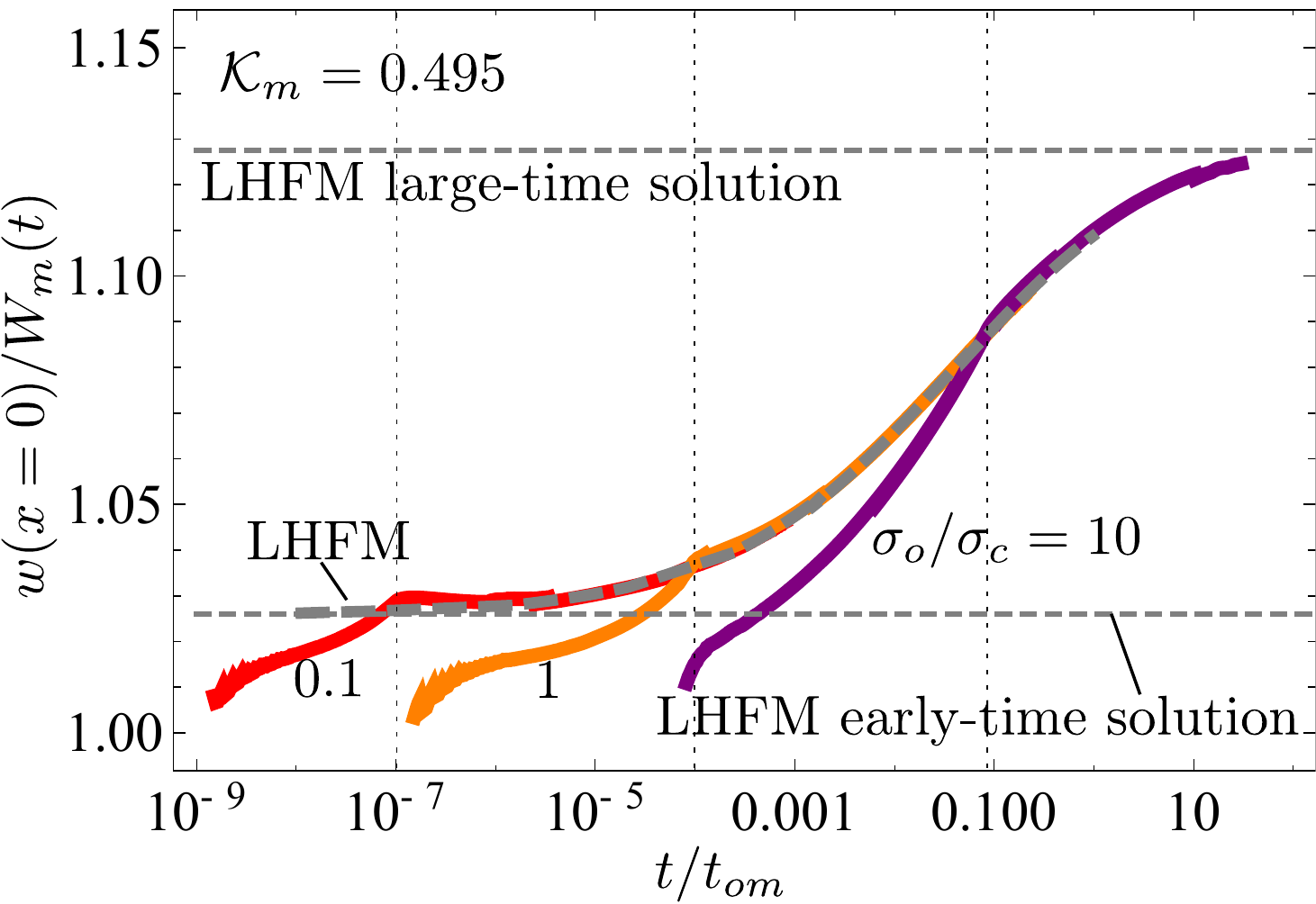}&
         \includegraphics[width=0.46\linewidth]{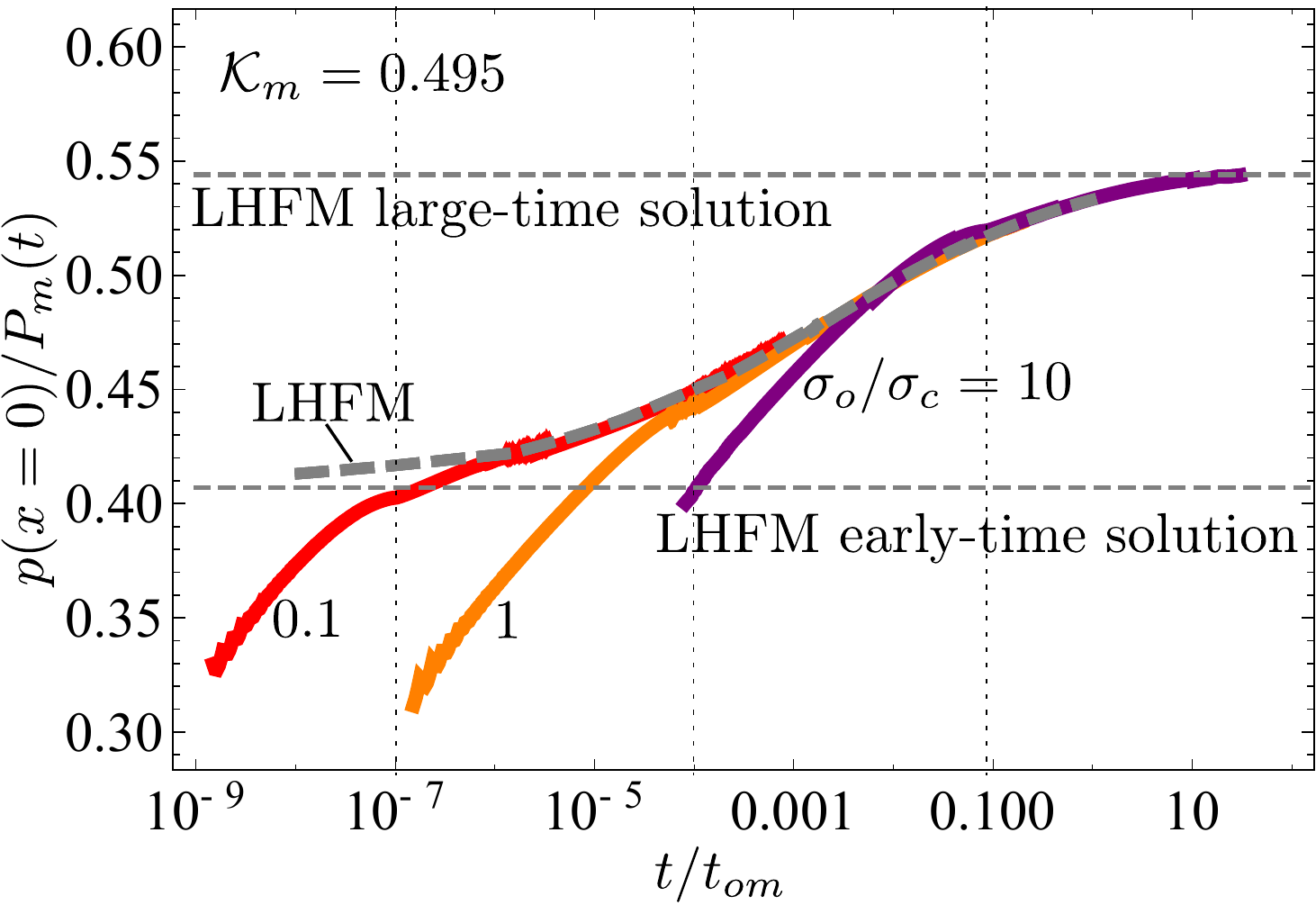}\\
         c) & d)\\
    \end{tabular}
    \caption{$\mathcal{K}_m=0.495$: evolution of a) the fracture half length, b) fluid fraction, c) inlet width, and d) inlet net pressure with $t/t_{om}$. The red, orange and purple curves correspond to different confining stress $\sigma_o/\sigma_c=0.1, 1, 10$ in a smooth cohesive HF with the dotted vertical lines as their corresponding cohesive zone nucleation period. The gray dashed curves indicate LHFM numerical results with a lag. The two gray horizontal lines correspond respectively to the LHFM early-time solutions with a lag \citep{Gara2006} and large-time solutions without a lag \citep{GaDe2005}. The time evolution of the cohesive zone length and the ratio between the lag and cohesive zone sizes, fracture apparent energy and ratio of energy dissipation in viscous flow to that in fracture surface creation is shown in Supplemental Materials.}
\label{fig:Km0d5}
\end{figure}

Fig.~\ref{fig:Km0d5} displays the dimensionless fracture length, fluid fraction, inlet width and pressure  for a small dimensionless toughness case ($\mathcal{K}_m=0.495$). We have plotted these time evolution as function of $t/t_{om}$ for better comparison with the LHFM solution accounting for a fluid lag \citep{LeDe2007}. For  low dimensionless toughness $\mathcal{K}_m$, the response converges well to the LHFM lag solution  \citep{LeDe2007} relatively quickly after nucleation  (contrary to the case of large $\mathcal{K}_m$).
On Fig.~\ref{fig:Km0d5}, the convergence occurs earlier for smaller $\sigma_o/\sigma_c$ in term of $t/t_{om}$ - actually later for smaller $\sigma_o/\sigma_c$ in term of $t/t_{cm}=t/t_{om}\times (\sigma_o/\sigma_c)^{-3}$ (in line with observations for larger $\mathcal{K}_m$).

%%%%%%%%%%%%%%%%%%%%%%%%%%%%%%%%%%%%%
% here the key points to strengthen are:
%  t_cm -> is indeed nucleation time scale 
%  t_om -> lag coalescece time scale in the LHFM is here modulated by so/sc
% for small Km, all s_o/sc coalesces at t~t_{om} - coalescence toward LHFM occurs earlier for lower t/t_cm (larger t/t_om) for larger so/sc
% for large Km, all s_o/sc coalesces at t~t_cm.
% coalesence toward LHFM occurs at t*>>t_cm and t* increases with K_m

\begin{figure}
\centering
\begin{tabular}{cc}
\includegraphics[width=0.48\linewidth]{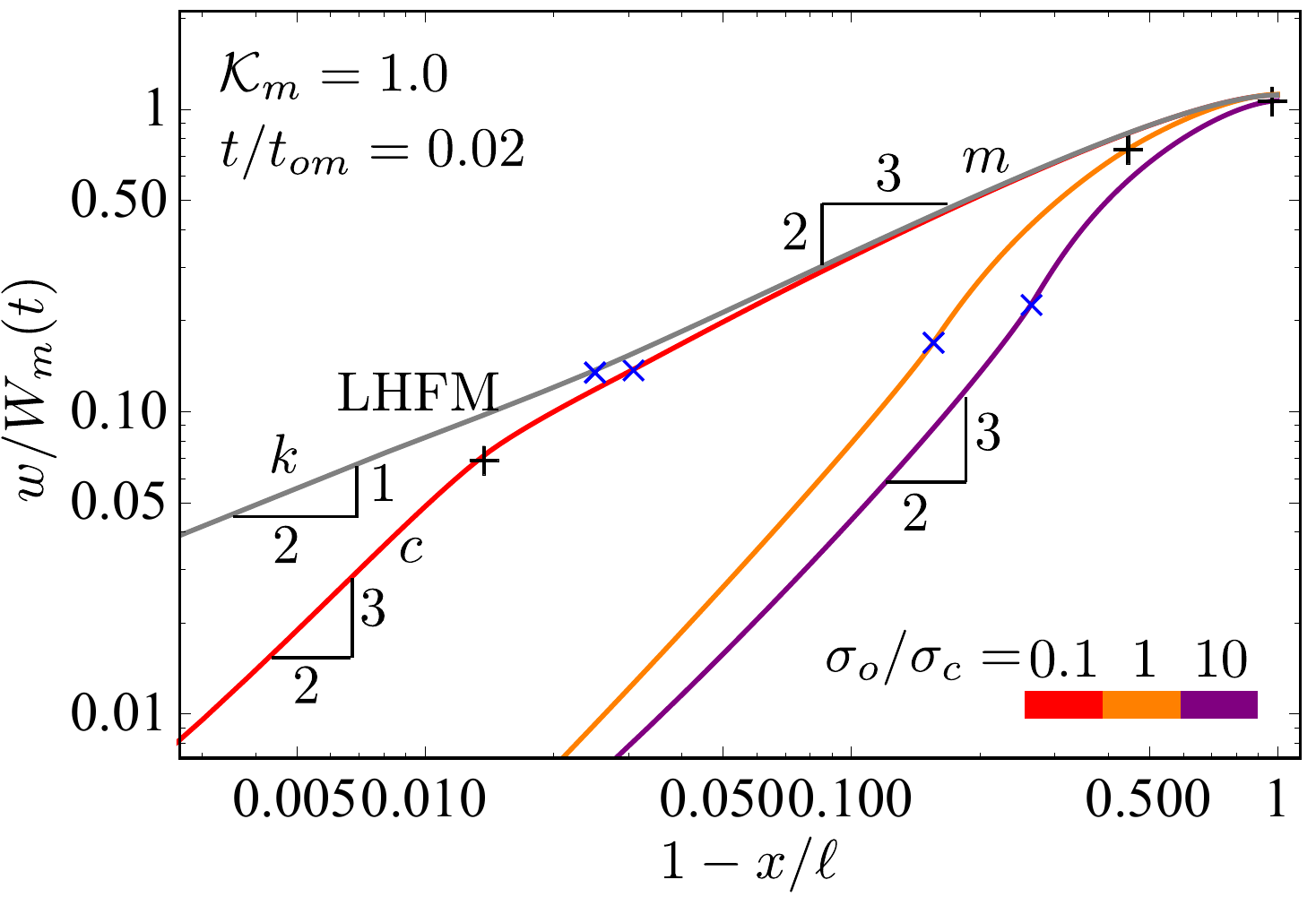} &
\includegraphics[width=0.48\linewidth]{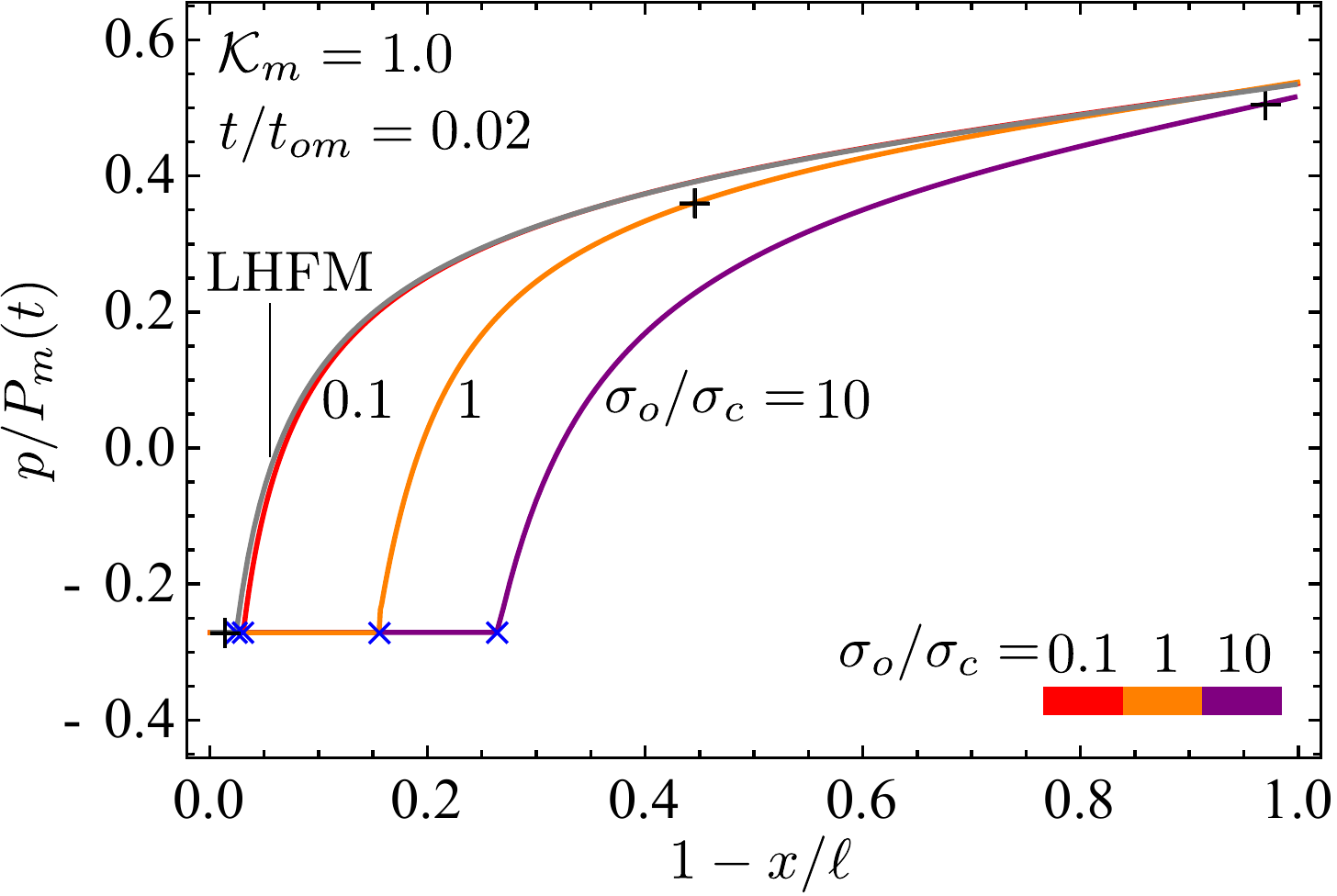} \\
    a) & b)
\end{tabular}
\caption{a) Dimensionless opening, and b) net pressure profiles at $t/t_{om}=0.02$ for $\mathcal{K}_{m}=1.009$. “$+$” indicates the boundary of the cohesive zone and “$\times$” indicates the fluid front location. The red, orange, and purple curves represent different confining stress level $\sigma_o/\sigma_c=0.1, 1.0, 10$. The gray curves represent the LHFM solutions with a lag at the same time $t/t_{om}=0.02$.}
\label{fig:Tip-Asymptote}
\end{figure}

\paragraph{Tip asymptotes}
The width and net pressure profiles in the tip reference frame for $\mathcal{K}_m=1$ is displayed on Fig.~\ref{fig:Tip-Asymptote} at time $t/t_{om}=0.02$ for different $\sigma_o/\sigma_c$ (thus at different $t/t_{cm}$ for the different $\sigma_o/\sigma_c$ and different ratio $\ell_{coh}/\ell$).
On can observe  different asymptotic behavior as function of distance from the tip on  Fig.~\ref{fig:Tip-Asymptote}. In the far-field, the 2/3 viscosity 'm' asymptote \citep{DeDe1994} is visible in the low confinement case - for which at this time, the fluid front is actually outside the cohesive zone. Closer to the tip, the 3/2 cohesive zone 'c' asymptote is visible. 
These results are in line with the cohesive tip solution of \cite{Gara2019}, although here the cohesive zone is not necessarily small compared to the overall fracture length. This induces a significant offset compared to the semi-infinite results reported in \cite{Gara2019} (see Supplemental Materials for details).

%%%%%%%%%%%%%%%%%%
\subsection{A rough cohesive fracture ($\alpha_e=2$)}
%%%%%%%%%%%%%%%%%%

\begin{figure}
\centering
\includegraphics[width=0.46\linewidth]{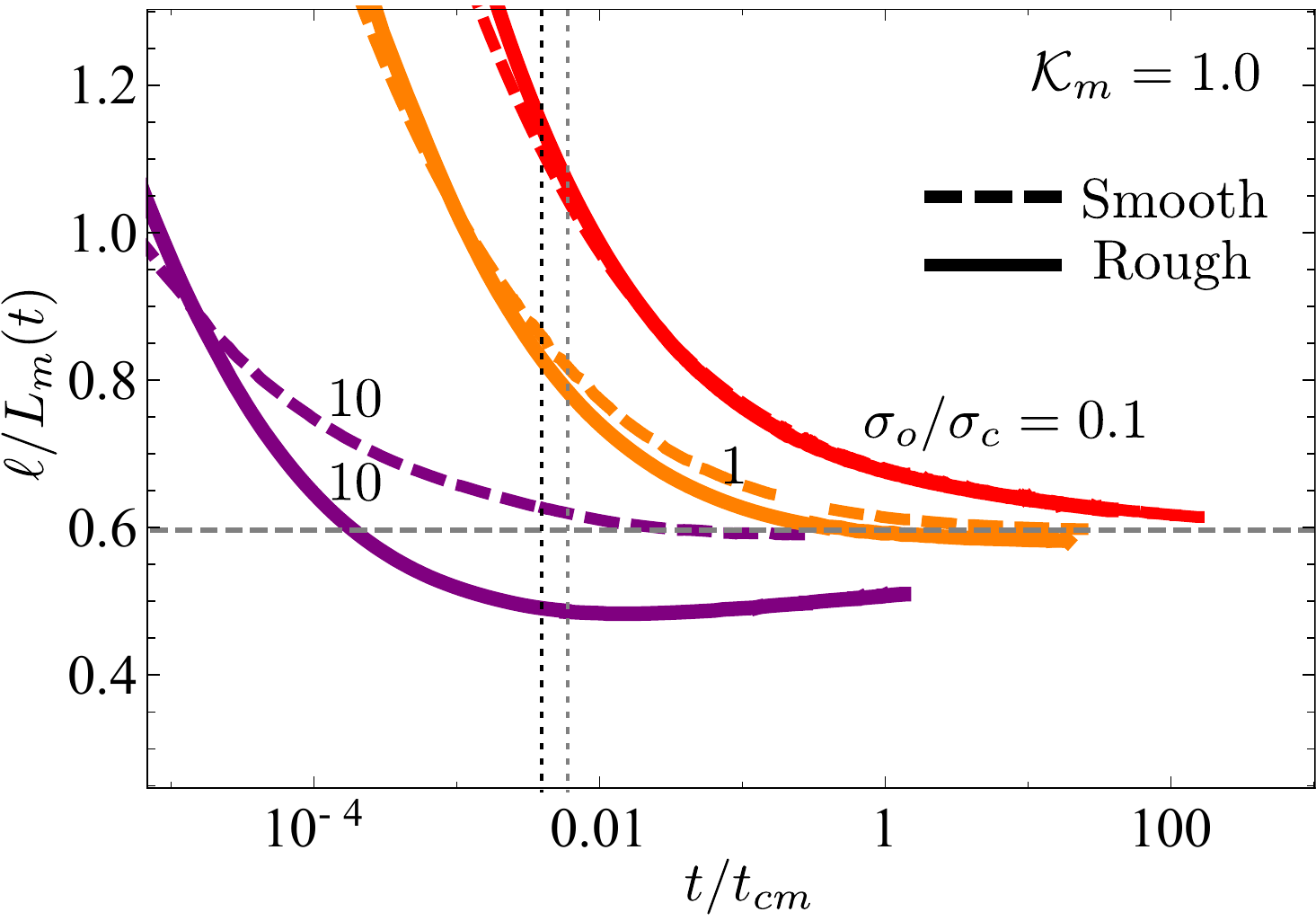}
\includegraphics[width=0.46\linewidth]{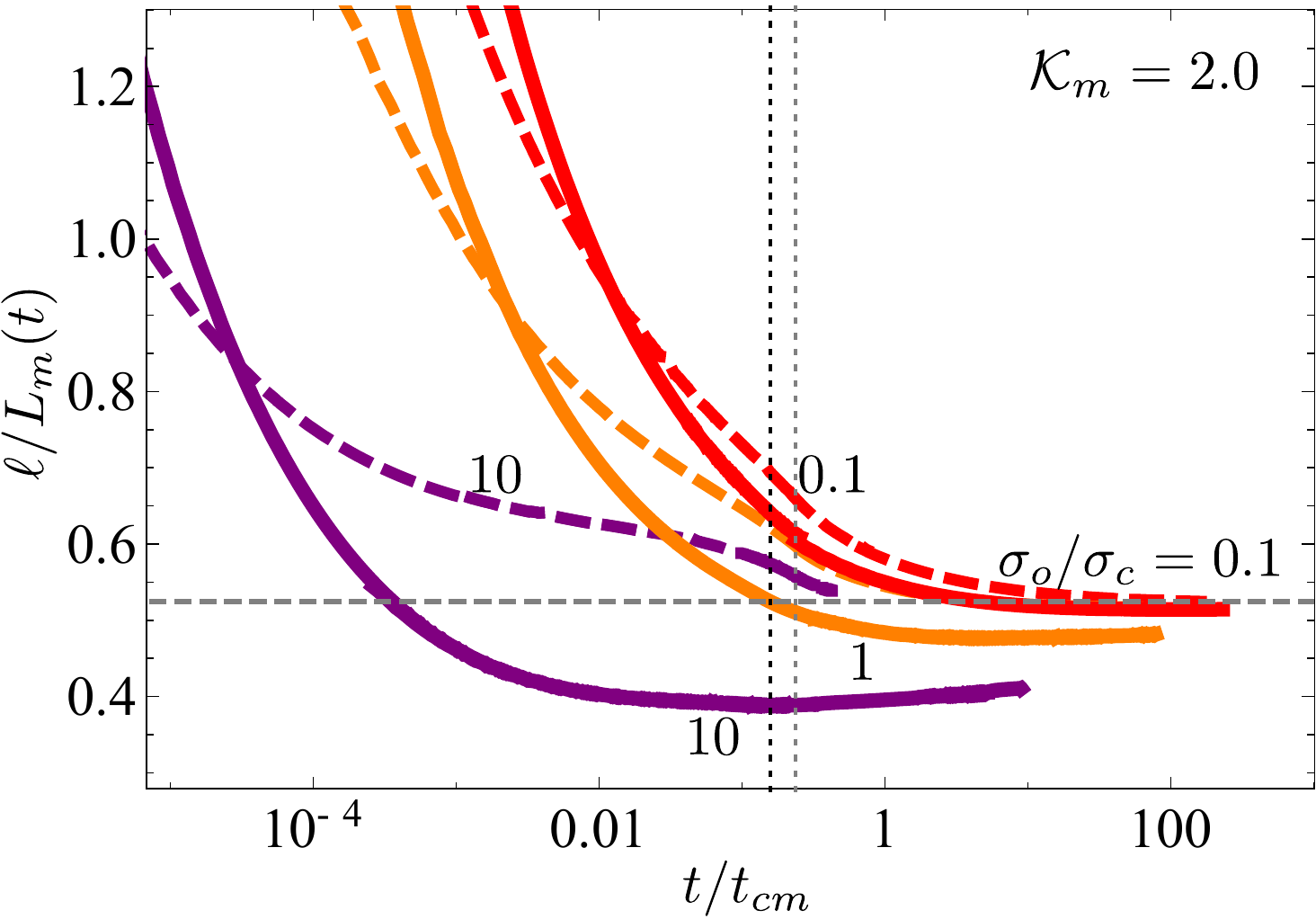}\\
\includegraphics[width=0.46\linewidth]{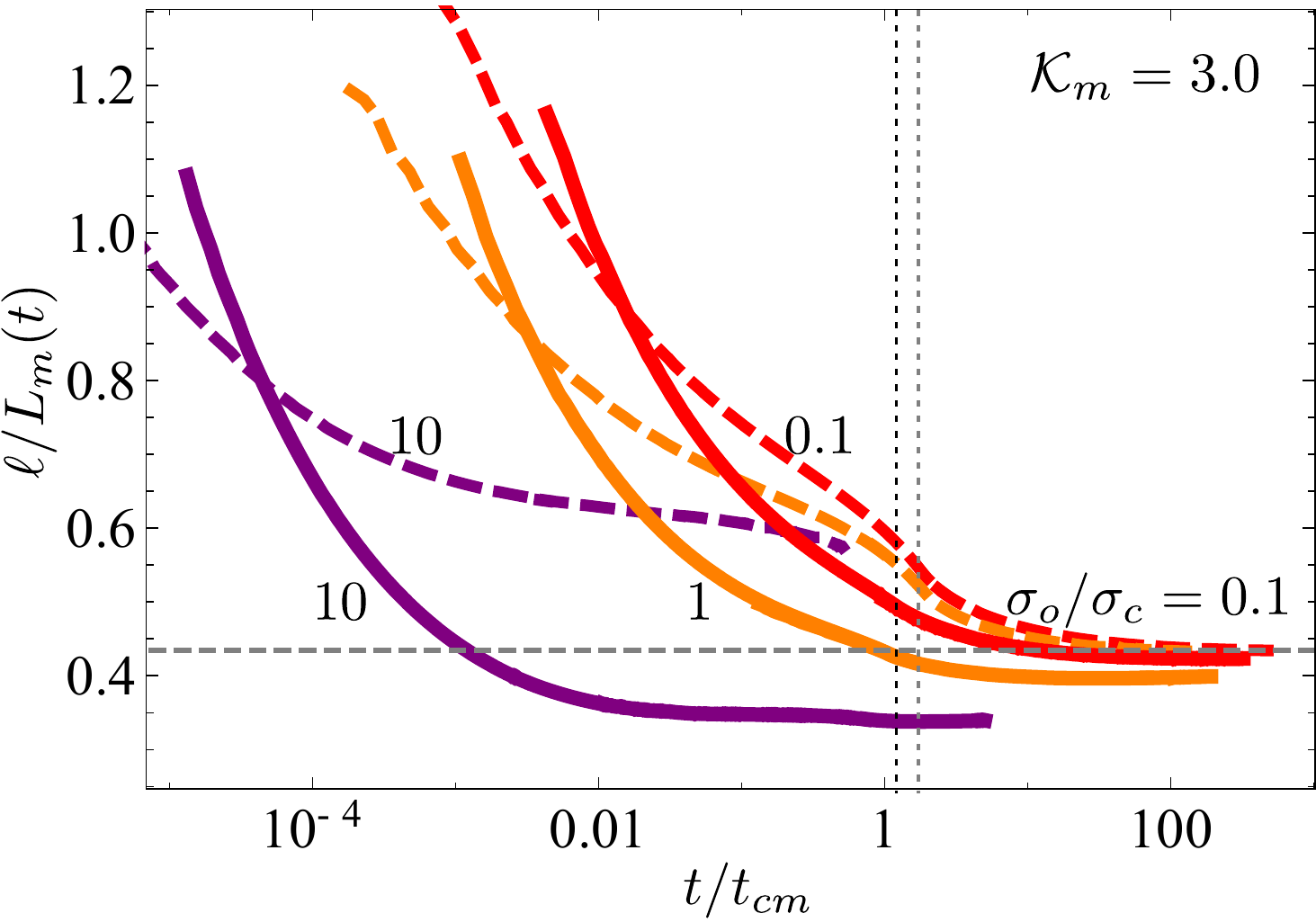}
\includegraphics[width=0.46\linewidth]{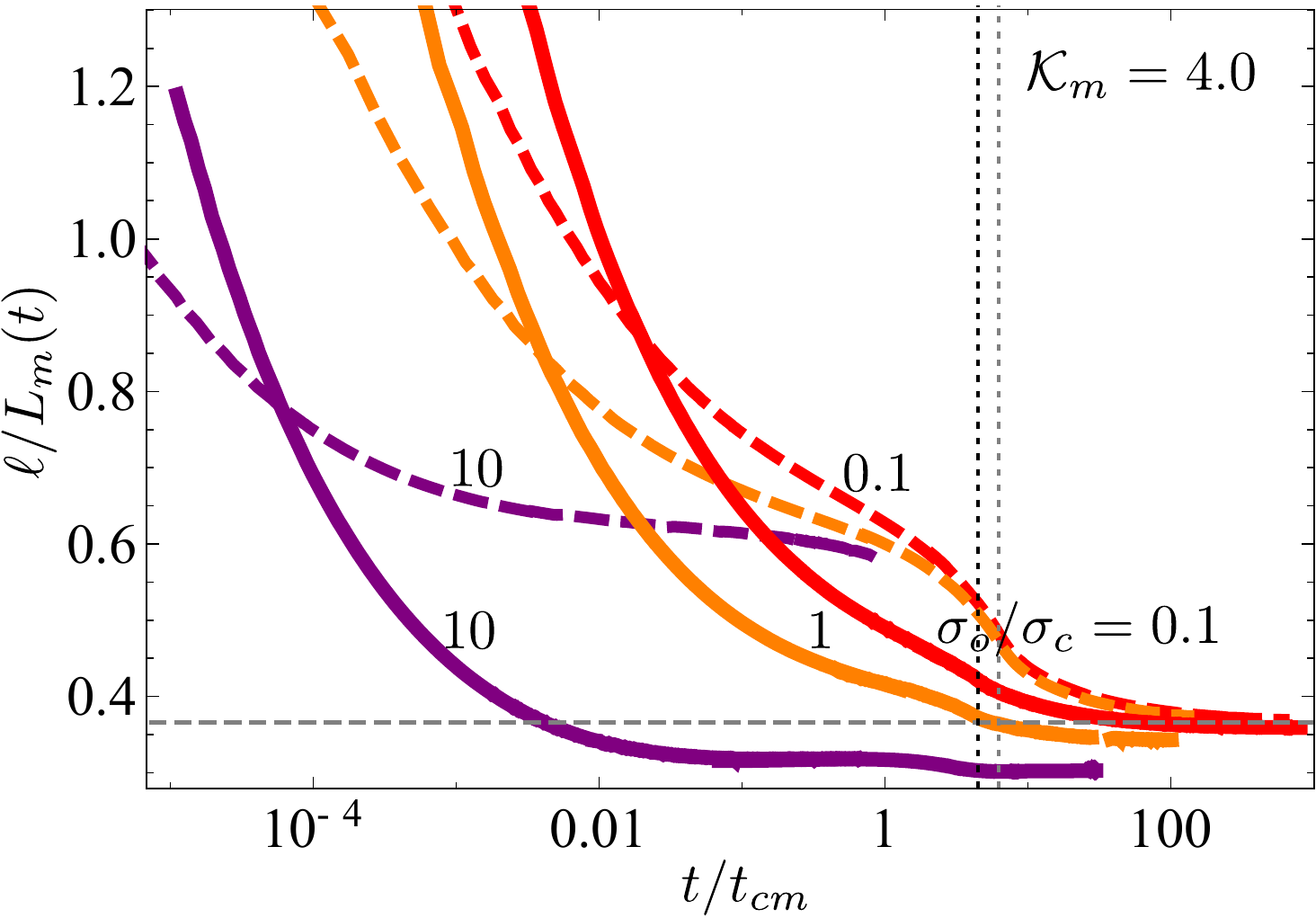}
\caption{Evolution of the dimensionless fracture half length $\ell/L_m(t)$ with $t/t_{cm}$ for $\mathcal{K}_m = 1-4$. The red, orange, and purple curves correspond respectively to $\sigma_o/\sigma_c=0.1, 1.0, 10$ and the solid and dashed curves correspond respectively to a rough ($\alpha_e=2$) and smooth fracture ($\alpha_e=0$). The dotted vertical lines indicate the cohesive zone nucleation period of $\sigma_o/\sigma_c=0.1$ for a smooth (gray) and a rough (black) fracture. The gray horizontal lines indicate the LHFM solutions in the zero fluid lag limit.}
\label{fig:lengthcm}
\end{figure}

\begin{figure}
\centering
\includegraphics[width=0.46\linewidth]{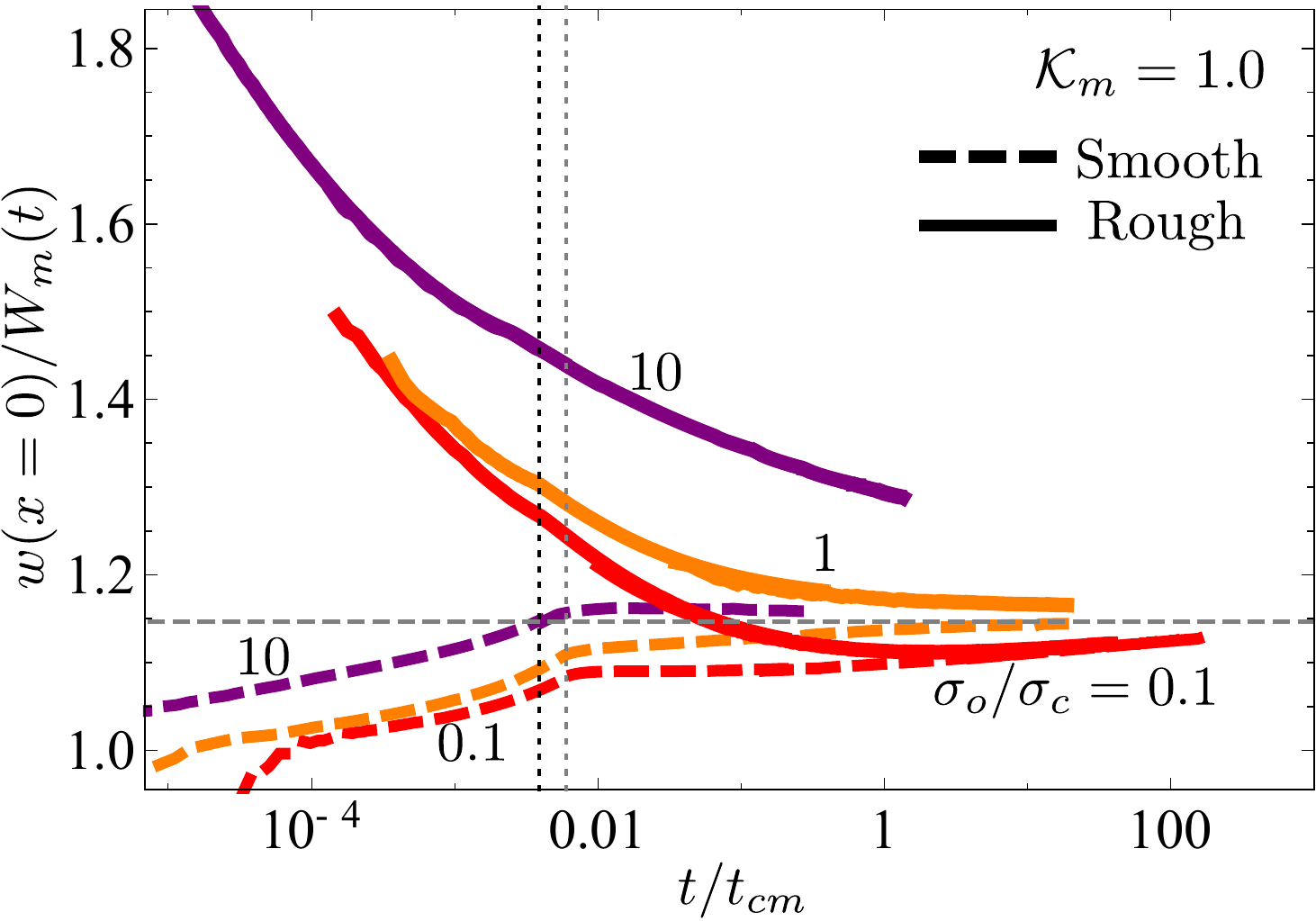}
\includegraphics[width=0.46\linewidth]{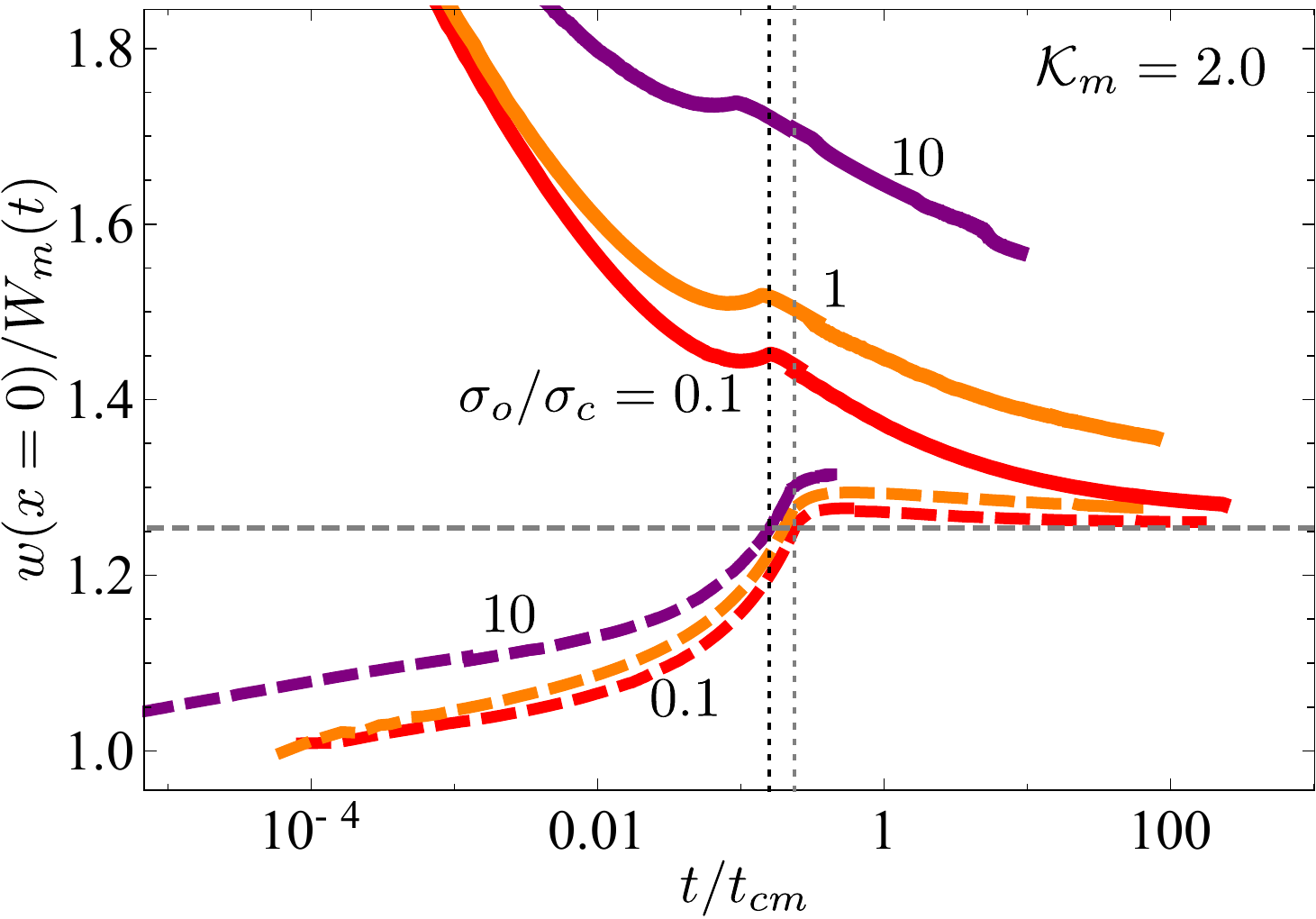}\\
\includegraphics[width=0.46\linewidth]{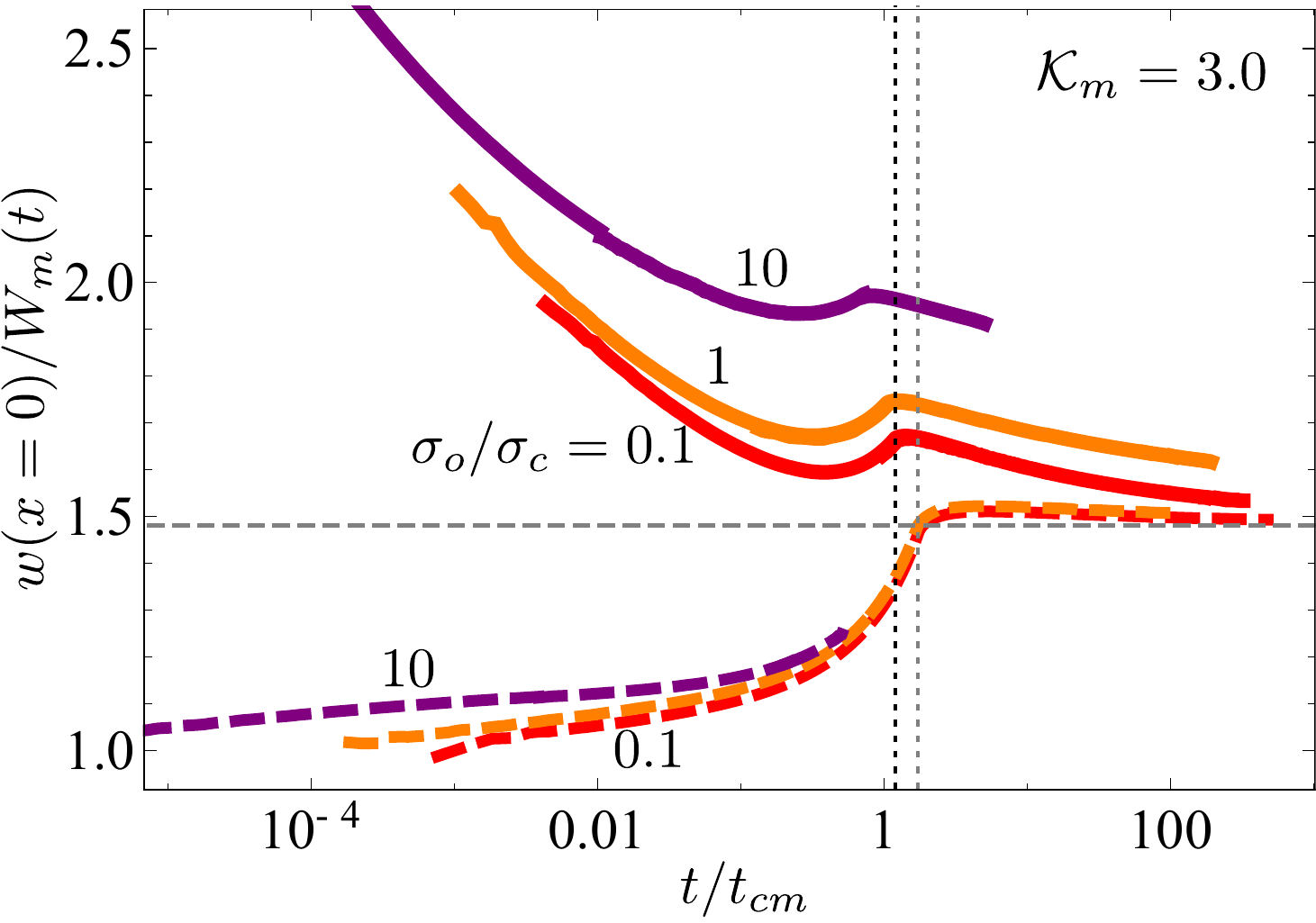}
\includegraphics[width=0.46\linewidth]{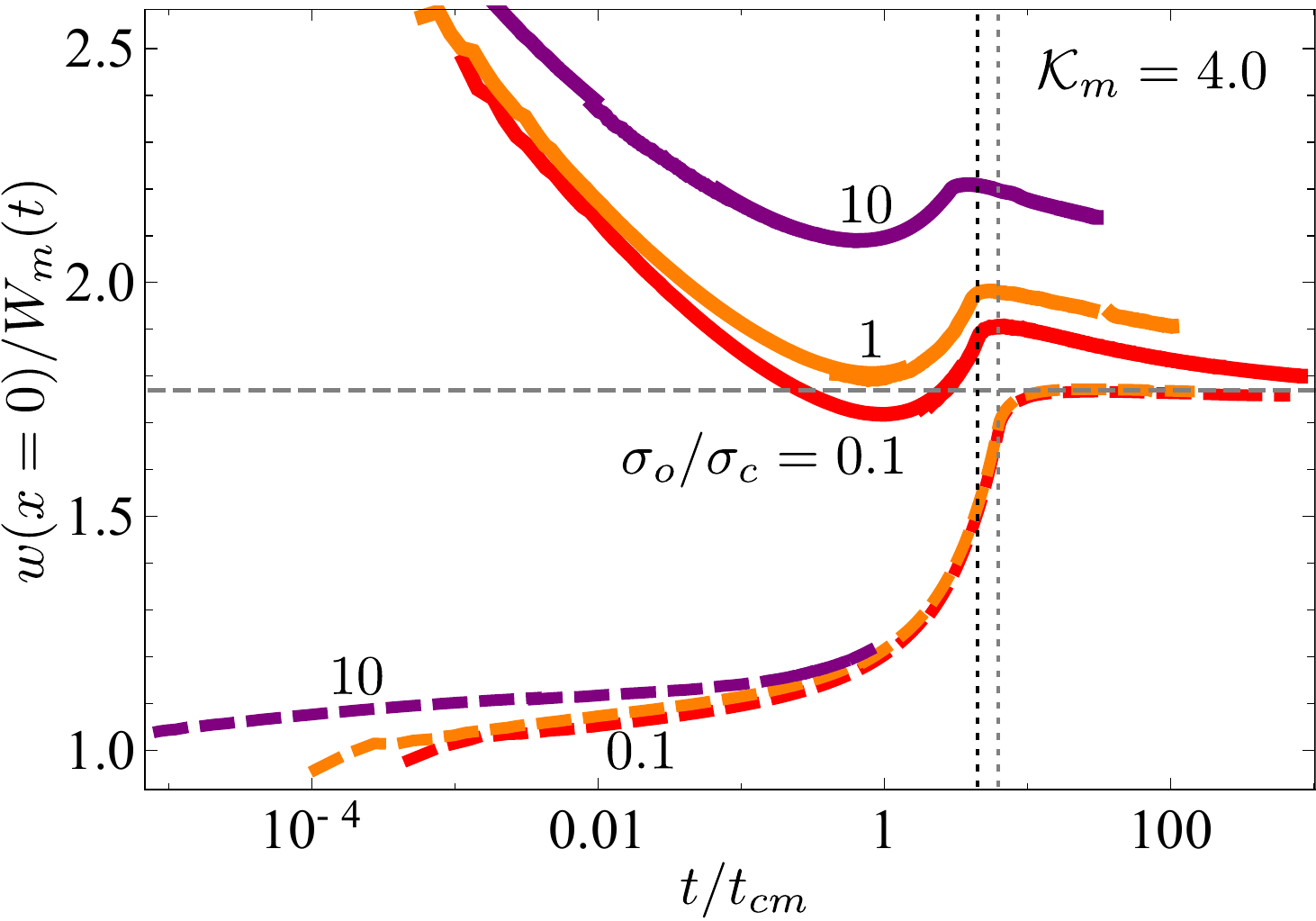}
\caption{Evolution of the inlet width $w(x=0)/W_m(t)$ with dimensionless time $t/t_{cm}$ for $\mathcal{K}_m = 1-4$. The red, orange, and purple curves correspond respectively to $\sigma_o/\sigma_c=0.1, 1.0, 10$ and the solid and dashed curves correspond respectively to a rough ($\alpha_e=2$) and smooth fracture ($\alpha_e=0$). The dotted vertical lines indicate the cohesive zone nucleation period of $\sigma_o/\sigma_c=0.1$ for a smooth (gray) and a rough (black) fracture. The gray horizontal lines indicate the LHFM solutions in the zero fluid lag limit.}
\label{fig:widthcm}
\end{figure}

\begin{figure}
\centering
\includegraphics[width=0.46\linewidth]{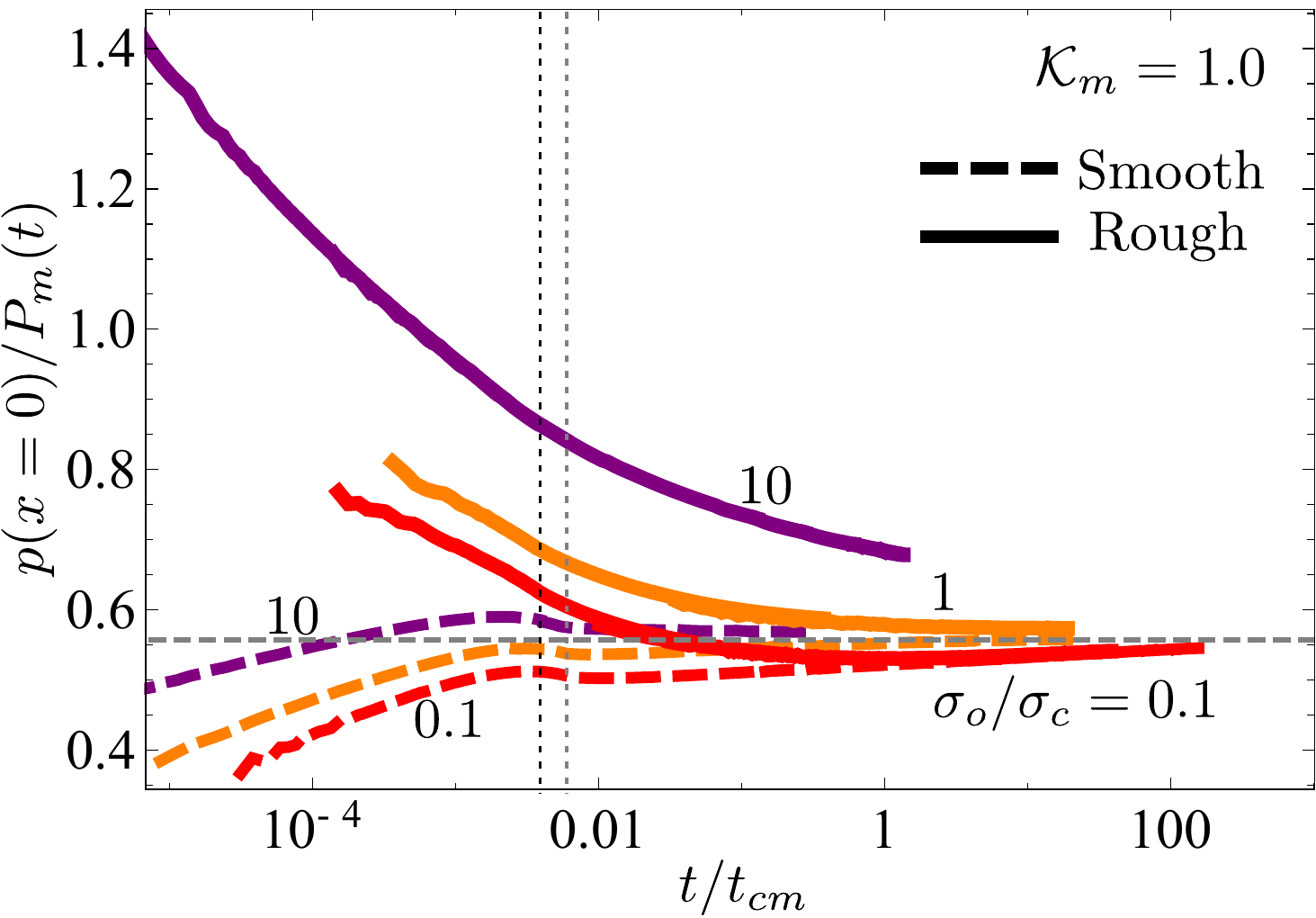}
\includegraphics[width=0.46\linewidth]{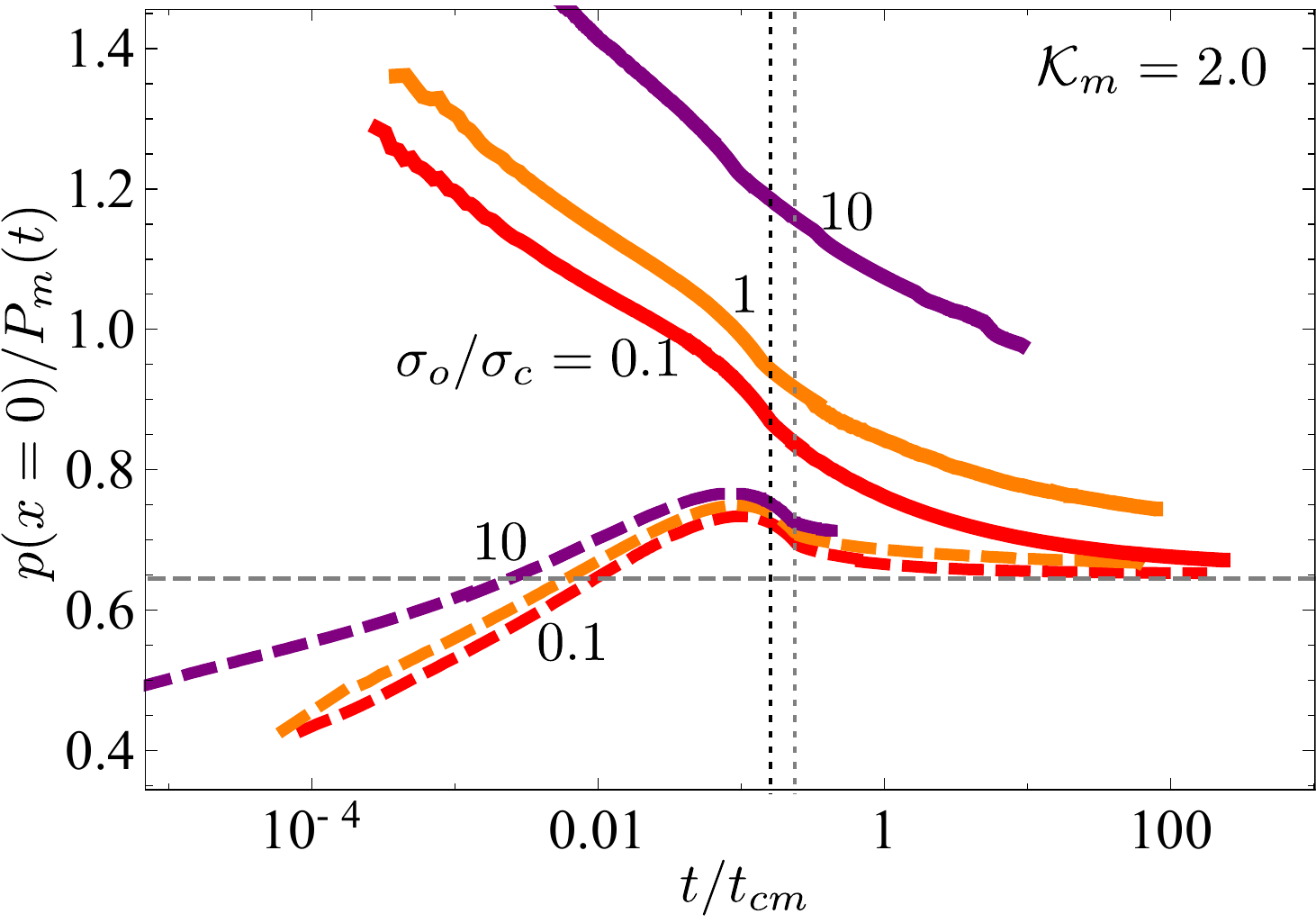}\\
\includegraphics[width=0.46\linewidth]{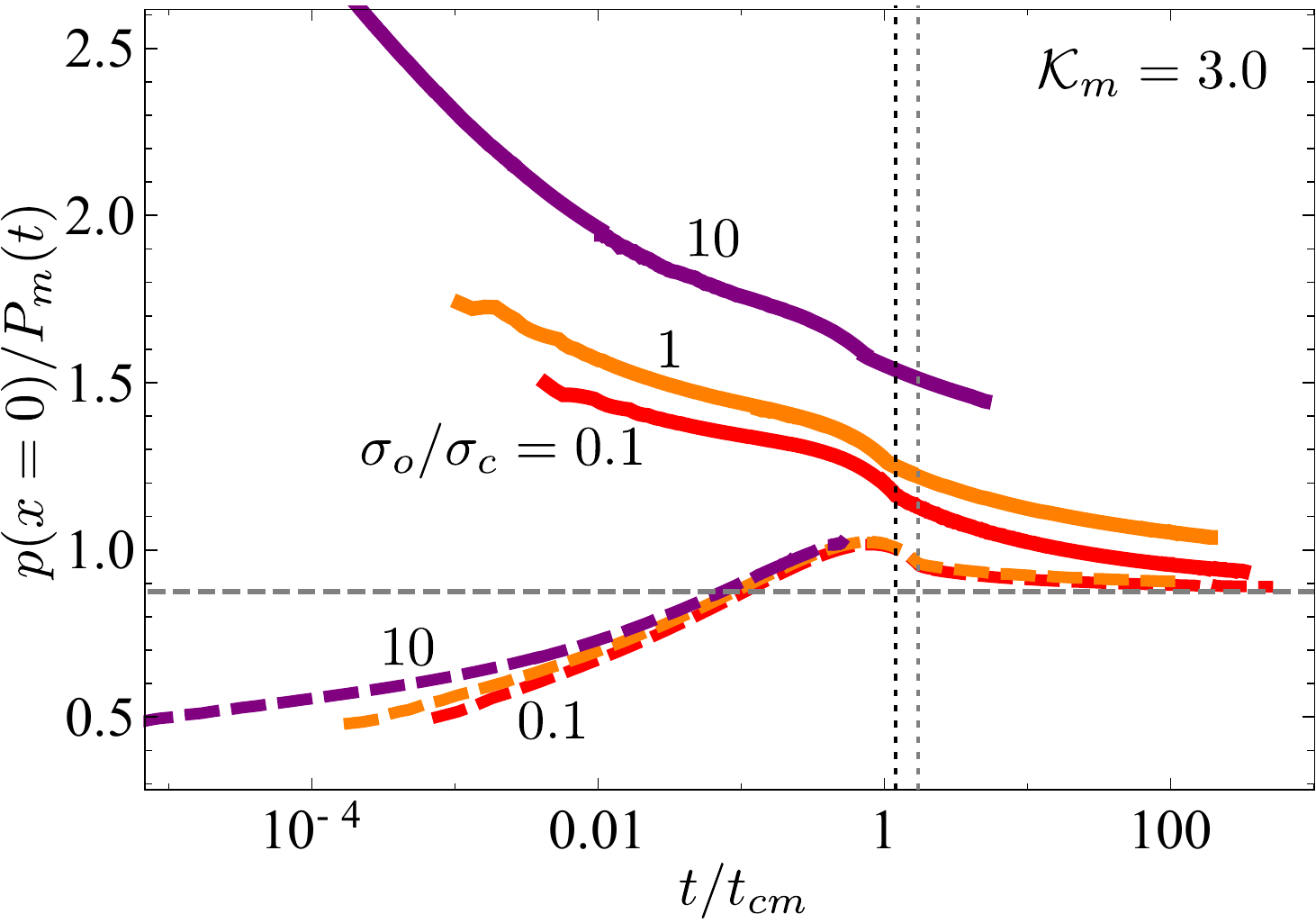}
\includegraphics[width=0.46\linewidth]{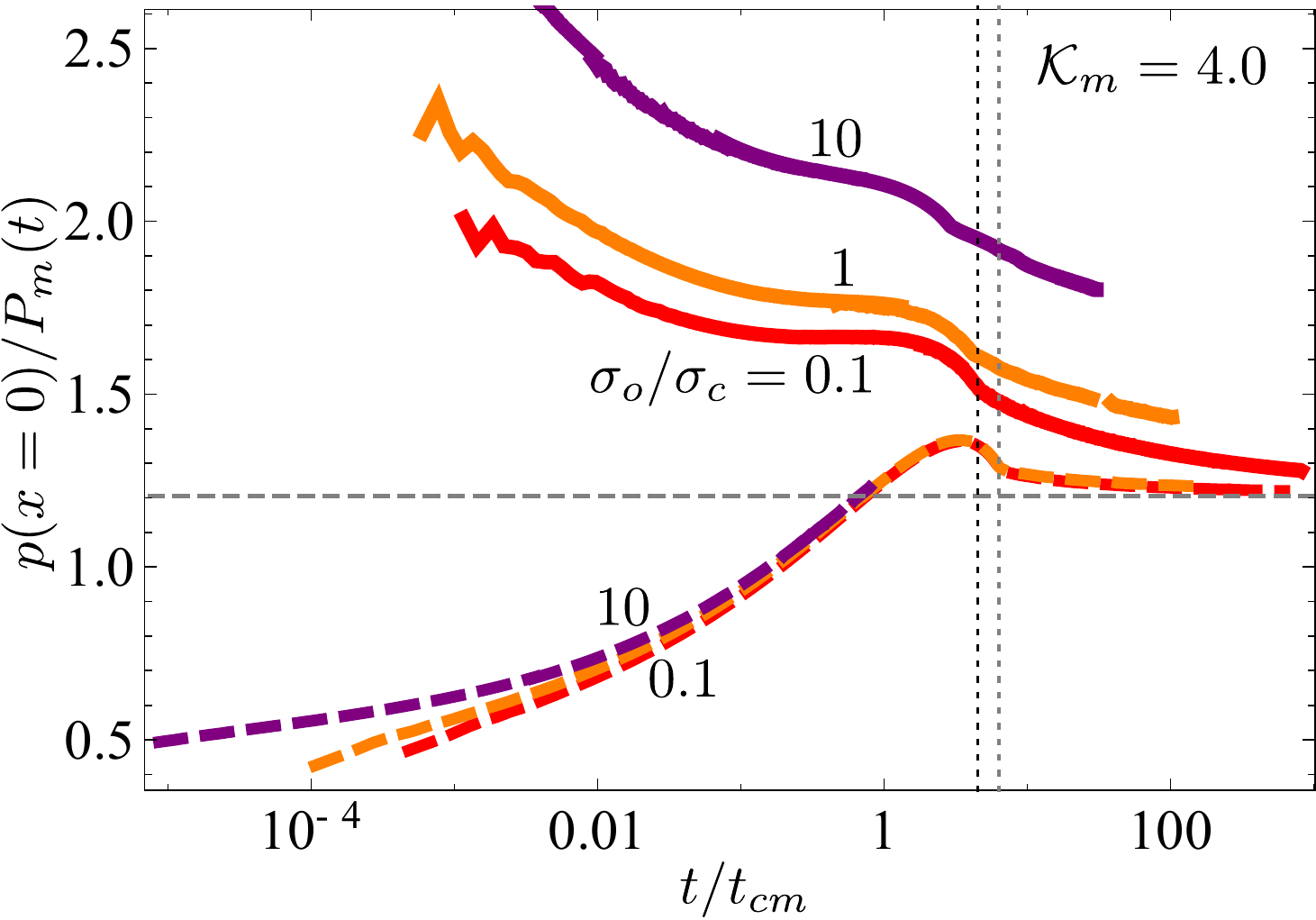}
\caption{Evolution of the inlet net pressure $p(x=0)/P_m(t)$ with $t/t_{cm}$ for $\mathcal{K}_m = 1-4$. The red, orange, and purple curves correspond respectively to $\sigma_o/\sigma_c=0.1, 1.0, 10$ and the solid and dashed curves correspond respectively to a rough ($\alpha_e=2$) and smooth fracture. The dotted vertical lines indicate the cohesive zone nucleation period of $\sigma_o/\sigma_c=0.1$ for a smooth (gray) and a rough (black) fracture. The gray horizontal lines indicate the LHFM solutions in the zero fluid lag limit.}
\label{fig:pressurecm}
\end{figure}

The additional resistance to fluid flow associated with fracture aperture roughness has a profound impact on  growth  both at the nucleation and intermediate stage. The effect is amplified for larger $\sigma_o/\sigma_c$ and larger $\mathcal{K}_m$. This can be well observed from the evolution of length, inlet width and net pressures displayed on Figures \ref{fig:lengthcm}, \ref{fig:widthcm} and \ref{fig:pressurecm} respectively. In particular the net pressure and width are significantly larger compared to the smooth cohesive zone and LHFM cases, while the dimensionless length is shorter after nucleation.

The convergence toward the LHFM solutions with zero lag 
are in some cases not fully achieved even at very large time ($t\gg t_{cm}$ especially for the large $\sigma_o/\sigma_c$ cases. As mentioned earlier, we actually stop these simulations when the fluid fraction $\xi_f=\ell_f/\ell$ reached 0.99 or the fracture length was within five percent of the LHFM solutions.

\paragraph{Faster nucleation of the cohesive zone}

\begin{figure}
\centering
\includegraphics[width=0.46\linewidth]{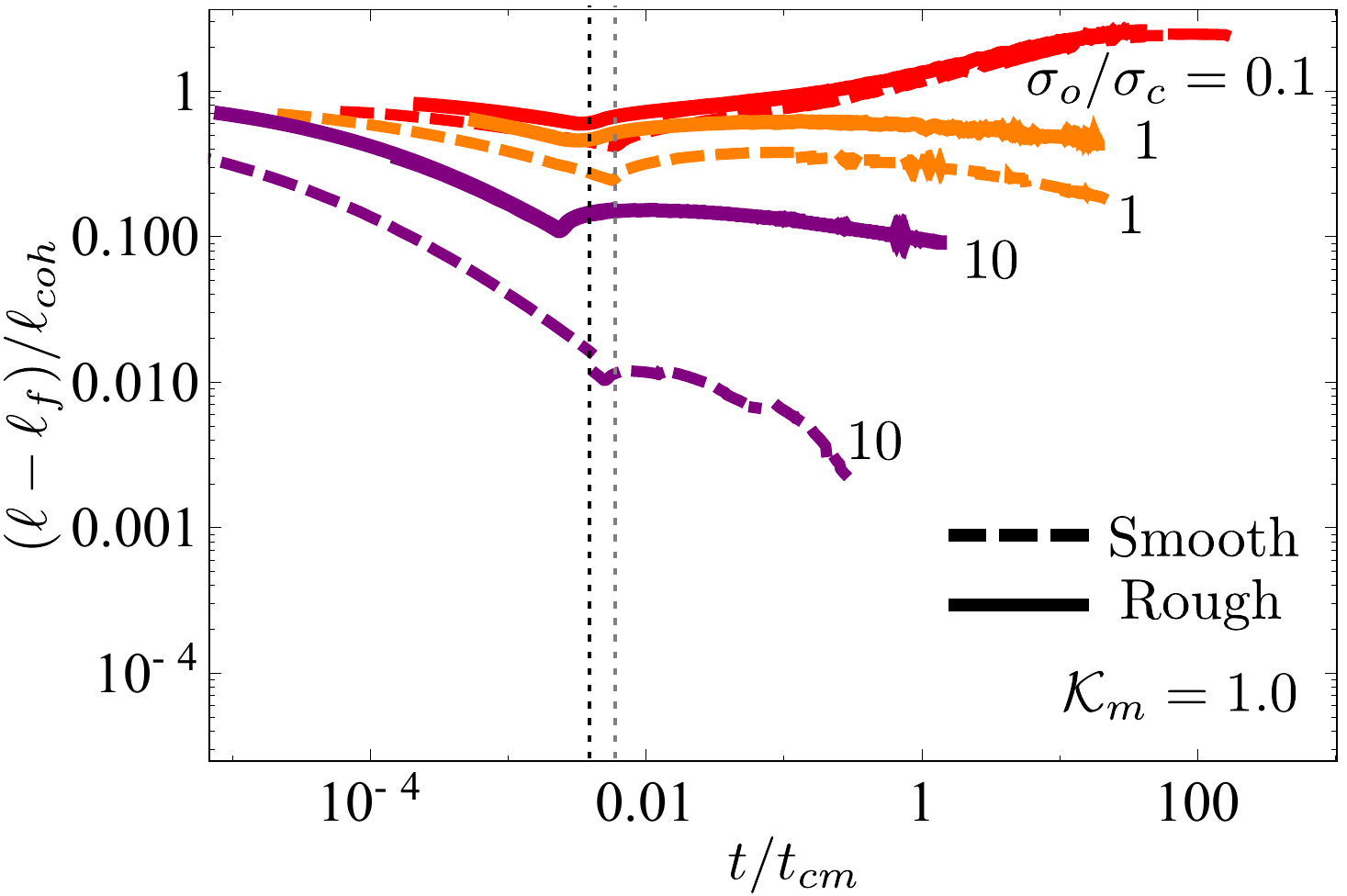}
\includegraphics[width=0.46\linewidth]{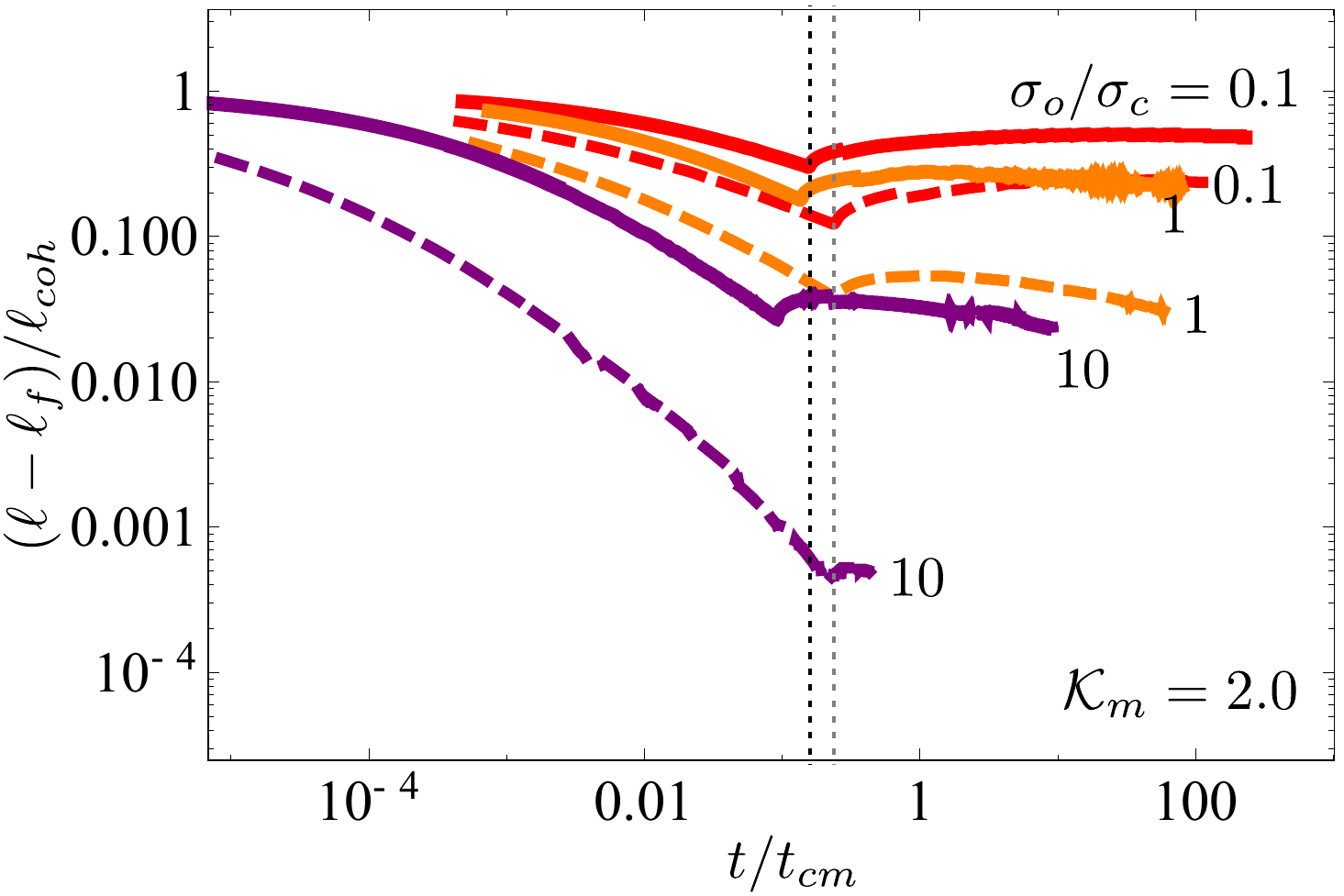}\\
\includegraphics[width=0.46\linewidth]{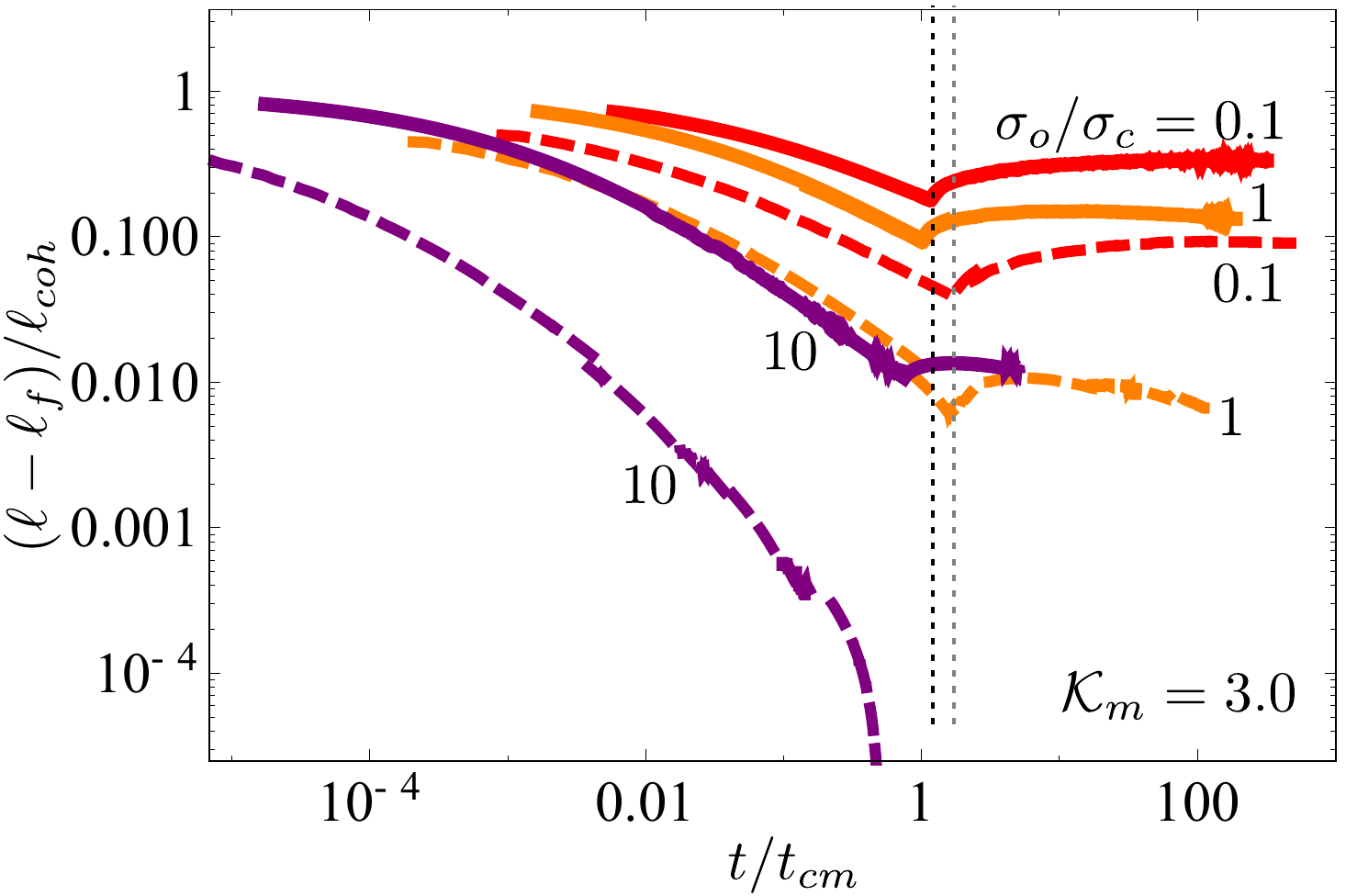}
\includegraphics[width=0.46\linewidth]{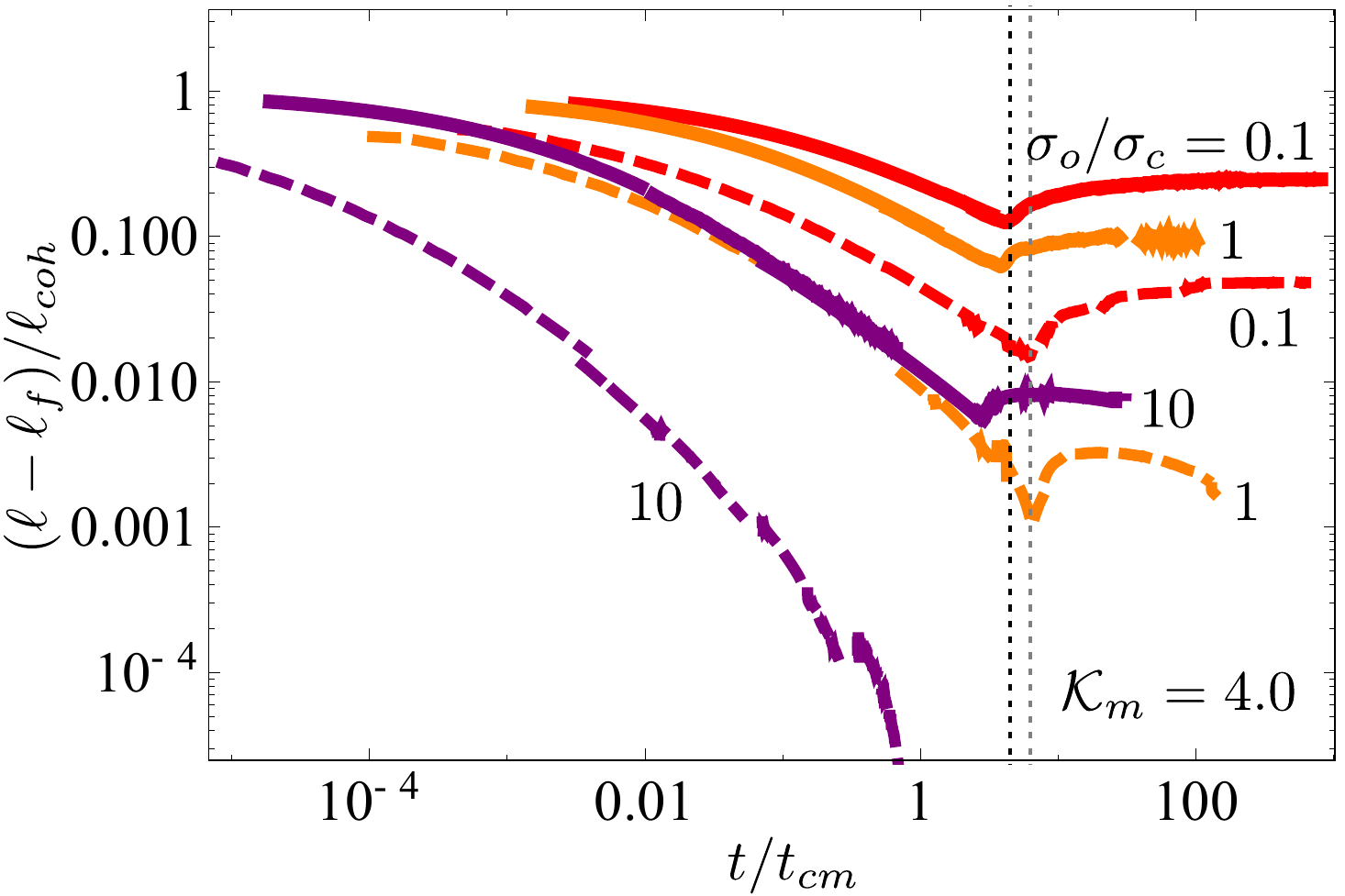}
\caption{Time evolution of the ratio between the lag and cohesive zone sizes $(\ell-\ell_f)/\ell_{coh}$ for $\mathcal{K}_m = 1-4$. The red, orange, and purple curves correspond to  $\sigma_o/\sigma_c=0.1, 1.0, 10$ respectively and the solid and dashed curves correspond respectively to a rough ($\alpha_e=2$) and smooth fracture ($\alpha_e=0$). The dotted vertical lines indicate the cohesive zone nucleation period of $\sigma_o/\sigma_c=0.1$ for a smooth (gray) and a rough (black) fracture.}
\label{fig:penetrationcm}
\end{figure}

As the fluid front is necessarily embedded in the cohesive zone during the nucleation stage, the effect of roughness is significant during nucleation.
For the same stress ratio $\sigma_o/\sigma_c$ and dimensionless toughness $\mathcal{K}_m$, 
 roughness influences the fracture growth by decreasing the fluid front penetration into the cohesive zone as illustrated by the evolution of the ratio between the lag and cohesive zone sizes in Fig.~\ref{fig:penetrationcm}.

\begin{figure}
\centering
\begin{tabular}{cc}
     \includegraphics[height=0.32\linewidth]{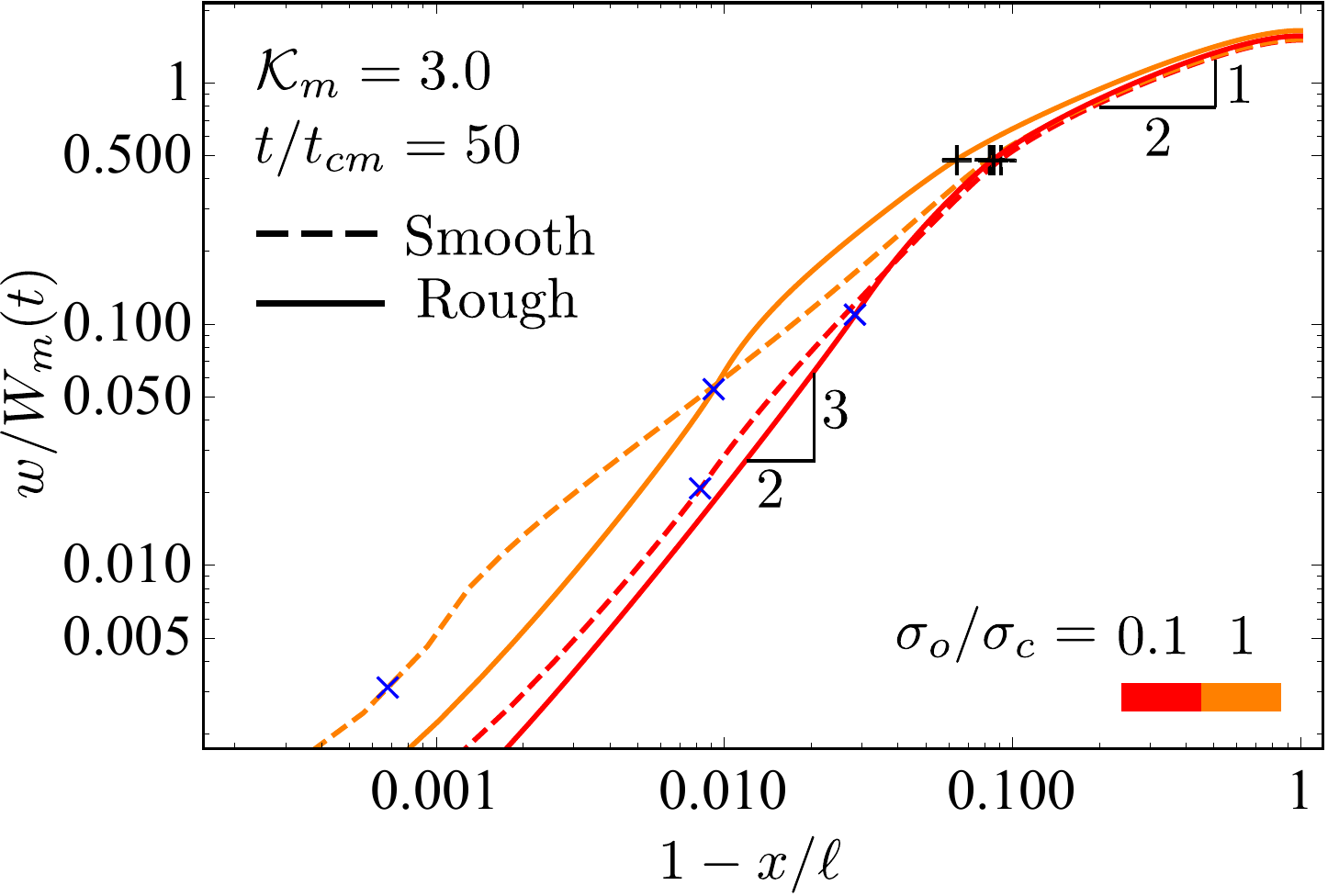}&
\includegraphics[height=0.32\linewidth]{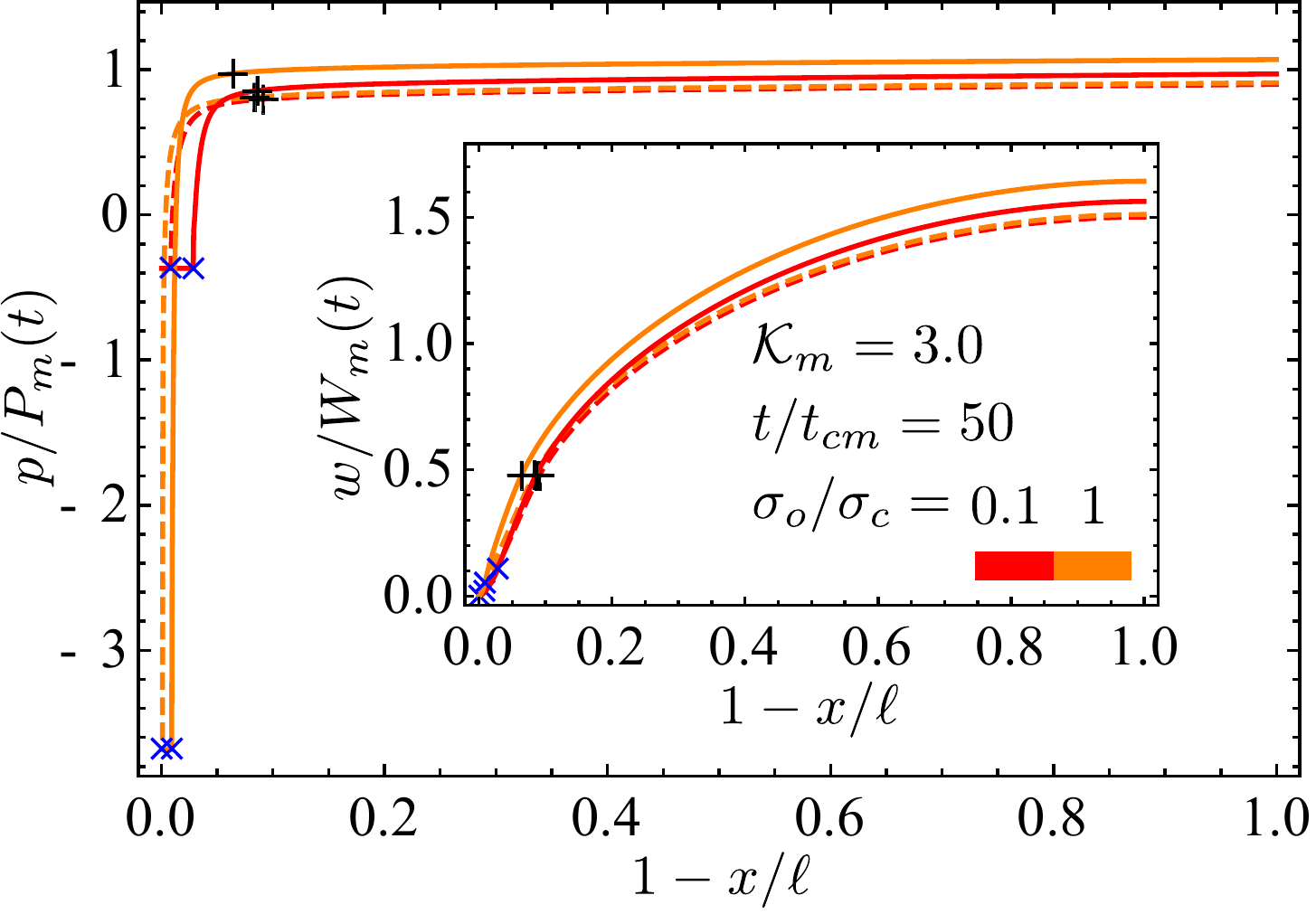} \\
     (a) & (b) 
\end{tabular}
\caption{a) Dimensionless opening and b) net pressure profiles at $t/t_{cm}=50$ for  $\mathcal{K}_{m}=3.0$. “$+$” indicates the boundary of the cohesive zone and “$\times$” indicates the fluid front location. The red and orange curves correspond to $\sigma_o/\sigma_c=0.1, 1.0$ respectively. The solid and dashed curves indicate respectively a rough ($\alpha_e=2$) and smooth fracture ($\alpha_e=0$). }
\label{fig:RTip-Asymptote}
\end{figure}

\begin{figure}
\centering
\includegraphics[width=0.46\linewidth]{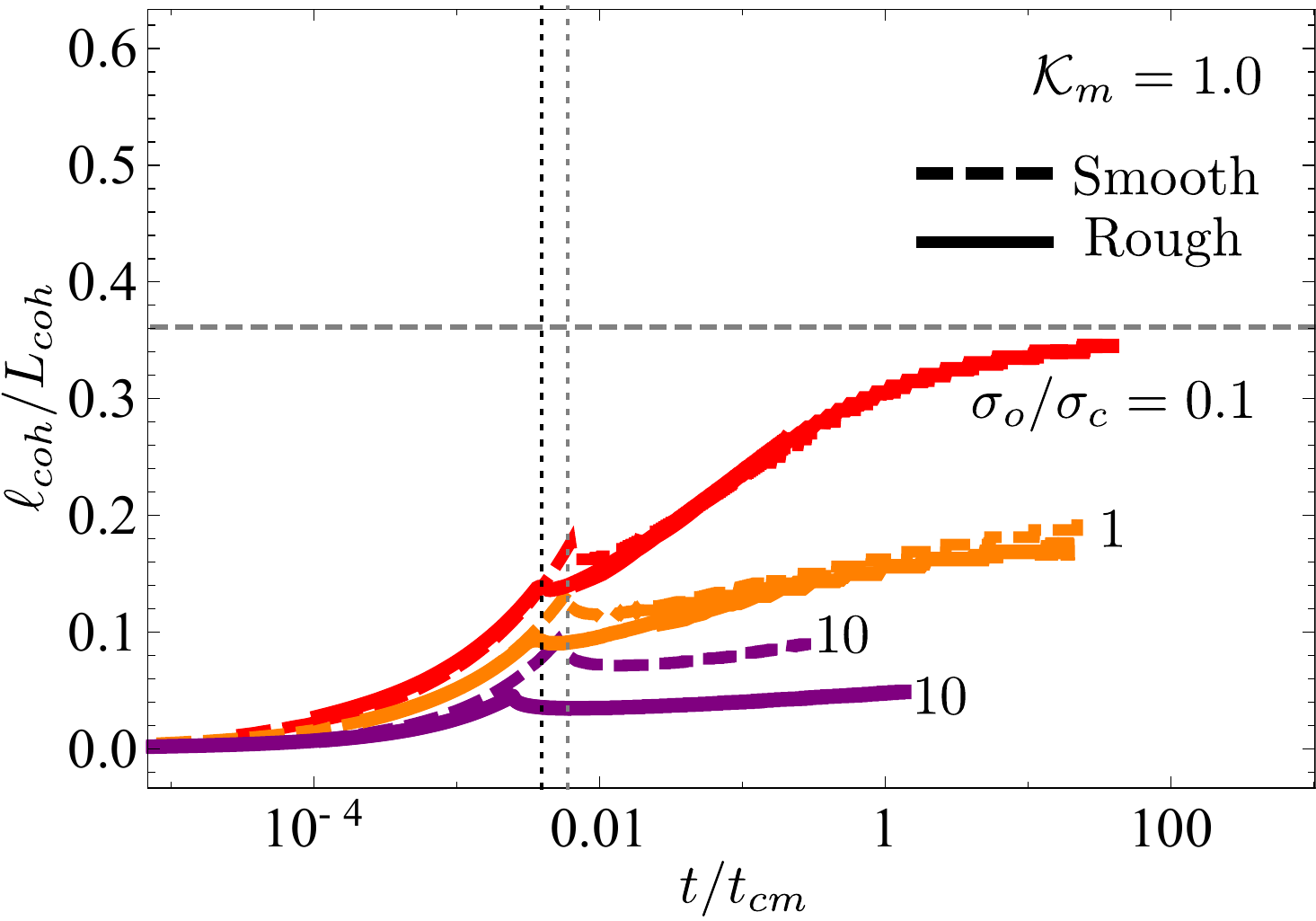}
\includegraphics[width=0.46\linewidth]{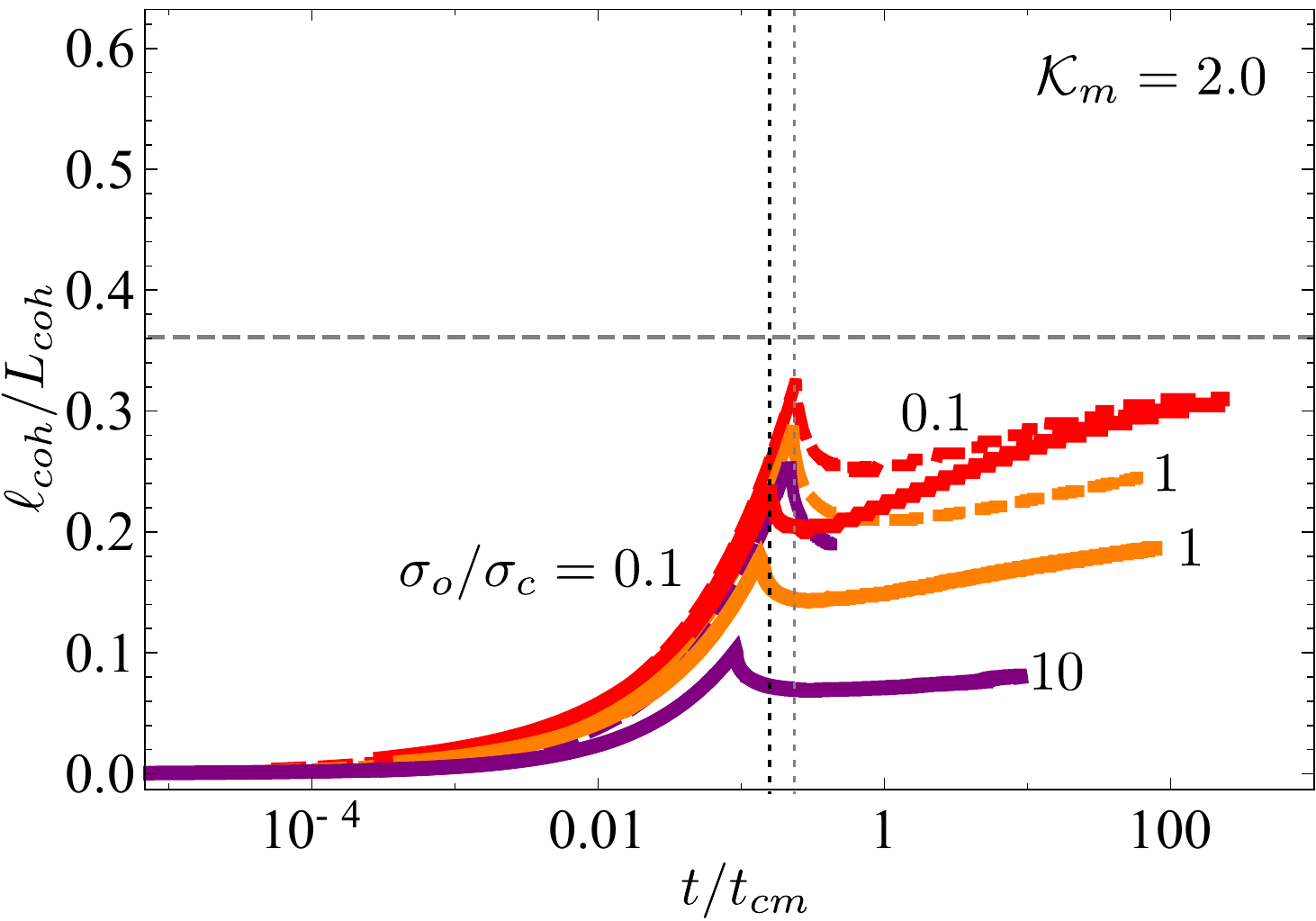}\\
\includegraphics[width=0.46\linewidth]{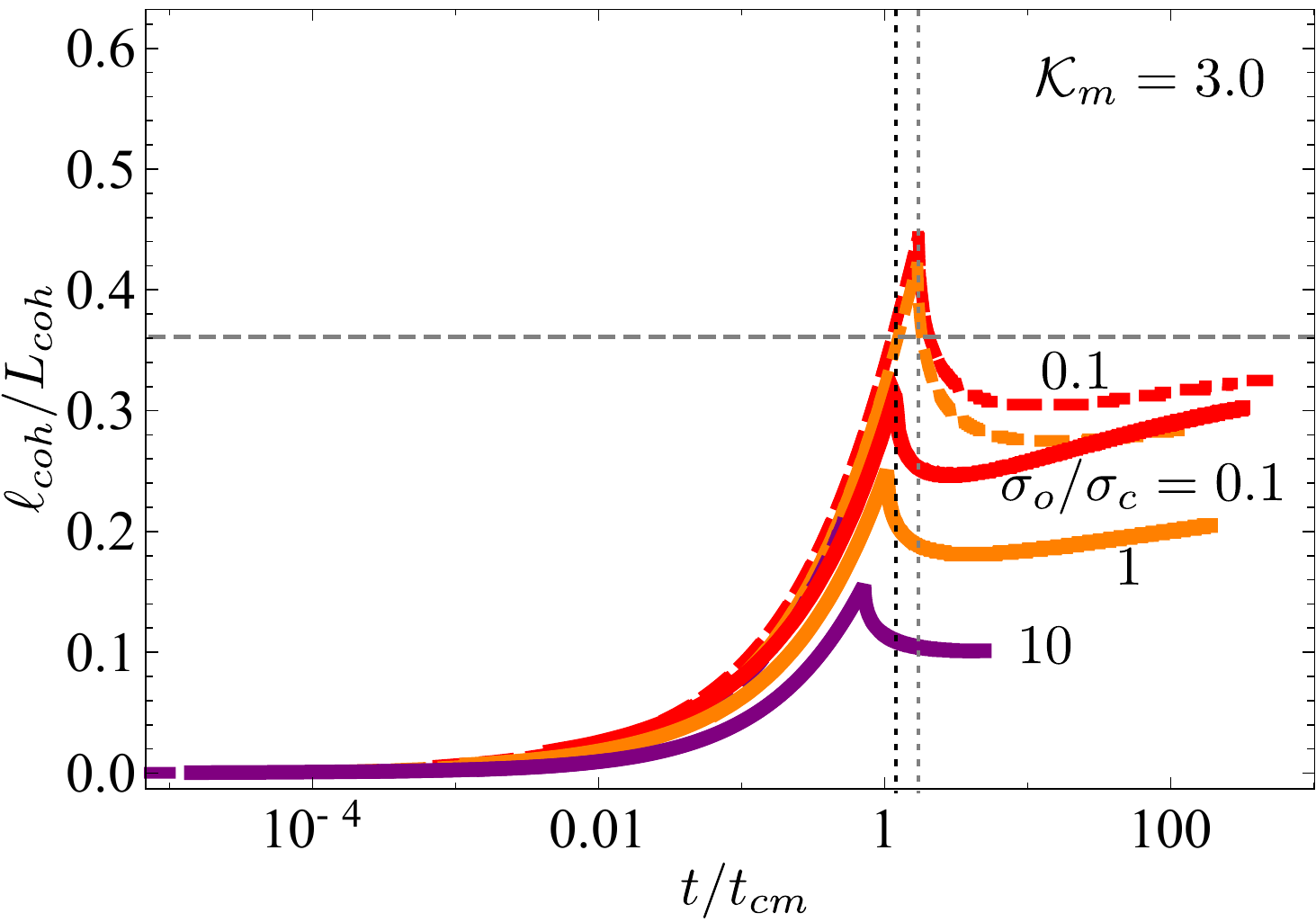}
\includegraphics[width=0.46\linewidth]{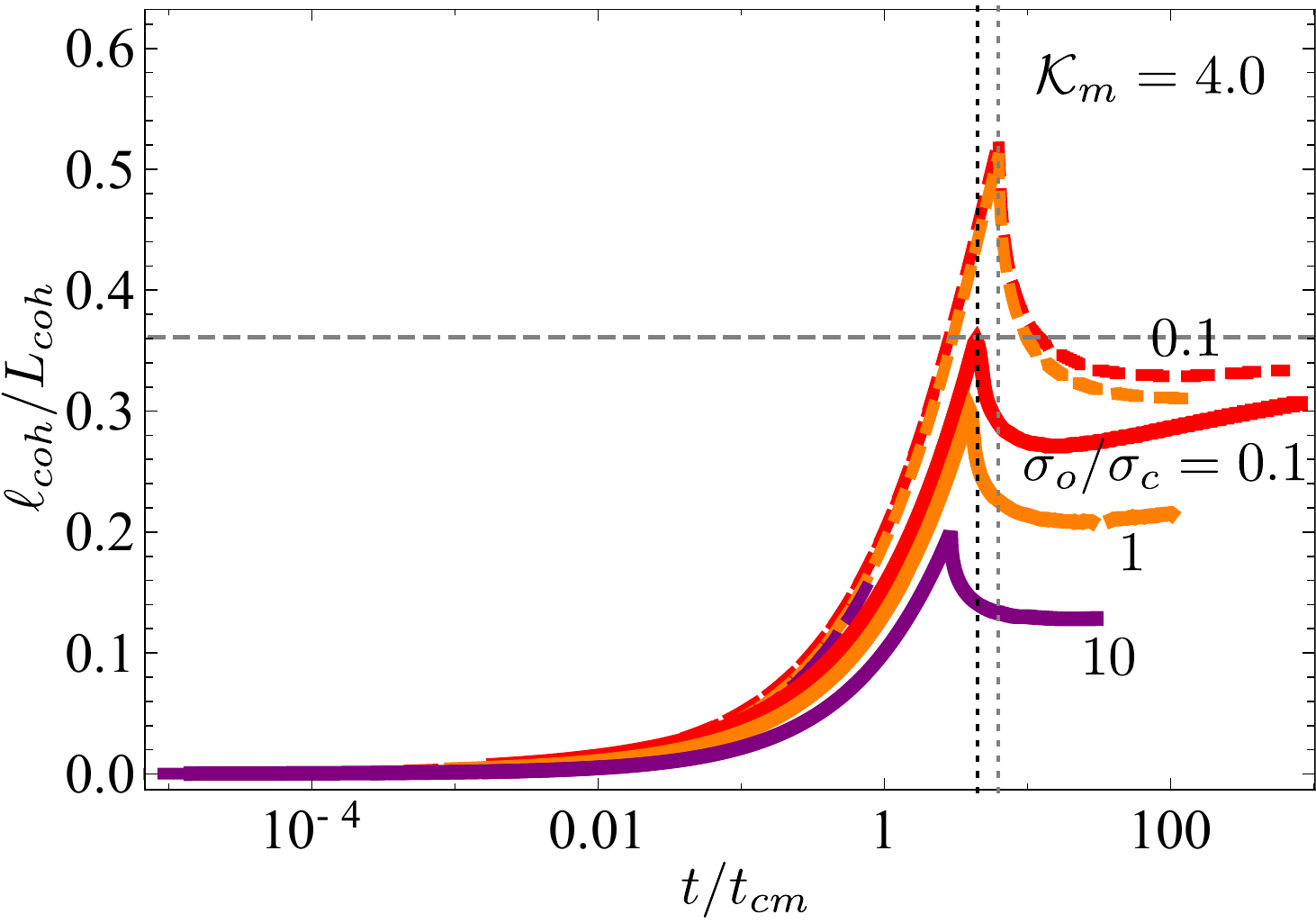}
\caption{Time evolution of the cohesive length $\ell_{coh}/L_{coh}$ for $\mathcal{K}_m = 1-4$ . The red, orange, and purple curves correspond respectively to $\sigma_o/\sigma_c=0.1, 1.0, 10$ and the solid and dashed curves correspond respectively to a rough ($\alpha_e=2$) and smooth fracture ($\alpha_e=0$). The dotted vertical lines indicate the cohesive zone nucleation period of $\sigma_o/\sigma_c=0.1$ for a smooth (gray) and a rough (black) fracture. The dashed horizontal line represents the small-scale yielding asymptote ($\approx 0.115 \pi$) of the cohesive zone length for the linear-softening cohesive model \citep{DeTa2010}.}
\label{fig:cohlengthcm}
\end{figure}

\begin{figure}
\centering
\includegraphics[width=0.46\linewidth]{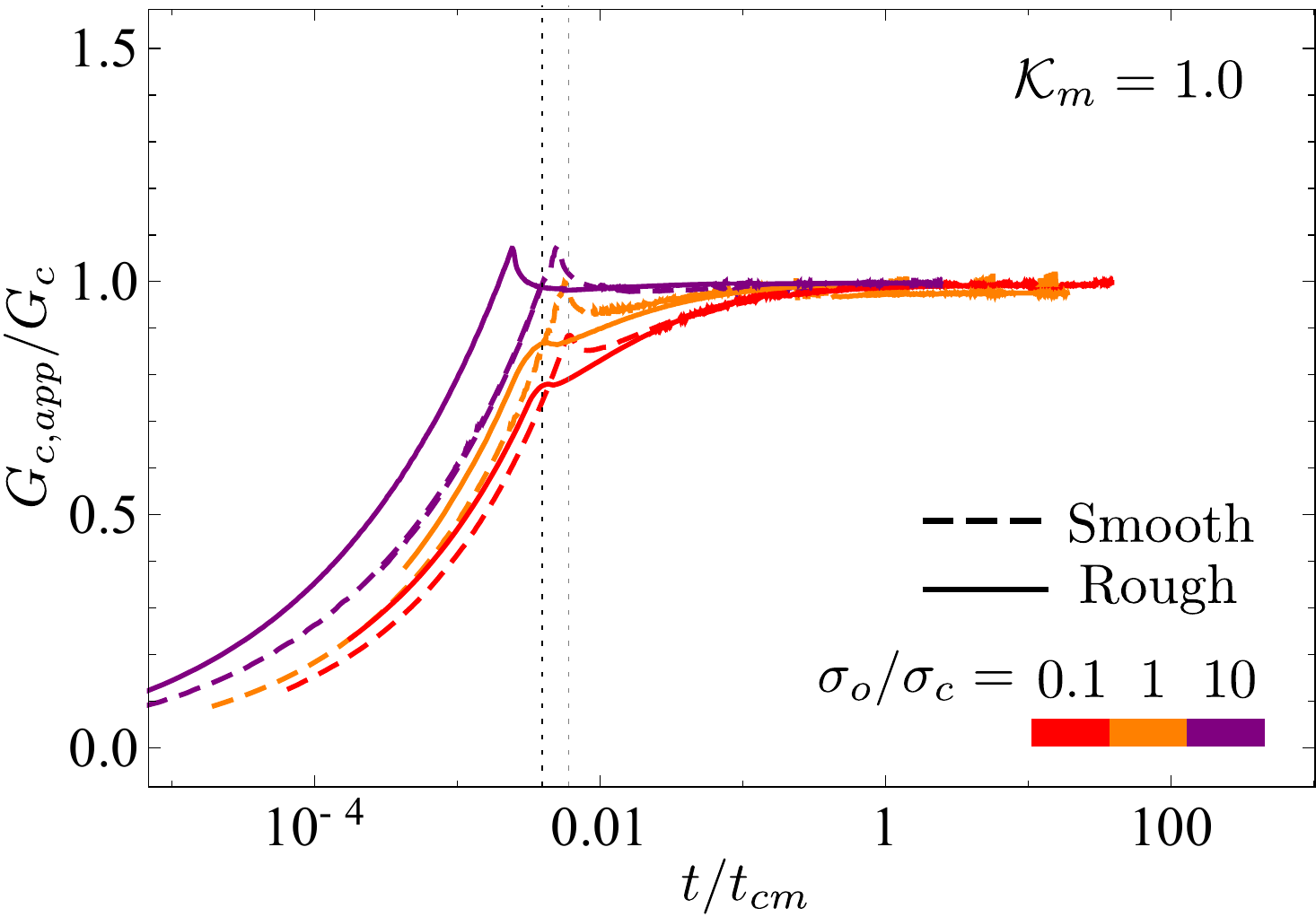} 
\includegraphics[width=0.46\linewidth]{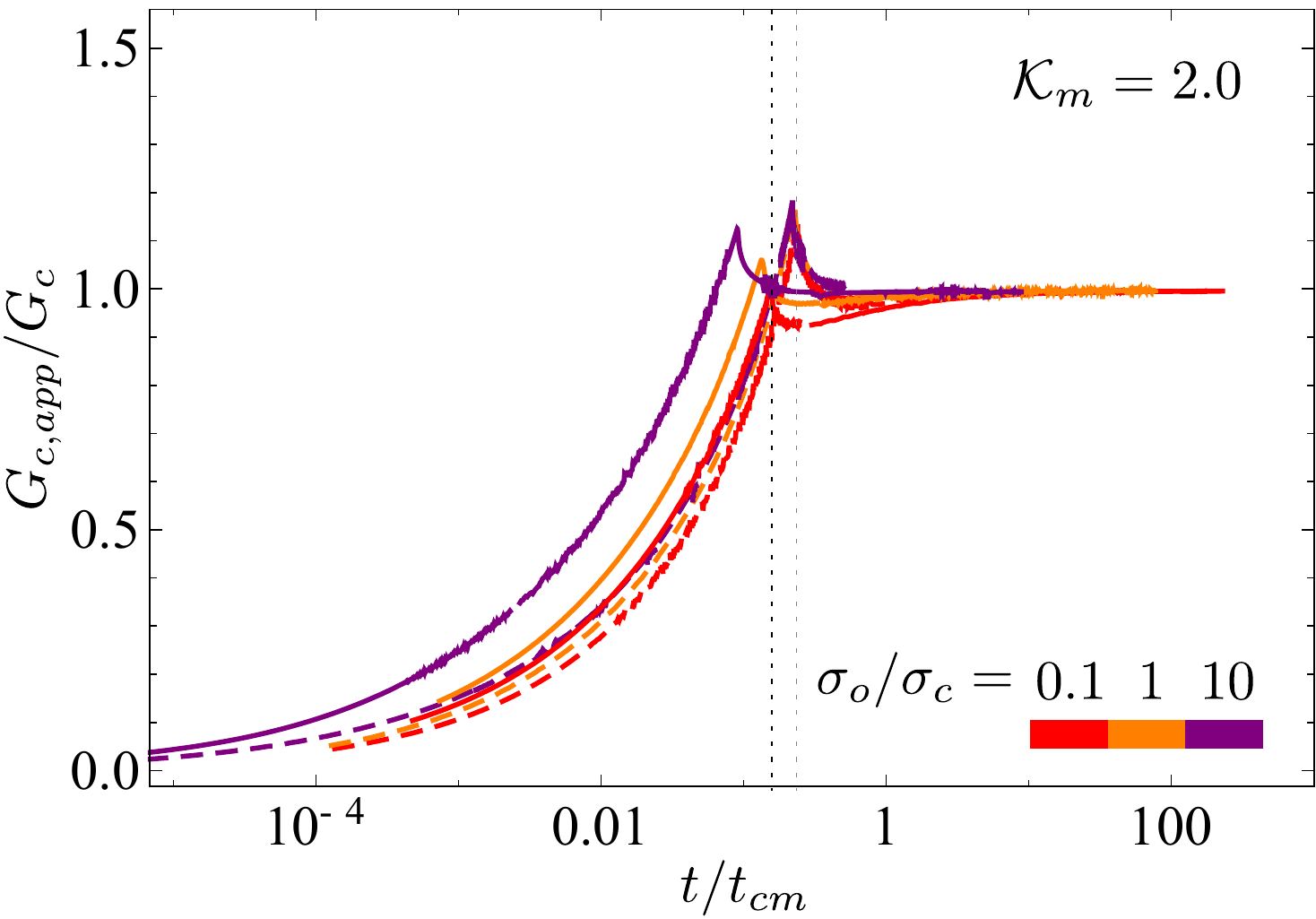} \\
\includegraphics[width=0.46\linewidth]{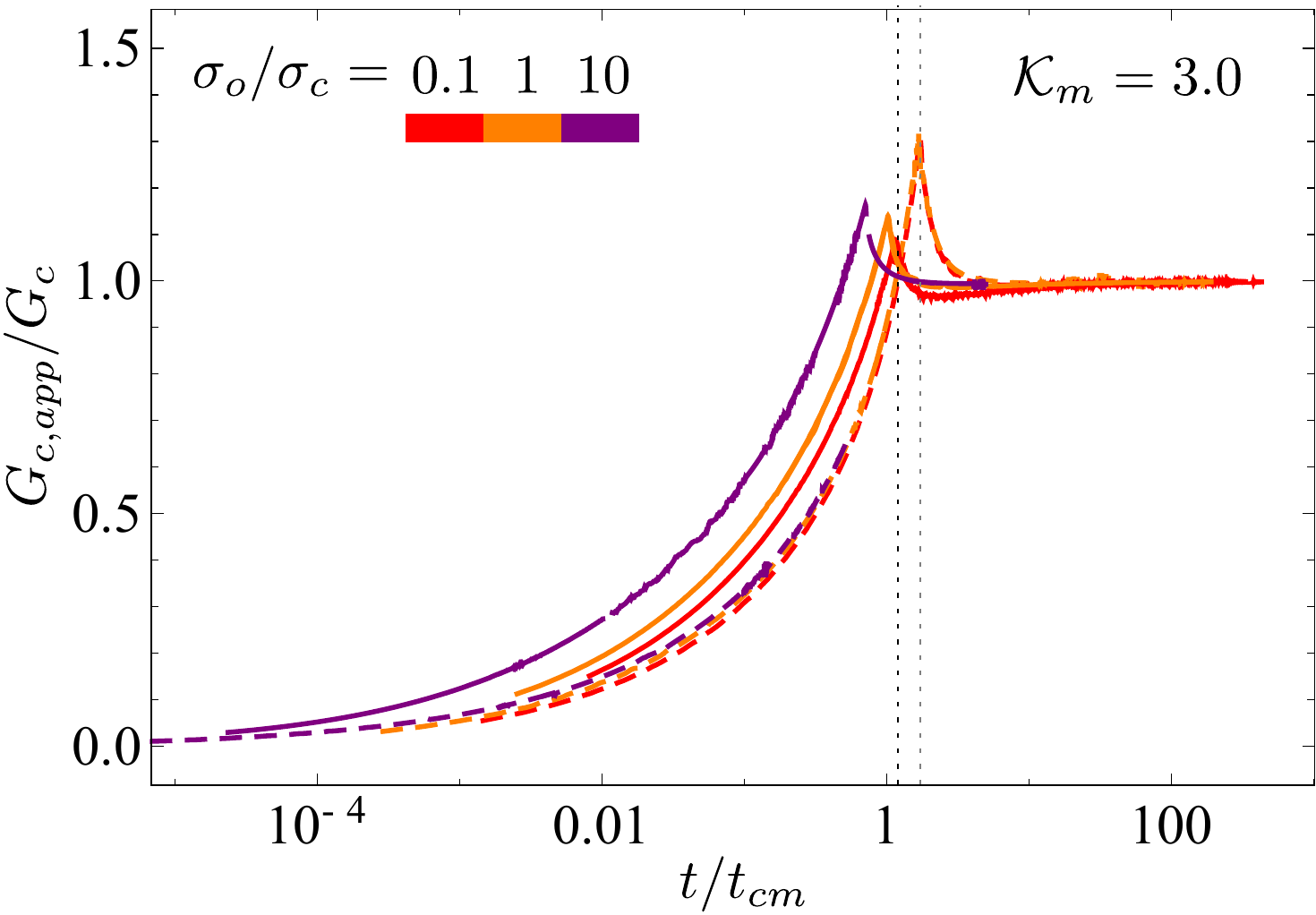} 
\includegraphics[width=0.46\linewidth]{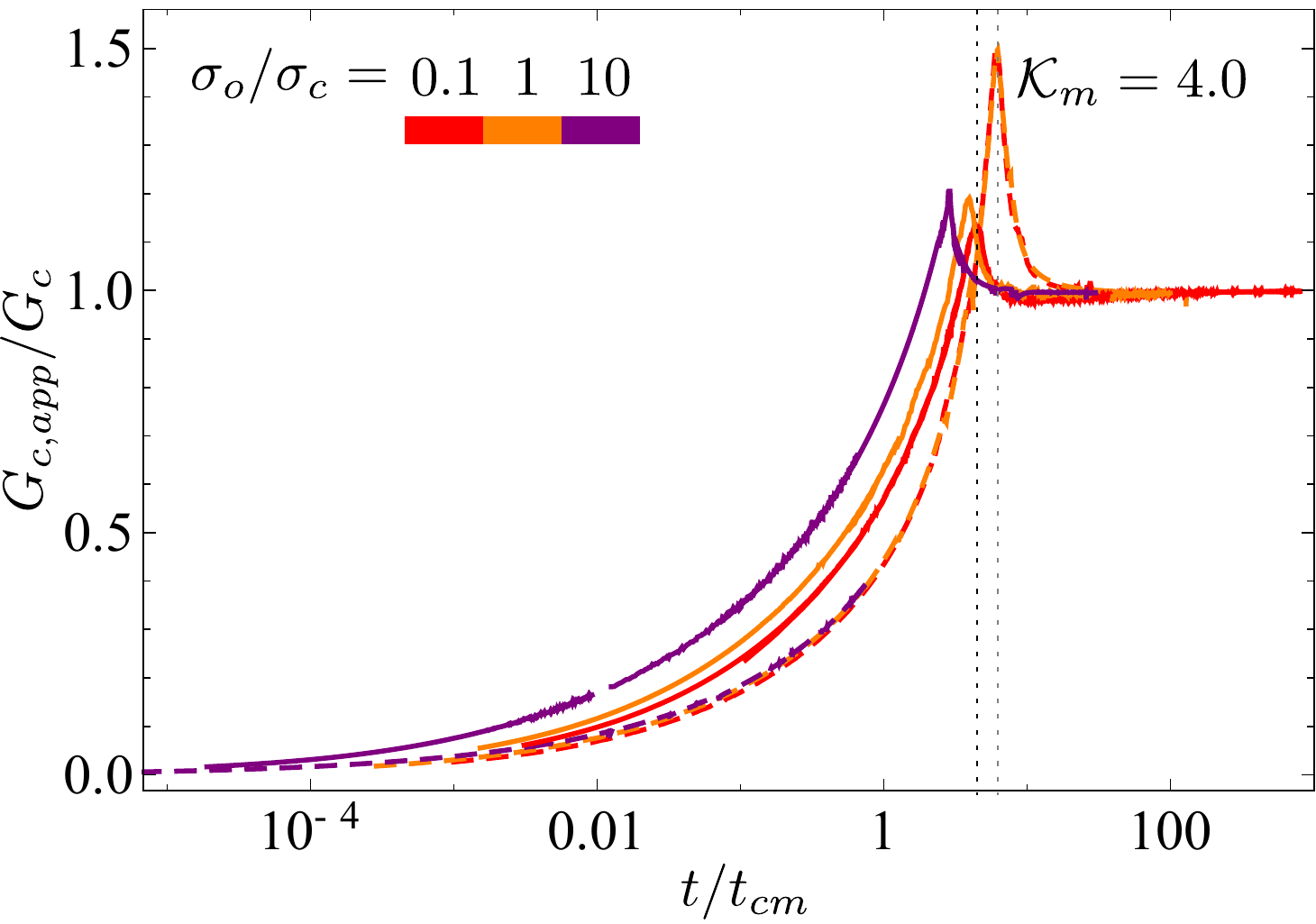} 
\caption{Time evolution of the apparent fracture energy $G_{c,app}/G_c$ for $\mathcal{K}_m = 1-4$. The red, orange, and purple curves correspond respectively to $\sigma_o/\sigma_c=0.1, 1.0, 10$ and the solid and dashed curves correspond respectively to a rough ($\alpha_e=2$) and smooth fracture ($\alpha_e=0$). The dotted vertical lines indicate the cohesive zone nucleation period of $\sigma_o/\sigma_c=0.1$ for a smooth (gray) and a rough (black) fracture.}
\label{fig:apparentdissipationcm}
\end{figure}

The increase of the fluid flow resistance brought by roughness can also be observed on the net pressure and width profiles (see Fig.~\ref{fig:RTip-Asymptote}). 
The steeper pressure gradient near the fluid front results in a wider opening in the fluid-filled part of the fracture, ultimately making it easier to completely debond the cohesive tractions ($w>w_c$) near the tip. The nucleation process is therefore accelerated as shown in  Figs.~\ref{fig:cohlengthcm},~\ref{fig:apparentdissipationcm}. 
The cohesive length is shorter at nucleation compared to the smooth case, but tends to converge to the same value as the smooth case at late time at least for smaller dimensionless toughness. In spite of the lack of stabilized cohesive zone length for the large dimensionless toughness / large $\sigma_o/\sigma_c$ cases, the trend for $\sigma_o/\sigma_c=1$ hints a similar behavior for larger confinement albeit at a much later dimensionless time. 

\paragraph{Additional energy dissipation}

\begin{figure}
\centering
\includegraphics[width=0.46\linewidth]{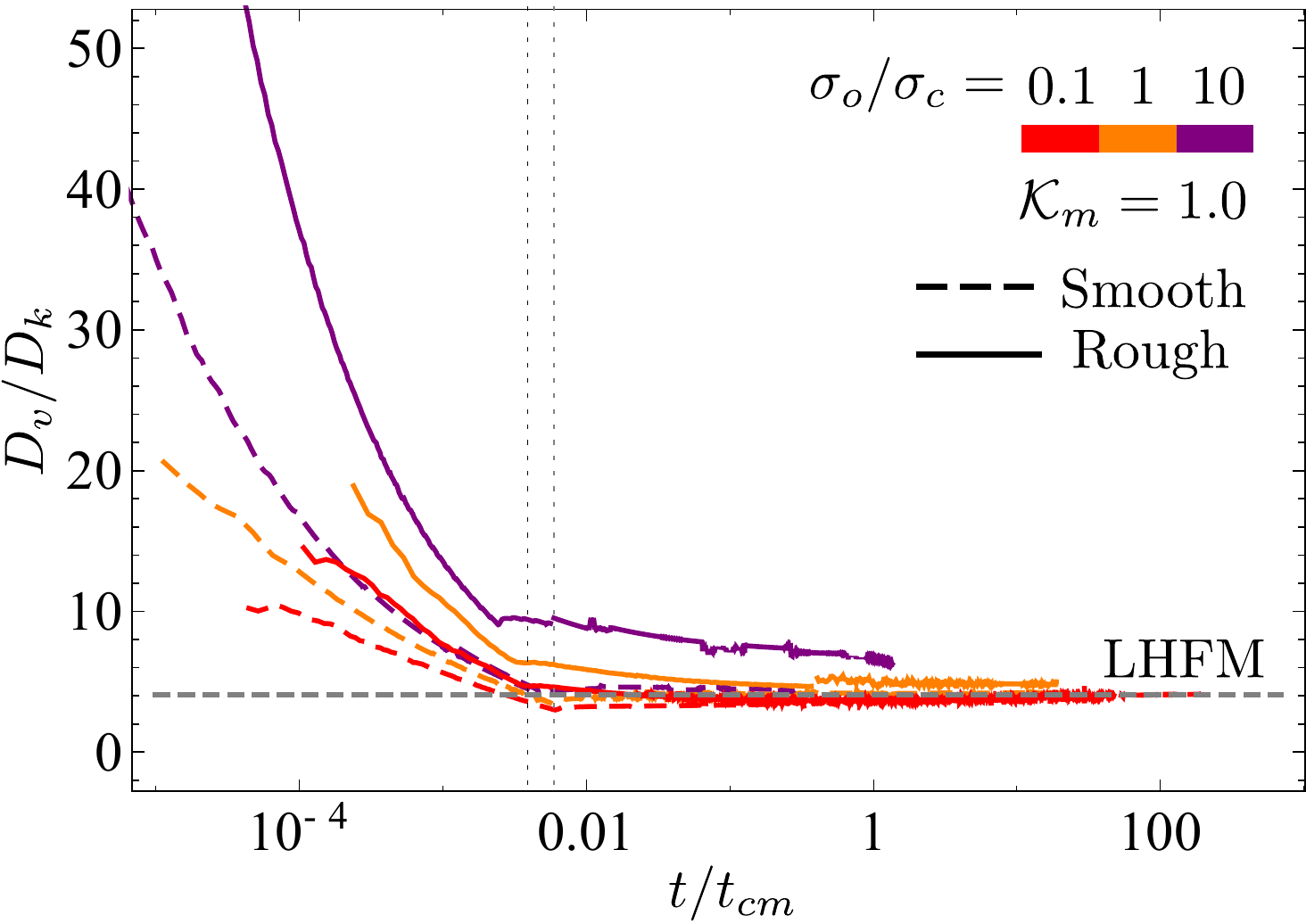}
\includegraphics[width=0.46\linewidth]{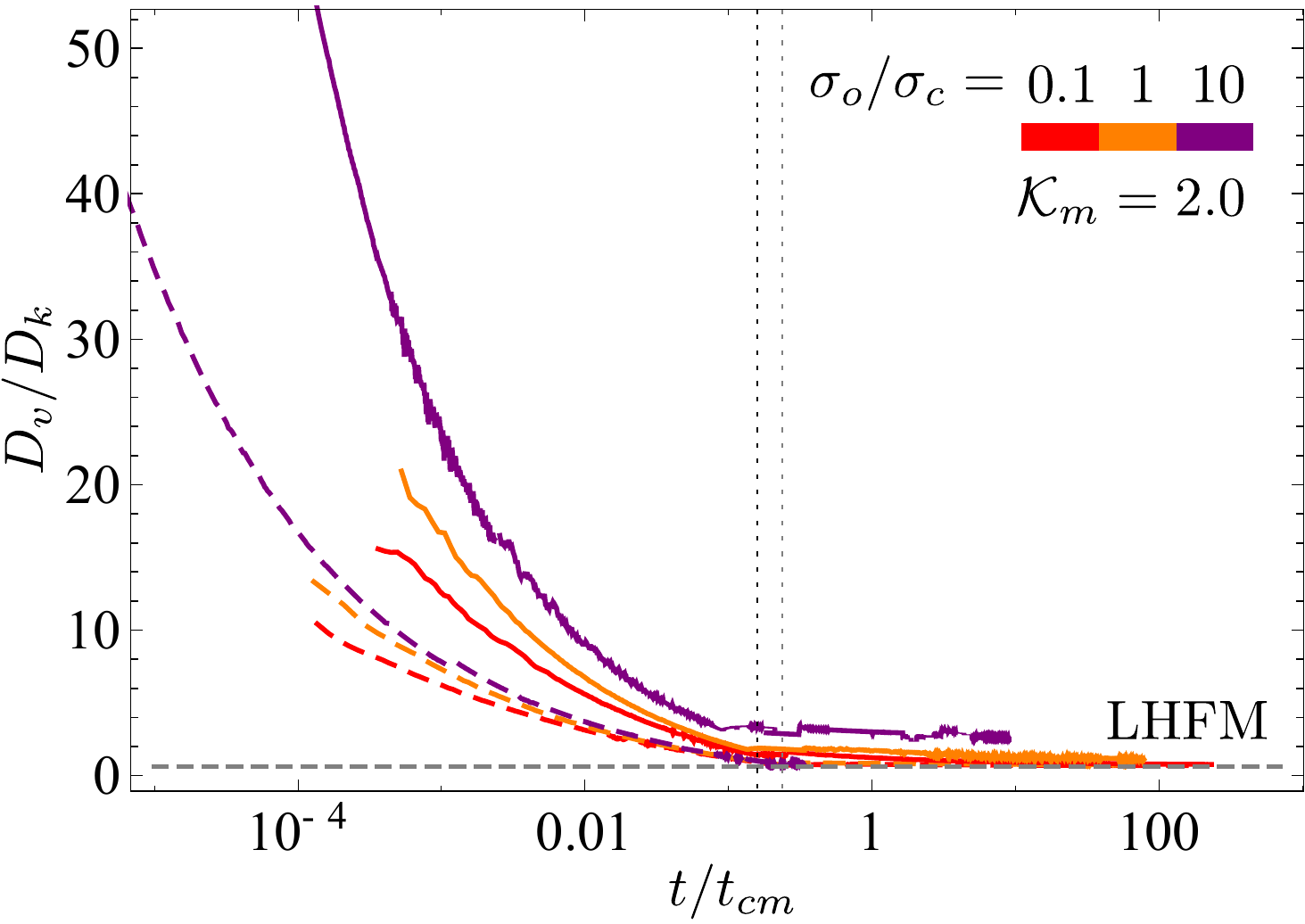}\\
\includegraphics[width=0.46\linewidth]{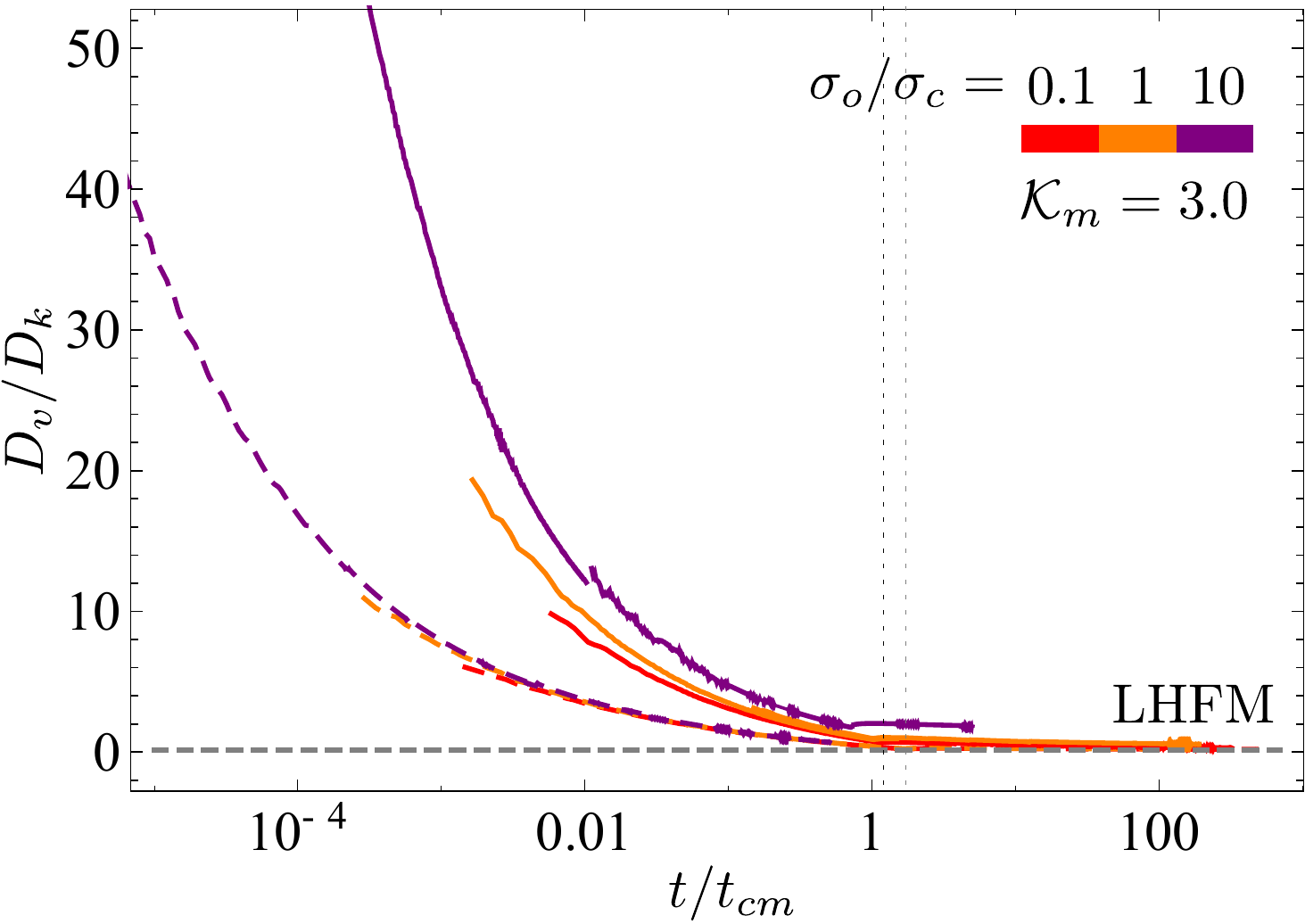}
\includegraphics[width=0.46\linewidth]{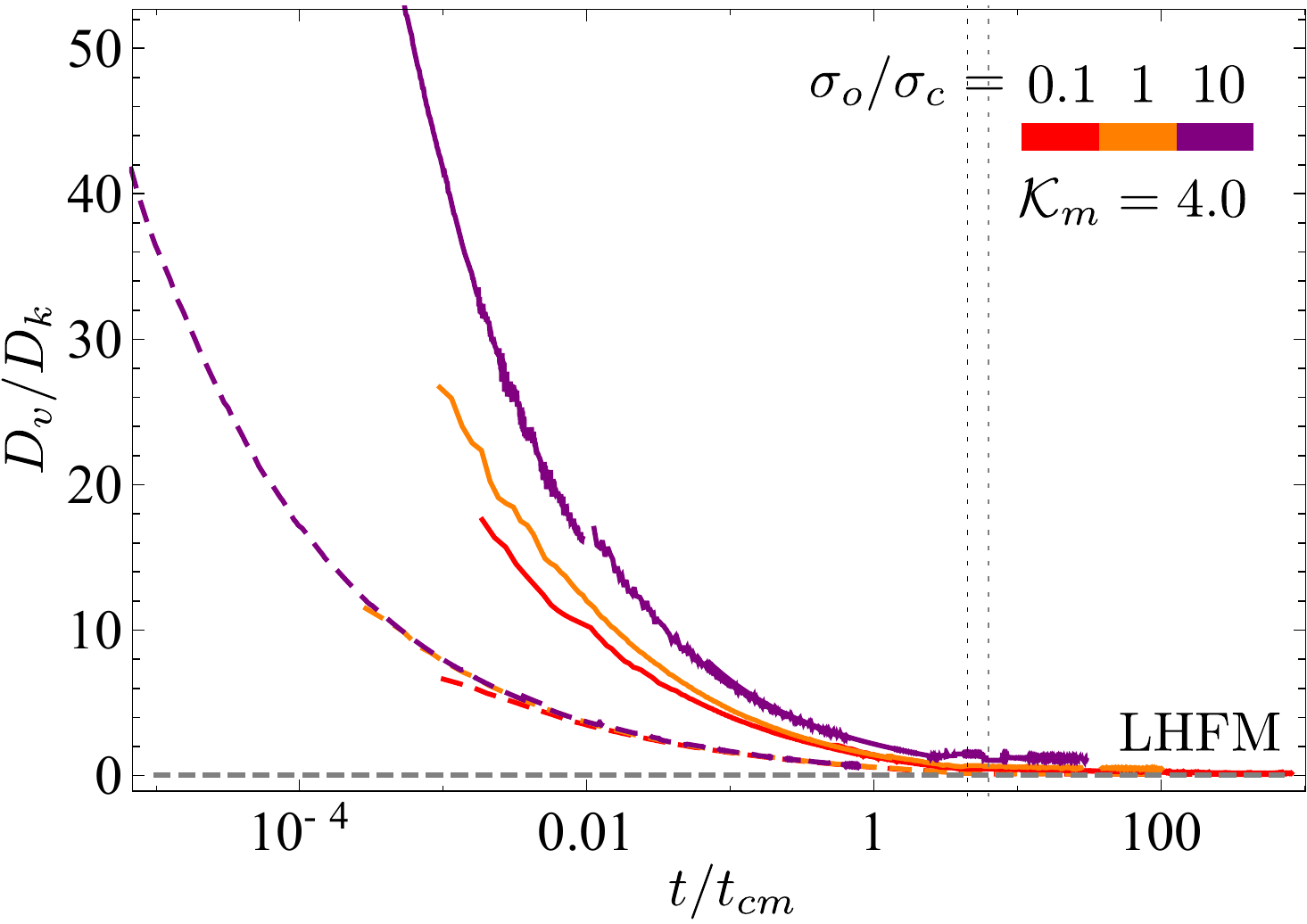}
\caption{Time evolution of the ratio of the energies dissipated in fluid viscous flow and in the creation of new fracture surfaces $D_v/D_k$ for $\mathcal{K}_m = 1-4$. The red, orange, and purple curves correspond respectively to $\sigma_o/\sigma_c=0.1, 1.0, 10$ and the solid and dashed curves correspond respectively to a rough ($\alpha_e=2$) and smooth fracture ($\alpha_e=0$). The dotted vertical lines indicate the cohesive zone nucleation period of $\sigma_o/\sigma_c=0.1$ for a smooth (gray) and a rough (black) fracture. The gray horizontal lines are the corresponding LHFM limits with zero fluid lag.}
\label{fig:viscousdissipationcm}
\end{figure}

These observations indicate an increase of the overall energy dissipated in the hydraulic fracturing process in the rough cohesive zone case. As shown in Figs.~\ref{fig:apparentdissipationcm},~\ref{fig:viscousdissipationcm}, the extra energy dissipation comes from
viscous fluid flow inside the rough cohesive zone and not from additional energy requirement to create new fracture surfaces.  
The evolution $G_{c,app}$ is not fundamentally different, with actually a smaller maximum at nucleation compared to the smooth cohesive zone case (Fig.~\ref{fig:apparentdissipationcm}).
The ratio $D_v/D_k$ of the energies dissipated in fluid viscous flow and in the creation of new fracture surfaces is significantly larger than the smooth and LHFM cases in the nucleation and intermediate stages (Fig.~\ref{fig:viscousdissipationcm}), especially for larger $\sigma_o/\sigma_c$. However, the $D_v/D_k$ ratio converges toward the LHFM limit at very large time $(t \gg t_{cm})$.

Fracture aperture roughness has an impact on the fracture growth only when the fluid front is located within the cohesive zone ($\ell-\ell_f < \ell_{coh}$). 
For small dimensionless toughness and stress ratio, the fluid lag is larger or just slightly smaller than the cohesive zone length after nucleation (see for example the $\mathcal{K}_m=1, \sigma_o/\sigma_c=0.1$ case). As a result roughness has little effect and the growth is similar to the smooth case in the intermediate stage of growth.
A larger dimensionless confining stress level or/and larger dimensionless toughness facilitates the penetration of the fluid front into the cohesive zone and results in additional fluid viscous dissipation due to the roughness. 

At large time, the cohesive zone and fluid lag size becomes much smaller than the overall fracture length such that the effect of roughness on growth is significantly reduced. 
The large time trend for $\sigma_o/\sigma_c=1$ (for all toughness) both in terms of length, width, pressure (see Figs.~\ref{fig:lengthcm},~\ref{fig:widthcm},~\ref{fig:pressurecm}) as well as energy (Figs.~\ref{fig:apparentdissipationcm})
 hints that the growth of a rough cohesive fracture  tends to LHFM limits at sufficiently large time, similarly than for the smooth case. However, the time at which fracture growth finally follows the LHFM prediction appears much larger than $t_{cm}$ especially for larger $\mathcal{K}_m$ and $\sigma_o/\sigma_c$.

%%%%%%%%%%%%%%%%%%%%%%%%%%%%%%%%%%%%%%%
\section{Discussions}
%%%%%%%%%%%%%%%%%%%%%%%%%%%%%%%%%%%%%%%

\subsection{Implications for HF at laboratory and field scales}

\begin{table}[]
    \centering
    \begin{tabular}{cccccc}
    \hline
    & Fracturing fluid & $\mu^{\prime}$ (Pa.s) & $Q_o$ (m$^2$/s) & $\sigma_o$ & Injection duration (s) \\%$t_{om}$ (s)  \\
    \hline
    Lab injection (1) & Silicone oil  & $12 \times 1000$ & $1.0\times 10^{-9}$ & 3 & 600-1800\\%$4.0\times 10^5$ \\
Lab injection (2) & Glycerol & $12\times 0.6$ & $1.0\times 10^{-9}$ & 0.3 & 30-1800\\%$2.4\times 10^4$ \\
Micro-HF test & Slick water & $12\times 0.005$ & $1.0\times 10^{-5}$ & 30 & 60-240\\%0.002 \\
Well stimulation & Slick water & $12\times 0.005$ & $1.0\times 10^{-3}$ & 30 & 1800-7200\\%0.002 \\
    \hline
    & $\mathcal{K}_m$ & $\sigma_o/\sigma_c$  & $t_{cm}$ (s) & $t_c$ (s) & $L_{coh}$ (m)\\
    \hline
Lab injection (1) & 0.88 & 1.0 & $4.0\times 10^{5}$ & $\approx 1.6\times 10^{3}$ & 0.3 \\
Lab injection (2) & 5.6 & 0.1 & $2.4\times 10^{2}$ & $> 1.1\times 10^{3}$ & 0.3 \\
Micro-HF test & 1.8 & 10 & 2.0 & $\approx 0.19$ & 0.3 \\
Well stimulation & 0.6 & 10 & 2.0 & $< 4.9\times 10^{-3}$ & 0.3 \\
\hline
    \end{tabular}
    \caption{Examples of characteristic scales for laboratory and field scale HF injection. We report the corresponding time $t_c$ and length scale $L_{coh}$ for  nucleation in the rough cohesive zone case ($\alpha_e=2$).}
    \label{tab:labscalings}
\end{table}

To gauge the implications for real systems, we consider typical values relevant to  laboratory and field scales hydraulic fractures in oil/gas bearing shale/mudstone formation. 
These rocks have a large range of reported tensile strength ($2-12$ MPa - \cite{RyRe2015}), elastic modulus ($4-30$ GPa - \cite{RyRe2015}) and fracture toughness (0.18-1.43 MPa.m$^{1/2}$ - \cite{ChMe2016}). 
We assume in what follows $\sigma_c=3$ MPa, $G_c=45$ N/m, $w_c=30\mu$m and $E^{\prime}=30$ GPa. We report the corresponding characteristic scales and dimensionless numbers for different type of injection  in Table.~\ref{tab:labscalings}.

Laboratory HF tests are performed on finite size samples $L_s$ (typically with $L_s$ at most half a meter) with a minimum confining stresses either smaller or on par with the material cohesive stress ($\sigma_o/\sigma_c \approx 0.1-2$).
In the case where the sample dimension $L_s$ is smaller or of the order of the characteristic scale of the cohesive zone $L_{coh}$, laboratory HF tests will only span the nucleation and intermediate stages of growth,  and as a result will strongly deviate from LHFM predictions. 
If $L_s$ is sufficiently larger than $L_{coh}$ ($L_s \gtrsim 10 L_{coh}$), the fracture growth will possibly converge to LHFM solutions at late time for small $\mathcal{K}_m$ (Lab injection (1) case in Table.~\ref{tab:labscalings}). Nevertheless, it will still  present significant deviations from LHFM solutions in the inlet width and net pressure (see Figs.~\ref{fig:widthcm}, ~\ref{fig:pressurecm}) for larger $\mathcal{K}_m$ values (Lab injection (2) case in Table.~\ref{tab:labscalings}).  

In-situ HF operations are performed at depth (anything from 1.5 to 4 km), and as a result the ratio $\sigma_o/\sigma_c$ is always much larger than unity ($\sigma_o/\sigma_c \sim 10$ or even larger).
We evaluate the characteristic scales by assuming injection of slick water in micro-HF tests and well stimulation operations (see Table.~\ref{tab:labscalings}). A micro-HF test (typically performed at a small injection rate) is characterized by a dimensionless toughness $\mathcal{K}_m$ around two. 
Based on the results presented previously,
significant deviations from  LHFM predictions are expected in that case with a fracture length shorter by about 15\% (see Fig.~\ref{fig:lengthcm}), a fracture opening larger by about 20\% (Fig.~\ref{fig:widthcm}), and a net pressure larger by about 40\% (Fig.~\ref{fig:pressurecm}) after less than a minute of propagation ($t \sim 100 t_{cm}$). For well stimulation applications, the fracture growth will converge toward the LHFM predictions after few minutes  thanks to the smaller dimensionless toughness resulting from the larger injection rate. 
This convergence will be delayed for deeper injections / larger $\sigma_o/\sigma_c$. 
One should bare in mind that very different responses can be encountered as function of rocks properties (notably of $w_c$, $\sigma_c$) and in-situ stress conditions.

\subsection{Limitations and possible extensions of the current study}
We have used a simple linear-weakening cohesive zone model to simulate the fracture process zone and a phenomenological correction to Poiseuille's law (assuming $w_R=w_c$) to account for the effect of aperture roughness on fracture hydraulic conductivity. 
These choices are actually the simplest possible ones, and 
may be oversimplified.
More advanced traction-separation relations 
with both a non-linear hardening and softening branch  
are often found to better reproduce experimental observations of fracture growth in quasi-brittle materials  \citep{PaPa2011,Need2014}. Similarly, the precise relation between the width scale of solid non-linearity $w_c$ and that of the one related to the flow deviation $w_R$ remains to be investigated from experiments.
These more detailed descriptions of the fracturing process will likely modify the hydraulic fracturing growth at the early and intermediate stages. However, the scaling and qualitative structure of HF growth presented here will remain similar.

Our results indicate a convergence of HF growth in quasi-brittle materials toward LHFM predictions at large time, even though the  investigation of the parametric space reported here is only partial due to the extremely significant numerical cost of the simulation in the vanishing lag size limits as time increases.
The numerical difficulty results from the requirement of a sufficiently fine mesh to resolve the shrinking fluid lag at large time as well as the small tensile zone ahead of the tip which significantly decreases for large $\sigma_{o}/\sigma_{c}$. 
An adaptive mesh refinement scheme must be developed to ensure a sufficiently fine resolution of the process zone and fluid lag in order to further investigate fracture growth for large $\sigma_o/\sigma_c$ cases.

% radial HF
The discussions and results presented here pertain to a plane-strain geometry,  but can be extended to the axisymmetric geometry \citep{LiLe2019b,Gara2019}.
For a radial cohesive HF, the energy dissipation in the creation of fracture surfaces increases with the extent of fracture perimeter.
In particular, the dimensionless toughness $\mathcal{K}_m$ 
increases with time as $\mathcal{K}_m = (t/t_{mk})^{1/9}$ \citep{SaDe2002}, with 
\begin{equation}
    t_{mk}=\frac{E^{\prime 13/2}Q_o^{3/2} \mu^{\prime 5/2}}{K^{\prime 9}}
\end{equation}
This introduces another time-scale into the problem besides $t_{om}$ and $t_{cm}=t_{om}\times (\sigma_o/\sigma_c)^3$.   
As a result, the exact growth of a radial cohesive HF will be impacted by the ratio $t_{cm}/t_{mk}$, or in other words by the competition between hydro-mechanical effects associated with nucleation and the overall transition toward the late-time toughness dominated regime.
The results of \cite{Gara2019} obtained using an equation of motion based on the solution of a steadily moving HF provides an estimate of the propagation, but should be taken with caution as this approach does not necessarily ensure that the cohesive zone length is smaller than the fracture length. Additional quantitative investigation of the radial cohesive HF are left for further studies.

\section{Conclusions}
 
We have investigated the growth of a plane-strain HF in a quasi-brittle material using a cohesive zone model including the effect of aperture roughness on fluid flow. In parallel to the cohesive zone, it is necessary to account for the presence of a fluid lag to ensure that both the fluid pressure and stresses in the near tip region remain finite. Resolving with sufficient accuracy these potentially small regions near the fracture tip renders the problem extremely challenging numerically.

We have shown that a plane-strain cohesive HF presents three distinct stages of growth: a nucleation phase, an intermediate phase during which the results slowly converge toward linear hydraulic fracture mechanics (LHFM) predictions in a third stage. The overall solution is characterized by a cohesive zone nucleation time scale $t_{cm}=E^{\prime 2}\mu^{\prime}/\sigma_{c}^{3}$, a dimensionless fracture toughness $\mathcal{K}_{m}$ (whose definition is similar to the LHFM case) and the ratio between in-situ and material cohesive stress $\sigma_{o}/\sigma_{c}$. In addition, the enhanced flow dissipation associated with fracture roughness significantly influences the solution as it re-inforces the hydro-mechanical coupling in the near tip region.

After the nucleation stage, for large $\mathcal{K}_{m}$, the effect of $\sigma_{o}/\sigma_{c}$ for a smooth cohesive zone case is not significant when the solutions tend toward the LHFM predictions. This convergence toward LHFM occurs at later $t/t_{cm}$ for larger $\mathcal{K}_{m}$. For small $\mathcal{K}_{m}$, the fluid lag diminishes faster for larger $\sigma_{o}/\sigma_{c}$ and the convergence to LHFM occurs for smaller $t/t_{cm}$ as a result.

Roughness significantly modifies the convergence toward LHFM notably for dimensionless toughness larger than 1. In addition, for these large toughness cases, larger $\sigma_{o}/\sigma_{c}$ results in larger deviations and a much slower convergence toward the LHFM predictions (which now occur for orders of magnitude of the nucleation time scale $t_{cm}$). Fracture roughness leads to additional energy dissipation in the viscous fluid flow associated with the fluid penetration in the cohesive zone. This ultimately results in larger openings, larger net pressures, shorter fracture extension and thus larger input energy. This additional viscous dissipation is further amplified for larger $\sigma_{o}/\sigma_{c}$, which facilitates the penetration of the fluid in the rough cohesive zone.
It is also worth noting that counter-intuitively the effect is stronger and remains in effect longer for larger dimensionless toughness: the viscous pressure drop localizes to an even smaller region near the tip for larger $\mathcal{K}_{m}$  such that viscous flow dissipation increases as a result.

The theoretical predictions presented here now need to be tested experimentally on well characterized quasi-brittle materials. This is particularly challenging as one must ensure that the sample size is at least ten times larger than the characteristic cohesive zone length $L_{coh}=E^\prime w_c /\sigma_c$ in order to hope capturing the convergence toward LHFM predictions. It is actually worth noting that so far all the quantitative experimental validations of linear hydraulic fracture mechanics have been obtained on transparent and/or model materials - all with very small process zone sizes (see \cite{LeDe2017} and references therein). HF experiments in rocks need to ensure a quantitative measurement of the time evolution of the fracture and fluid fronts, as well as fracture opening. This is possible via active acoustic imaging \citep{LiLe2020}. However, the accurate spatiotemporal imaging of the process zone of a growing hydraulic fracture under realistic stress conditions
remains truly challenging.

\subsection*{Acknowledgement}
The authors would like to thank Dmitry I. Garagash for insightful discussions at the early stage of this research.
\subsection*{CRediT Authors contributions}
\noindent Dong Liu: Conceptualization, Methodology, Formal analysis, Software, Investigation, Validation, Visualization, Writing – original draft. \\
Brice Lecampion: Conceptualization, Methodology, Supervision, Writing – review \& editing.

\appendix

\section{Energy balance\label{sec:energy}}
Following \cite{LeDe2007}, we write the energy balance of a propagating cohesive HF by combining the energy dissipation in the fluid and solid.
The external power provided by injecting fluid at a flow rate $Q_o$, under the inlet pressure $p_{f}(x=0, t)$, is balanced by the rate of work expended by the fluid on the walls of the fracture and by viscous dissipation. Hence,
\begin{equation}
Q_o p_{f}(x=0, t)=  2\int_0^{\ell_f} p_f \frac{\partial w}{\partial t} \text{d}x - 2\int_0^{\ell_f}  q \frac{\partial p}{\partial x} \text{d}x, \quad q=-\frac{w^3}{\mu^{\prime}  f}\frac{\partial p}{\partial x}
\label{Fluidenergy}
\end{equation}
where the cavity pressure in the lag zone is neglected in the above expression. By differentiating the global continuity equation with time, 
\begin{equation}
Q_o=2\int_0^{\ell_f}  \frac{\partial w}{\partial t} \text{d}x+2\dot{\ell_f}  w(\ell_f)
\end{equation}
After multiplying the above expression by $\sigma_o$ and subtracting it from Eq.~(\ref{Fluidenergy}), we obtain an alternative form of the energy balance in the fluid,
\begin{equation}
Q_o p_{f0}= Q_o \sigma_o +2\int_0^{\ell_f} p \frac{\partial w}{\partial t} \text{d}x- 2\int_0^{\ell_f}  q \frac{\partial p}{\partial x} \text{d}x-2\sigma_o \dot{\ell_f}  w(\ell_f) 
\label{Fluidenergy2}
\end{equation}

For a fracture propagating quasi-statically in limit equilibrium in the solid, 
the fracture energy release rate is then written as the decrease of the strain energy rate and the work rate of the external forces \citep{KeSi1996}.% (Keatings et al. 1996).

\begin{equation}
\int_0^{\ell_f} p \frac{\partial w}{\partial t}  \text{d}x- \int_0^{\ell_f}  w \frac{\partial p}{\partial t}  \text{d}x- \int_{\ell_f}^{\ell}   \sigma_o \frac{\partial w}{\partial t} \text{d}x+\int_0^{\ell} \sigma_{coh} \frac{\partial w}{\partial t}  \text{d}x- \int_0^{\ell}  w \frac{\partial \sigma_{coh}}{\partial t}  \text{d}x=0
\label{Solidenergy}
\end{equation}

Eqs.~ (\ref{Fluidenergy2}) and (\ref{Solidenergy}) can be combined to yield an energy balance for the whole system. 
\begin{equation}
    P_e=Q_o p_{f0}=\dot{W}_o+\dot{W}_e+\dot{W}_l+D_k+D_v
\end{equation}
where 
\begin{equation}
\begin{aligned}
& \dot{W}_o=Q_o \sigma_o,\\
&\dot{W}_e=\int_0^{\ell_f} p \frac{\partial w}{\partial t} \text{d}x+\int_0^{\ell_f}  w \frac{\partial p}{\partial t} \text{d}x- \sigma_o \int_{\ell_f}^{\ell}  \frac{\partial w}{\partial t}\text{d}x,
\quad \\
&\dot{W}_l=2 \int_{\ell_f}^{\ell} \sigma_o \frac{\partial w}{\partial t}\text{d}x-2\sigma_o \dot{\ell_f} w(\ell_f)=2\sigma_o \frac{\text{d}}{\text{d}t} \int_{\ell_f}^{\ell} w \text{d}x,\\
&D_k=-
\int_0^{\ell}  w \frac{\partial \sigma_{coh}}{\partial t}  \text{d}x+\int_0^{\ell} \sigma_{coh} \frac{\partial w}{\partial t}  \text{d}x\\
&D_v=-2 \int_0^{\ell_f}  q \frac{\partial p}{\partial x} \text{d}x\\
\end{aligned}
\end{equation}
Using the linear-softening cohesive traction-separation law, we rewrite $D_k$ in the coordinates of a moving tip 
\begin{equation}
    D_k=\int_{\ell-\ell_{coh}}^{\ell}\sigma_c \frac{w}{w_c} \frac{\partial w}{\partial t}\text{d}x+\int_{\ell-\ell_{coh}}^{\ell}\sigma_c \left(1-\frac{w}{w_c}\right)\frac{\partial w}{\partial t}\text{d}x=\sigma_c \int_{\ell-\ell_{coh}}^{\ell}\frac{\partial w}{\partial t}\text{d}x=\sigma_c \int_0^{\ell_{coh}}\frac{\partial w}{\partial t}\text{d}\hat{x}\\
\end{equation}
where
\begin{equation}
    \hat{x}=\ell-x, \quad \frac{\partial w}{\partial t}=\left. \frac{\partial w}{\partial t}\right|_{\hat{x}}-(-\dot{\ell})\frac{\partial w}{\partial \hat{x}}
\end{equation}
The energy dissipation during the fracturing process $D_k$ can be thus simplified as follows
\begin{equation}
    D_k=\sigma_c \int_0^{\ell_{coh}}\left. \frac{\partial w}{\partial t}\right|_{\hat{x}}\text{d}\hat{x}+\sigma_c \dot{\ell}\int_0^{\ell_{coh}}\frac{\partial w}{\partial \hat{x}}\text{d}\hat{x}=\sigma_c \int_0^{\ell_{coh}}\left. \frac{\partial w}{\partial t}\right|_{\hat{x}}\text{d}\hat{x}+  2\dot{\ell}(\frac{1}{2}\sigma_c  w(\hat{x}=\ell_{coh}))
\end{equation}
\section{Numerical scheme accounting for the nucleation of a cohesive zone and a fluid lag\label{sec:algorithm}}
As suggested in \cite{LiLe2019}, the problem is solved numerically via a fully implicit scheme based on the boundary element method. We automatically nucleate the fluid lag using the Elrod-Adams lubrication cavitation model at the early stage of fracture growth \citep{MoSh2018}. We then switch to a level-set algorithm for computational efficiency by precisely tracking the fluid front \citep{GoDe2011}.

\subsection{Fluid-lag-nucleation algorithm}
We initiate the fracture aperture from the solution of a static elastic fracture under a uniform fluid pressure slightly larger than $\sigma_o$. For a given fracture length increment, the solution is obtained using three nested iterative loops: we start from a trial time step and solve the fluid pressure for all elements inside the fracture using a quasi-Newton method. Such a procedure is repeated until each element in the fracture reaches a consistent state: either fluid or vapor. A converged estimate of the cohesive forces is then updated using fixed-point iterations with under-relaxation. The time step is finally adjusted in an outer loop using a bi-section and secant method to fulfill the propagation criterion.

\paragraph{Elasticity}
\begin{equation}
\mathbb{A}\mathbf{w}=\mathbf{p}_{f}-\mathbf{\sigma}_{coh}(\mathbf{w})-\sigma_{o}
\end{equation}
where $\mathbb{A}$ is the elastic matrix obtained via the discretization of the elastic operator using the displacement discontinuity method with piece-wise constant elements, and $\mathbf{p}_{f}, \sigma_{o}, \sigma_{coh}$ are respectively vectors of the fluid pressure, minimum compressive stress and cohesive forces.

\paragraph{Elrod-Adams lubrication}

A state variable $\theta$ is introduced in the mass conservation, characterizing the percentage of liquid occupying the fracture within one element. All the elements inside the fracture fulfil the condition $p_f(1-\theta)=0$ and can be classified into three domains according to the filling condition of the element: $\eta_{p}$ (elements fully filled with fluid),  $\eta_{\theta}$  (elements partially filled with fluid) and $\eta_{ex}$ (empty or vapor elements). 
\begin{equation}
\begin{aligned}
&\eta_p=\{i\in \eta_{\Gamma} \mid  \theta_i=1, \quad p_{fi}>0\} \\
&\eta_{\theta}=\{i \in \eta_{\Gamma}\mid  0<\theta_i<1, \quad p_{fi}=0\}\\
&\eta_{ex}=\{i \in \eta_{\Gamma}\mid  i\not\in (\eta_p \cup \eta_{\theta}), \quad p_{fi}=0, \quad \theta_i=0\}
\end{aligned}
\label{eq:cavitationdomain}
\end{equation}
where $\eta_p \cap \eta_{\theta}=\emptyset$ and $ \eta_{\Gamma}=  \eta_p \cup \eta_{\theta} \cup \eta_{ex}$. We integrate the lubrication equation over element $i$:
\begin{equation}
\underbrace{\int_{i}\frac{\partial(\theta w)}{\partial t}\text{d}x}_{1}+\underbrace{\int_{i}\frac{\partial}{\partial x}\left(-\frac{w^{3}}{\mu^\prime}\frac{\partial p_{f}}{\partial x}\right)\text{d}x}_{2}-\underbrace{\frac{Q_{o}}{2}\delta_{(i,1)}}_{3}=0
\end{equation}
The first and the second terms are respectively discretized as follows,
\begin{equation}
\int_{i}\frac{\partial\theta w}{\partial t}\text{d}x=\frac{1}{\Delta t}h(\theta_{i}w_{i}-\theta_{i}^{o}w_{i}^{o})
\label{eq:l1}
\end{equation}
\begin{equation}
\begin{aligned} & \int_{i}\frac{\partial}{\partial x}\left(-\frac{w^{3}}{\mu^\prime f}\frac{\partial p_{f}}{\partial x}\right)\text{d}x=\left[-\frac{w^{3}}{\mu^\prime f}\frac{\partial p_{f}}{\partial x}\right]_{i-1/2}^{i+1/2}\\
 & =\frac{1}{\mu^\prime f_{i-1/2}}w_{i-1/2}^{3}\left(\frac{p_{f,i}-p_{f,i-1}}{h}\right)-\frac{1}{\mu^\prime f_{i+1/2}}w_{i+1/2}^{3}\left(\frac{p_{f,i+1}-p_{f,i}}{h}\right), \\
& w_{i-1/2}=\frac{w_{i}+w_{i-1}}{2},\quad w_{i+1/2}=\frac{w_{i}+w_{i+1}}{2}
\end{aligned}
\label{eq:l2}
\end{equation}
% \begin{equation}
% w_{i-1/2}=\frac{w_{i}+w_{i-1}}{2},\quad w_{i+1/2}=\frac{w_{i}+w_{i+1}}{2}
% \end{equation}
where $h$ is the element size and the superscript $o$ denotes the solution at the previous time step. 

\begin{table}
\centering
\begin{tabular}{l}
\hline
\textbf{Repeat} solving for pressure $p_{fi}$, $\theta_i$  for $i\in \eta_p \cup \eta_{\theta}$ using Newton's method;\\
$\quad$\textbf{for} $i\in \eta_{\Gamma}$ \textbf{do}\\
$\quad \quad$ \textbf{if} $p_{f,i}<0$ \textbf{then} set $p_{f,i}=0$,
$\eta_p \leftarrow \eta_p \setminus\{i\}$, $\eta_{\theta}  \leftarrow \eta_{\theta}  \cup\{i\}$, $\eta_{ex} \leftarrow \eta_{\Gamma} \setminus (\eta_p \cup \eta_{\theta})$\\
$\quad \quad$ \textbf{if} $\theta_i>1$ \textbf{then} set $\theta_i=1$,
$\eta_{\theta} \leftarrow \eta_{\theta}  \setminus\{i\}$, $\eta_p \leftarrow \eta_p \cup\{i\}$, $\eta_{ex} \leftarrow \eta_{\Gamma} \setminus (\eta_p \cup \eta_{\theta})$\\ 
$\quad \quad$ \textbf{if} $\theta_i<0$ \textbf{then} set $\theta_i=0$,
$\eta_{\theta} \leftarrow \eta_{\theta}  \setminus\{i\}$,  $\eta_{ex} \leftarrow \eta_{\Gamma} \setminus (\eta_p \cup \eta_{\theta})$\\ 
$\quad$\textbf{end}\\
\textbf{until} all constraints $p_{f,i}\geq 0$, $0\leq \theta_i \leq 1$ for $i \in \eta_{\Gamma}$ are satisfied, in other words, $p_{f,i}(1-\theta_i)=0$.\\
\hline
\end{tabular}
\caption{Algorithm using the Elrod-Adams model (adapted from \cite{MoSh2018}) within one iteration with a given cohesive force vector }
\label{lubricationalgorithm}
\end{table}

We back-substitute the elasticity into the lubrication equation and use the quasi-Newton method to solve the non-linear problem. We set the solution of the previous time step as an initial guess and
solve iteratively for $p_{f,i}(i\in\eta_{p})$ and $\theta_{i}(i\in\eta_{\theta})$.  The lag-nucleation algorithm then updates the sets of $\eta_{p}$ and $\eta_{\theta}$ as demonstrated in Table~\ref{lubricationalgorithm}.

\paragraph{Propagation condition}
In the context of a cohesive zone, we check the equality of the tensile stress component ahead of the fracture tip with the material peak strength: 
\begin{equation}
\sigma_{yy,n+1}=A_{n+1,j}w_{j}-\sigma_{o}=\sigma_{c},\quad j=1...n
\label{eq:propagation}
\end{equation}
where $n$ is the number of elements inside the fracture at the current time step.

\subsection{Fluid-front-tracking algorithm}
The fluid-front tracking algorithm \citep{GoDe2011} assumes a clear boundary between the fluid and cavity. The $n$ elements inside the fracture is divided into $m$ fluid channel elements fully-filled with fluid ($p_f>0$), $(n-m-1)$ fluid lag elements with a negligible cavitation pressure ($p_f=0$) and one partially filled element ($p_f=0$) where locates the fluid front.
By introducing a filling fraction $\phi$, we estimate the fluid front position using the solution of the lag-nucleation  / Elrod-Adams based algorithm. 
We assume that fluid-front-tracking algorithm initializes with a solution ($w^{O}, p_{f}^{O}, V^{O}, m^{O}, \phi^{O}, \ell_{f}^{O}$) obtained from the lag-nucleation / Elrod-Adams based algorithm at a chosen time step $k$. $m^{O}$ is the number of elements in the domain $\eta_{p}$. $\phi^{O}$ is the filling fraction obtained by gathering the fluid mass of all lag elements from the lag-nucleation algorithm in the partially-filled element (the $(m^{O}+1)^{\text{th}}$ element) of the fluid-front-tracking algorithm. 
\begin{equation}
\phi^{O}=\sum_{i}\theta_{i}^{k}w_{i}^{k}/w_{m^{O}+1},\quad i\in\eta_{\theta}
\end{equation}
We then obtain the fluid front position $\ell_f^O$ and the fluid front velocity $V^{O}$ for a chosen time step $k$.
\begin{equation}
\begin{aligned}
    &\ell_f^{O}=(m^O+\phi^O)h,\\
    &V^{O}=(\ell_{f,k+1}-\ell_{f,k-1})/(t_{k+1}-t_{k-1})
\end{aligned}
\end{equation}
where and $t_{k-1}$ and $t_{k+1}$ are respectively propagation time at the $(k-1)^{\text{{th}}}$ and $(k+1)^{\text{{th}}}$ time step in the lag-nucleation algorithm.

Based on this initial estimation of the fluid front, we solve iteratively the increment of the opening in the channel elements for a given fracture front through three nested loops in the fluid-front-tracking algorithm. One loop tracks the fluid front, one updates the time step to fulfill the propagation condition and another solves the non-linear system due to the cohesive forces and lubricated fluid flow through a fixed-point scheme. We present in the following the discretization of the non-linear system. 

\paragraph{Elasticity}
\begin{equation}
\mathbf{p}_{c}-\sigma_o-\mathbf{\sigma}_{cohc}=\mathbb{A}_{cw} \mathbf{w}+ \mathbb{A}_{ol} (-\sigma_o-\mathbf{\sigma}_{cohl})
\end{equation}
where $\mathbf{p}_{c}$ is the vector net pressures in the channel part of the fracture; $\mathbf{\sigma}_{cohc}$ and $\mathbf{\sigma}_{cohl}$ cohesive forces applied in the fluid channel and fluid lag. 
\begin{equation}
\begin{aligned}
\mathbb{A}_{cw}&=\mathbb{A}_{cc}-\mathbb{A}_{cl} \mathbb{A}_{ll}^{-1} \mathbb{A}_{lc}\\
\mathbb{A}_{ol}&=\mathbb{A}_{cl} \mathbb{A}_{ll}^{-1}
\end{aligned}
\end{equation}
$\mathbb{A}_{cc}, \mathbb{A}_{cl}, \mathbb{A}_{lc}, \mathbb{A}_{ll}$ are sub-matrix of the elastic matrix $\mathbb{A}$ associated with elements inside the fluid channel and lag.
\paragraph{Lubrication flow}
For fluid channel elements ($1\leq i\leq m$), 
\begin{equation}
\begin{aligned}\Delta w_{i}= & \frac{\Delta t}{\mu^\prime h^{2}}\left(\frac{1}{f_{i-1/2}}w_{i-1/2}^{3}p_{c,i-1}+\frac{1}{f_{i+1/2}}w_{i+1/2}^{3}p_{c,i+1}\right)\\
 & -\frac{\Delta t}{\mu^\prime h^{2}}(\frac{1}{f_{i-1/2}}w_{i-1/2}^{3}+\frac{1}{f_{i+1/2}}w_{i+1/2}^{3})p_{c,i}+\delta_{(i,1)}\frac{Q_{o}\Delta t}{2h}\\
 & -\delta_{(i,m)}F_{m}-H(i-m^{o})\sum_{k=m^{o}+1}^{m}\delta_{(i,k)}F_{k}
\end{aligned}
\end{equation}
The second term on the second line represents the contribution due to a constant injection rate and the two terms on the third line are mass corrections due to the partially-filled element where the fluid front locates. $H(\cdot)$ is the Heaviside step function.

\begin{equation}
F_{m}=
\begin{cases}
\phi w_{m+1}-\phi^{o}w_{m+1}^{o},\quad m=m^{o}\\
\phi w_{m+1}-\phi^{o}w_{m^{o}+1}^{o}-\sum_{i=m+1}^{m^{o}}w_{i},\quad m<m^{o}
\end{cases}
\end{equation}

\begin{equation}
F_{k}=\begin{cases}
(1-\phi^{o})w_{k}^{o},\quad k=m^{o}+1\\
w_{k}^{o},\quad k>m^{o}+1
\end{cases}
\end{equation}
where the superscript $o$ refers to the solutions at the previous time step. The lubrication equation can be thus arranged as 
\begin{equation}
\Delta\mathbf{w}=\mathbb{L}\cdot\mathbf{p}_{c}+\mathbf{S}_{1}-\mathbf{S}_{m}-\mathbf{S}_{m^{o}}
\end{equation}

\paragraph{Coupled system of equations}
We back-substitute the elasticity and write the coupled system as in Eq.~(\ref{eq:Coupledsys}). For a given fracture front and a trial time step, we solve for incremental apertures $\Delta\mathbf{w}$ using fixed-point iterations. The tangent linear system reads:
\begin{equation}
(\mathbb{I}-\mathbb{L}(\Delta\mathbf{w}^{(s-1)})\mathbb{A}_{cw})\Delta\mathbf{w}^{s}=\mathbb{L}(\Delta\mathbf{w}^{(s-1)})\mathbb{A}_{cw}\mathbf{w}^{o}\\+\mathbb{L}(\Delta\mathbf{w}^{(s-1)})\mathbb{A}_{ol}(-\sigma_{o}-\sigma_{cohl}(\Delta\mathbf{w}^{(s-1)}))
\label{eq:Coupledsys}
\end{equation}
where $s$ refers to the solution at the previous iteration.
\paragraph{Update of the fluid front position}
The fluid front position is estimated as
\begin{equation}
\ell_{f}^{(s)}=(m^{o}+\phi^{o})h+V^{(s-1)}\Delta t, m^{(s)}=\text{floor[}\ell_{f}^{(s)}/h], \phi^{(s)}=\ell_{f}^{(s)}/h-m^{(s)}
\label{eq:fluidfront}
\end{equation}
where $V$ is the fluid front velocity and it can be obtained through lubrication theories,
\begin{equation}
\begin{aligned} & V=\frac{1}{2}\left(V^{o}-\frac{1}{\mu^\prime f_m}w_{m}^{2}\frac{\partial p}{\partial x}\right),\\
 & \frac{\partial p}{\partial x}=\left(p_{c,m}-\frac{p_{c,m}+\sigma_{o}}{\phi+1/2}-p_{c,m-1}\right)/(2h),\quad m>1
\end{aligned}
\end{equation}
The iteration starts with $V^{(0)}=V^{o}$ and continues until $|(\ell_{f}^{(s)}-\ell_{f}^{(s-1)})/\ell_{f}^{(s-1)}|$ is within a set tolerance.

\paragraph{Control of overestimation of the fluid front position}
We may possibly overestimate the fluid front position using Eq.~(\ref{eq:fluidfront}) especially  when the fracture front advances too much compared to the previous time step. As a result, negative pressure may be detected in the channel elements near the fluid front.

In order to better locate the fluid front, we adopt a strategy similar to the one in \cite{GoAb2019}. Once the scheme detects a negative fluid pressure in the channel elements (where the elements are fully-filled with fluid) during the $s^{th}$ iteration at the current time step, we utilizes the bi-section algorithm to estimate the fluid front position \citep{LiLe2019b}. We set the fluid front position at the previous time step as the lower bound $\ell_{f-}=\ell_{f}^{o}$ and the current position obtained from the previous iteration as the upper bound $\ell_{f+}=\ell_{f}^{(s-1)}$. As long as the fluid front advances during the fracture growth, the trial fluid front position for the next iteration can be estimated from 
\begin{equation}
\ell_f^{(s)}=(\ell_{f+}+\ell_{f-})/2
\label{eq:bisecfluid}
\end{equation}
We iterate on $\ell_{f}$ until that $|(\ell_{f}^{(s)}-\ell_{f}^{(s-1)})/\ell_{f}^{(s-1)}|$ is within a set tolerance and that all fluid pressure in the channel elements remain positive.

\subsection{Benchmark of the growth of a linear elastic fracture}
\begin{figure}
\centering
\begin{tabular}{cc}
     \includegraphics[width=0.45\linewidth]{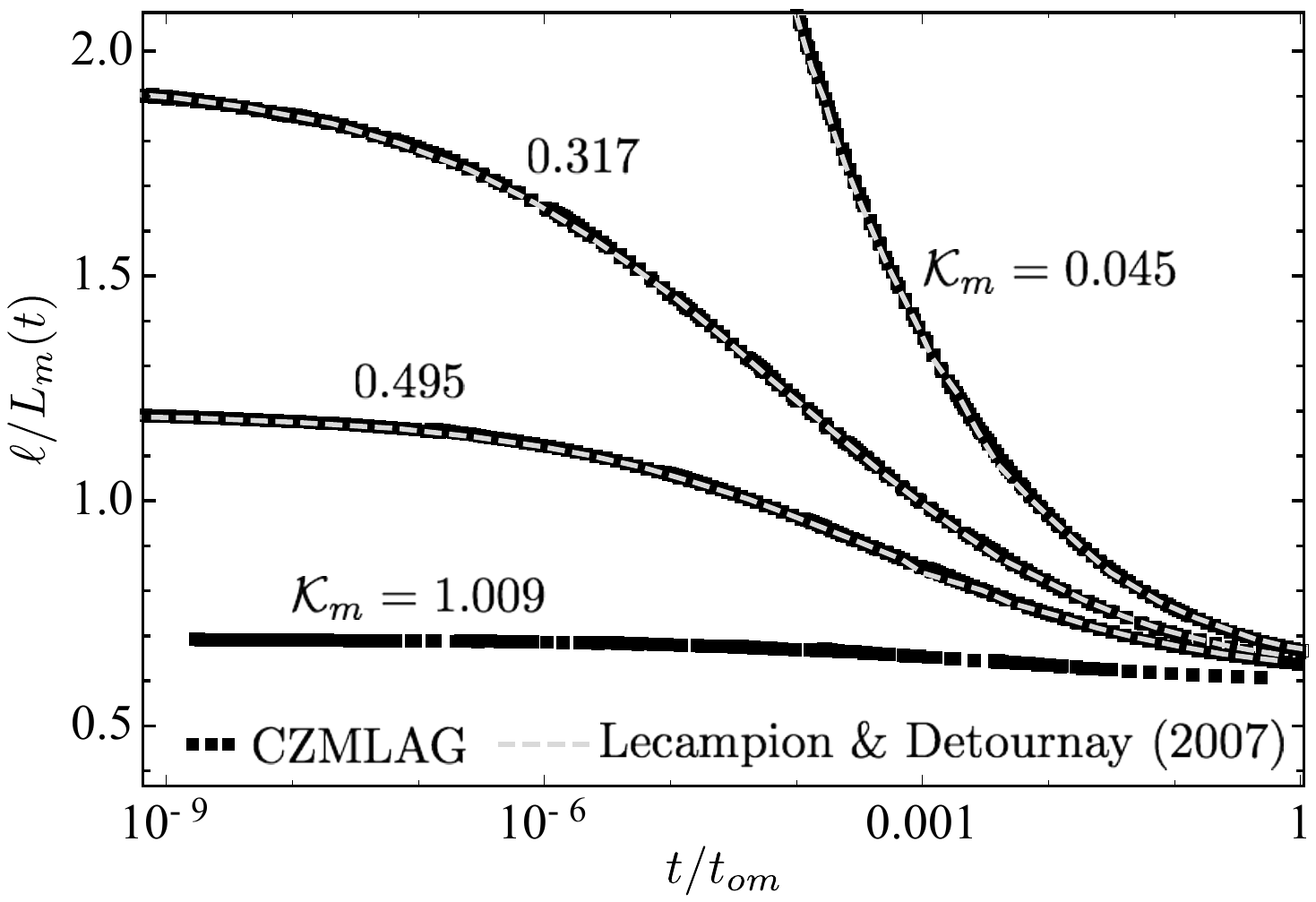} &
\includegraphics[width=0.45\linewidth]{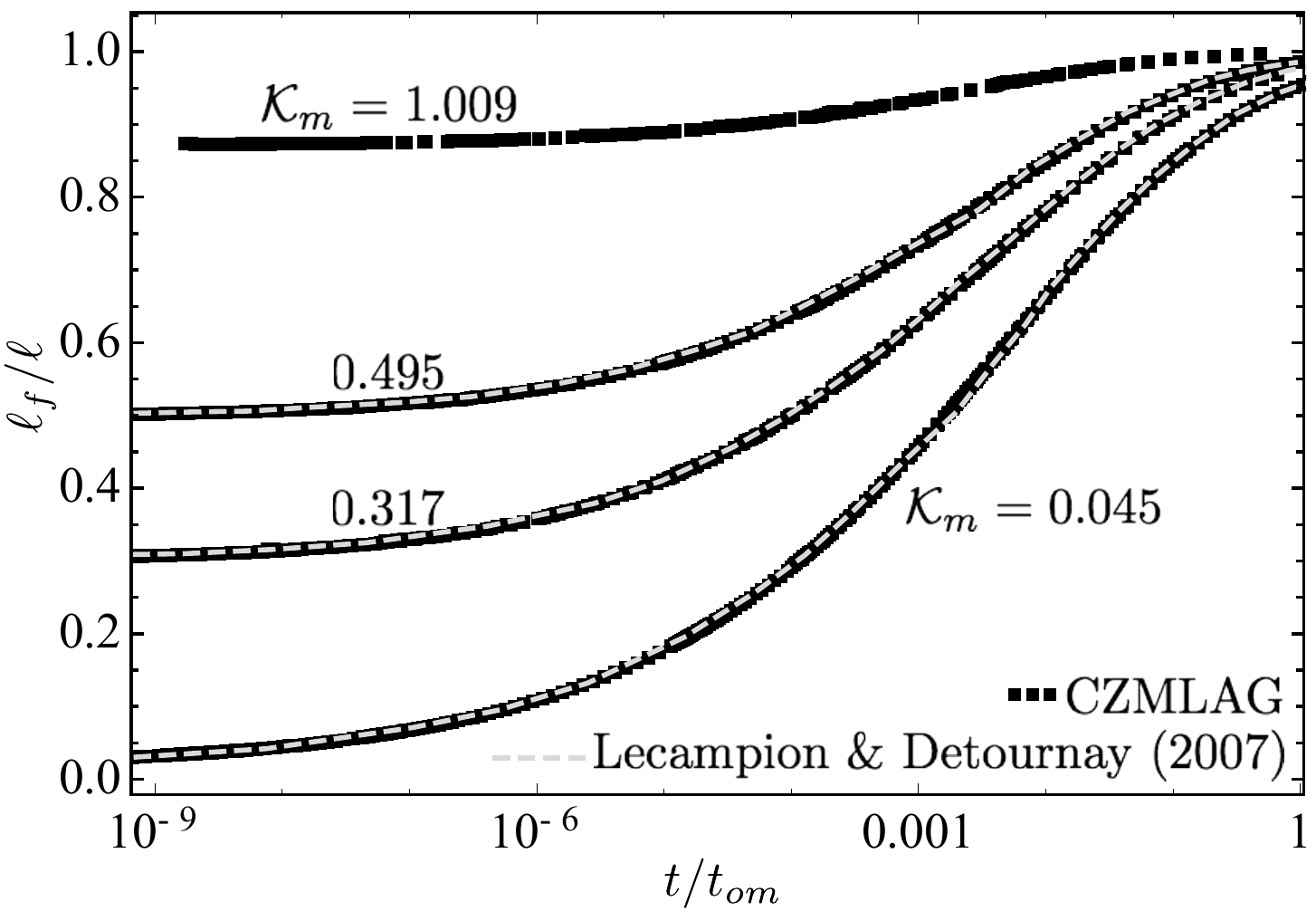}  \\
     a) & b) 
\end{tabular}
\caption{Time evolution of a) the half fracture length and b) fluid fraction in viscosity scaling for different dimensionless toughness $\mathcal{K}_{m}$.}
\label{fig:GammaXif}
\end{figure}

We simulate the growth of a plane-strain HF in a linear elastic medium by adapting the propagation condition as 
\begin{equation}
    w_{n}=\frac{2}{3}\frac{K^\prime \sqrt{h}}{E^\prime}
\end{equation}
where $w_n$ is the opening of the element closest to the fracture tip obtained by the integration of the tip asymptote. 
We benchmark our scheme using different $\mathcal{K}_{m}$ values and formulate the problem with the viscosity scaling in the time-domain $t/t_{om}$ similar to \cite{LeDe2007}. 
We show in Fig.~\ref{fig:GammaXif} that our results (CZMLAG) are in good agreement with the numerical solutions reported in \cite{LeDe2007}. 

%\bibliographystyle{elsarticle-num-names}
\bibliographystyle{model5-names}
\biboptions{authoryear} 
\bibliography{GestionbibArrange.bib}



%\bibitem{}

%%\end{thebibliography}
\end{document}